\documentstyle[11pt,aaspp4]{article}

\newcommand{\br}{$B\!-\!R$}
\newcommand{\eg}{e.g.\ }
\newcommand{\etal}{et al.\ }
\newcommand{\sn}[1]{$\times 10^{#1}$}
\newcommand{\steng}{${\rm S}_{10}({\rm V})_{\rm G2V}$ }
\newcommand{\stena}{${\rm S}_{10}({\rm V})_{\rm A0V}$ }
\newcommand{\sten}{${\rm S}_{10}({\rm V})_{\rm X}$ }
\newcommand{\bicds}{Bull.\ Inf.\ Centre Donn\'ees Stellaires}

\begin{document}

\title{Detection of Extended Red Emisson in the Diffuse Interstellar Medium}

\author{Karl D.\ Gordon\altaffilmark{1}, Adolf N.\ Witt, and 
Brian C.\ Friedmann}
\affil{Ritter Astrophysical Research Center, The University of Toledo \\
   Toledo, OH 43606}
\altaffiltext{1}{present address: Department of Physics \& Astronomy,
Louisiana State University, Baton Rouge, LA 70803}

\lefthead{Gordon, Witt, \& Friedmann}
\righthead{ERE in the Diffuse ISM}

\begin{abstract}
  Extended Red Emission (ERE) has been detected in many dusty
astrophysical objects and this raises the question: Is ERE present
only in discrete objects or is it an observational feature of all
dust, i.e.\ present in the diffuse interstellar medium?  In order to
answer this question, we determined the blue and red intensities of
the radiation from the diffuse interstellar medium (ISM) and examined
the red intensity for the presence of an excess above that expected
for scattered light.  The diffuse ISM blue and red intensities were
obtained by subtracting the integrated star and galaxy intensities
from the blue and red measurements made by the Imaging
Photopolarimeter (IPP) aboard the Pioneer 10 and 11 spacecraft.  The
unique characteristic of the Pioneer measurements is that they were
taken outside the zodiacal dust cloud and, therefore, are free from
zodiacal light.  The color of the diffuse ISM was found to be {\em
redder} than the Pioneer intensities.  If the diffuse ISM intensities
were entirely due to scattering from dust (i.e.\ Diffuse Galactic
Light or DGL), the color of the diffuse ISM would be {\em bluer} than
the Pioneer intensities.  Finding a {\em redder} color implies the
presence of an excess red intensity.  Using a model for the DGL, we
found the blue diffuse ISM intensity to be entirely attributable to
the DGL.  The red DGL was calculated using the blue diffuse ISM
intensities and the approximately invariant color of the DGL
calculated with the DGL model.  Subtracting the calculated red DGL
from the red diffuse ISM intensities resulted in the detection of an
excess red intensity with an average value of $\sim$10~\steng.  This
represents the likely detection of ERE in the diffuse ISM since
H$\alpha$ emission cannot account for the strength of this excess and
the only other known emission process applicable to the diffuse ISM is
ERE.  Thus, ERE appears to be a general characteristic of dust.  The
correlation between $N_{HI}$ and ERE intensity is $(1.43 \pm
0.31)$\sn{-29}~ergs~s$^{-1}$~\AA$^{-1}$~sr$^{-1}$~\mbox{H atom}$^{-1}$
from which the ERE photon conversion efficiency was estimated at $10
\pm 3$\%.
\end{abstract}

\section{Introduction}

\begin{figure}[thp]
\begin{center}
\plotone{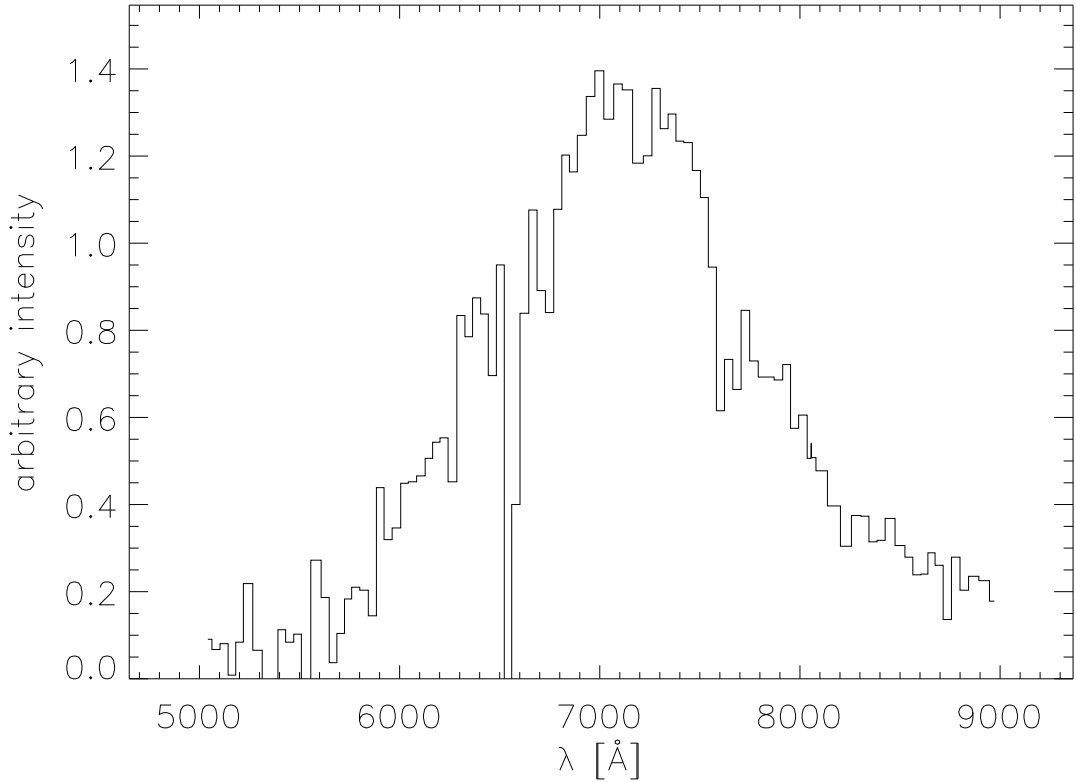}
\caption{An example of Extended Red Emission is plotted.  This
spectrum is for the reflection nebula NGC 2327 and has had the
underlying scattering continuum subtracted (Witt 1988).
\label{fig_ERE_NGC2327}}
\end{center}
\end{figure}

  Extended Red Emission (ERE) is a broad ($\Delta\lambda \sim
800$~\AA) emission band with a peak wavelength between 6500~\AA\ and
8000~\AA\ seen in many dusty astrophysical objects
(Figure~\ref{fig_ERE_NGC2327}).  For many years, the Red Rectangle was
the only object known to possess such an emission feature (see
Schmidt, Cohen, \& Margon [1980] for an excellent Red Rectangle
spectrum).  This all changed with the discovery that many reflection
nebulae possessed flux levels in the R and I bands in excess of that
expected from dust scattered starlight (\cite{wit84}; \cite{wit85},
1986).  The spectroscopic confirmation of the excess flux as ERE
(\cite{wit88}; \cite{wit89}; \cite{wit89b}; \cite{wit90}) proved that
ERE was a feature of many, but not all, reflection nebulae.  The
identification of the band as emission was strengthened by the imaging
polarimetry of NGC~7023 by Watkin, Geldhill, \& Scarrott (1991), which
showed a reduction in the R and I polarization where R and I excess
flux existed.  This convincingly proved that the excess flux is due to
an emission feature and not changes in the dust scattering properties.

   The detection of ERE in other dusty astrophysical objects quickly
followed the confirmation of ERE in reflection nebulae.  Up to the
date of this paper, ERE has been detected in the Red Rectangle
(\cite{sch80}), reflection nebulae (see above), a dark nebula
(\cite{mat79}; \cite{chl87}), Galactic cirrus clouds (\cite{guh89};
\cite{guh94}; \cite{szo98}), planetary nebulae (Furton \& Witt 1990,
1992), \ion{H}{2} regions (\cite{per92}; \cite{siv93}), a nova
(\cite{sco94}), the halo of the galaxy M82 (\cite{per95}), and the
30~Doradus nebula in the Large Magellanic Cloud (\cite{dar98}).

   Clues as to the identity of the material that produces ERE are
contained in the above observations.  The first clue comes from the
wide variety of objects that show ERE.  The material must be able to
survive in radically different environments: from cold, quiescent
environments (dark nebulae and Galactic cirrus clouds) to hot, dynamic
environments (reflection nebulae, planetary nebulae, and \ion{H}{2}
regions).  Second, the material most likely is carbonaceous since ERE
has been detected in carbon rich planetary nebulae but not oxygen rich
planetary nebulae (\cite{fur92}).  The robustness of carbonaceous dust
material is supported by the essentially constant C/H values found for
sight lines exhibiting different dust characteristics (\cite{sof97}),
implying that carbon is not easily exchanged between the gas and dust
phases of the ISM.  The third clue comes from the spatial distribution
of ERE in reflection nebulae (\cite{wit88}; \cite{wit89};
\cite{wit89b}; \cite{wit90}; \cite{rog95}; \cite{lem96}) and planetary
nebulae (\cite{fur90}).  The ERE is strongest in regions where H$_2$
is being dissociated, leading to the conclusion that warm atomic
hydrogen increases the efficiency of ERE luminescence.  So, the
material that produces ERE must be robust, contain carbon, and produce
ERE more efficiently in the presence of atomic hydrogen.

  A prime candidate for this material is hydrogenated amorphous carbon
(HAC), which was first proposed to explain the ERE in the Red
Rectangle (\cite{dul85}).  With the discovery of ERE in reflection
nebulae and other dusty objects, the identification of HAC with ERE
has strengthened (\cite{dul88}; \cite{wit88}; \cite{dul90};
\cite{wit94}).  The identification of HAC with ERE is strongly
supported by the laboratory work of Furton \& Witt (1993).  They have
shown that the low or non-existent photoluminescence of previously
annealed HAC or pure amorphous carbon can be greatly enhanced by
exposure to atomic hydrogen and/or ultraviolet radiation.  This
corresponds to the strengthening of ERE in H$_2$ dissociation regions
discussed above.  Other ERE producing materials have been proposed,
such as filmy quenched carbonaceous composite (\cite{sak92}) and
C$_{60}$ (\cite{web93}).  For a discussion of the similarities and
differences between these materials see Papoular \etal (1996).

  As ERE has been detected in a large range of dusty objects, the
question arises: Is ERE present in the diffuse interstellar medium
(ISM)?  If the answer is yes, this would make ERE a general
characteristic of dust.  A positive answer has been claimed by Duley
\& Whittet (1990), who identified the very broad structure (VBS) seen
in extinction curves (\eg \cite{van81}) as due to ERE.  Jenniskens
(1994) has pointed out that ERE cannot be the cause of the VBS for two
reasons.  First, spectra showing VBS have had the nearby sky
subtracted, effectively removing any emission from the diffuse ISM.
Second, the VBS strength does not depend on the size of the aperture
used.  Hence, ERE has not been detected in the diffuse ISM.  As a
result, dust models for the diffuse ISM have not used the ability to
produce ERE as a constraint on possible dust grain materials (\eg
\cite{kim96}; \cite{mat96}; \cite{zub96}; \cite{dwe97}; \cite{li97}).

  This investigation is aimed at determining whether ERE is present in
the diffuse ISM.  Detecting ERE in the diffuse ISM is much more
difficult than doing the same in a discrete object.  For a discrete
object, ERE detection is done by subtracting a nearby sky spectrum
from a spectrum of the object (\eg \cite{wit90}).  Subtracting a
nearby sky spectrum removes contributions to the object spectrum from
the Earth's atmosphere (airglow), zodiacal light (dust scattered
sunlight), and Galactic background light (diffuse ISM, faint stars,
and galaxies).  This results in a spectrum with contributions only
from the object being studied, and any ERE is directly attributable to
that object.  As the light from the diffuse ISM is part of the sky
spectrum, this method will not work for it.  A different method is
required.

  Two of the strongest (and most difficult to model) sources in a sky
spectrum are airglow and zodiacal light (\cite{tol81}).  Both of these
sources can be avoided by simply taking observations outside the
atmosphere (for airglow) and the zodiacal dust cloud (for zodiacal
light).  Such measurements have already been carried out by the
Imaging Photopolarimeters (IPP) carried aboard both Pioneer 10 and 11
(\cite{pel73}; \cite{wei74}).  The IPP measured the intensity of
almost the entire sky in the blue (437 nm) and the red (644 nm).  By
using only measurements taken when the Pioneer spacecraft were beyond
3.27~AU, contributions from the zodiacal light are avoided
(\cite{han74}).  Therefore, the only known sources contributing to the
IPP measurements are stars, galaxies, and the diffuse ISM.  Using
photometric star and galaxy catalogs, the contribution to the IPP
measurements from stars and galaxies can be removed.  The resulting
blue and red intensities are due only to the diffuse ISM.

  While the IPP measurements have given all-sky maps at two
wavelengths and not a spectrum, this is sufficient to detect ERE in
the diffuse ISM.  The presence of ERE in reflection nebulae was first
detected by observing that these objects had red fluxes in excess of
that expected from dust scattered starlight (\cite{wit84}).  The
diffuse ISM is a gigantic reflection nebula with the Galaxy's
starlight scattered by the Galaxy's dust.  Therefore, we can use the
same criterion, excess red flux, to detect ERE in the diffuse ISM.
The scattered light in the diffuse ISM is termed Diffuse Galactic
Light (DGL).  The DGL will have a bluer color than the integrated
starlight because scattering by dust is more efficient at shorter
wavelengths.  So, if the diffuse ISM color (red/blue ratio) is as red
as or redder than the integrated starlight and other sources of excess
red light can be positively excluded, ERE is present.

  Section~\ref{sec_pioneer} describes the Pioneer IPP measurements and
the construction of the blue and red all-sky maps.  The compilation of
a star and galaxy photometric catalog, complete to approximately 20th
magnitude, is detailed in section~\ref{sec_counts}.  The detection of
ERE in the diffuse ISM is contained in section~\ref{sec_detect_red}.
Section~\ref{sec_ere_correlations} presents the properties of the ERE
in the diffuse ISM.  Finally, section~\ref{sec_discussion} discusses
the implications of our results and summarizes our conclusions.

\section{Pioneer Data \label{sec_pioneer}}

  One of the instruments onboard the Pioneer 10 and 11 spacecraft was
the Imaging Photopolarimeter (IPP).  The primary objectives of the IPP
were to produce blue and red maps of the brightness and polarization
of the zodiacal dust cloud from 1 to 5 AU, the background light
outside the zodiacal dust cloud, and Jupiter (\cite{pel73}).  Of
these, we were concerned with only the all-sky blue and red surface
brightness maps taken outside the zodiacal dust cloud.

  The IPP was a Maksutov-type f/3.4 telescope with an aperture of 2.54
cm and a detector consisting of a Wollaston prism, multilayer filters,
and two dual-channel Bendix channeltrons (\cite{pel73}; \cite{wei74}).
Simultaneous measurements were made of the orthogonal components of
the electric field in both the blue and red.  The spectral bandpass
(half-power) was 3950--4850 \AA\ for the blue channel and 5900--6900
\AA\ for the red channel (\cite{pel73}).  See
subsection~\ref{sec_transformations} for more information on the
photometric characteristics of the IPP.  The IPP instantaneous field
of view (FOV) was $2\fdg 29 \times 2\fdg 29$ for the background light
measurements.  The IPP was mounted on a movable arm and 64
measurements were taken during a single rotation of the Pioneer
spacecraft.  The angle between the arm and the Pioneer spacecraft spin
axis (look angle = L) was changed in increments of $1\fdg 83$ to build
up a map of the sky.  The look angle ranged between $29\arcdeg$ and
$170\arcdeg$ (\cite{pel73}). The effective FOV of the measurements was
$2\fdg 29 \times (2\fdg 29 + 5\fdg 625\sin L)$, with a maximum of
$2\fdg 29 \times 7\fdg 92$ when the look angle was $90\arcdeg$ and a
minimum of $2\fdg 29 \times 3\fdg 27$ when the look angle was
$170\arcdeg$.  At each look angle, a 20 data roll (rotation)
measurement cycle was performed with 8 rolls for the background light,
1 for a radioisotope-activated phosphor source ($^{14}$C), 1 for
offset and dark current levels, and 10 for data readout (\cite{pel73};
\cite{wei74}; \cite{tol81}).

  The raw IPP background sky measurements were processed to produce
the Pioneer 10/11 Background Sky data set available from the National
Space Science Data Center (NSSDC).  The details of the processing can
be found elsewhere (\cite{wei74}; \cite{tol81}; \cite{wei81};
\cite{sch97}).  A brief description of the processing follows.  First,
the data were calibrated using the inflight measurements of the
radioisotope-activated phosphor source.  Second, the FOV center was
computed from the spacecraft spin axis direction, the look angle, and
the clock angle.  Third, the contribution from bright stars was
subtracted using the stars in the Bright Star Catalogue (\cite{hof91})
and stars with $m_V < 8$ (\cite{tol87}) in the Photoelectric Catalog
(\cite{bla68}; \cite{och74}).  Fourth, 37 resolved stars were used to
determine the time decay of the instrument sensitivity and corrections
to the telescope pointing.  The final error in the positions of the
FOVs was on the order of $0\fdg 15$--$0\fdg 40$.  The Pioneer 10 red
data have abnormally high noise, but as we are also using Pioneer 11
data, this did not adversely affect our results.  The final Pioneer
data are expressed in \steng units, the equivalent number of 10th
magnitude (V band) solar-type stars per square degree.  See
subsection~\ref{sec_transformations} for details of this unit.

  During the cruise portion of the Pioneer 10 and 11 missions, the IPP
mapped the background light a number of times.  In order to determine
the spatial extent of the zodiacal dust cloud, Hanner \etal (1974)
examined the brightness of two different regions of the sky as seen by
the IPP when Pioneer 10 was between 2.41 and 4.82~AU.  They found that
the brightness of these two regions stopped changing after Pioneer 10
passed 3.27~AU, making this distance the outermost detectable edge of
the zodiacal dust cloud.  Therefore, all the measurements taken beyond
3.27~AU are free from detectable zodiacal light and useful for this
investigation.

\begin{figure}[thp]
\begin{center}
\plottwo{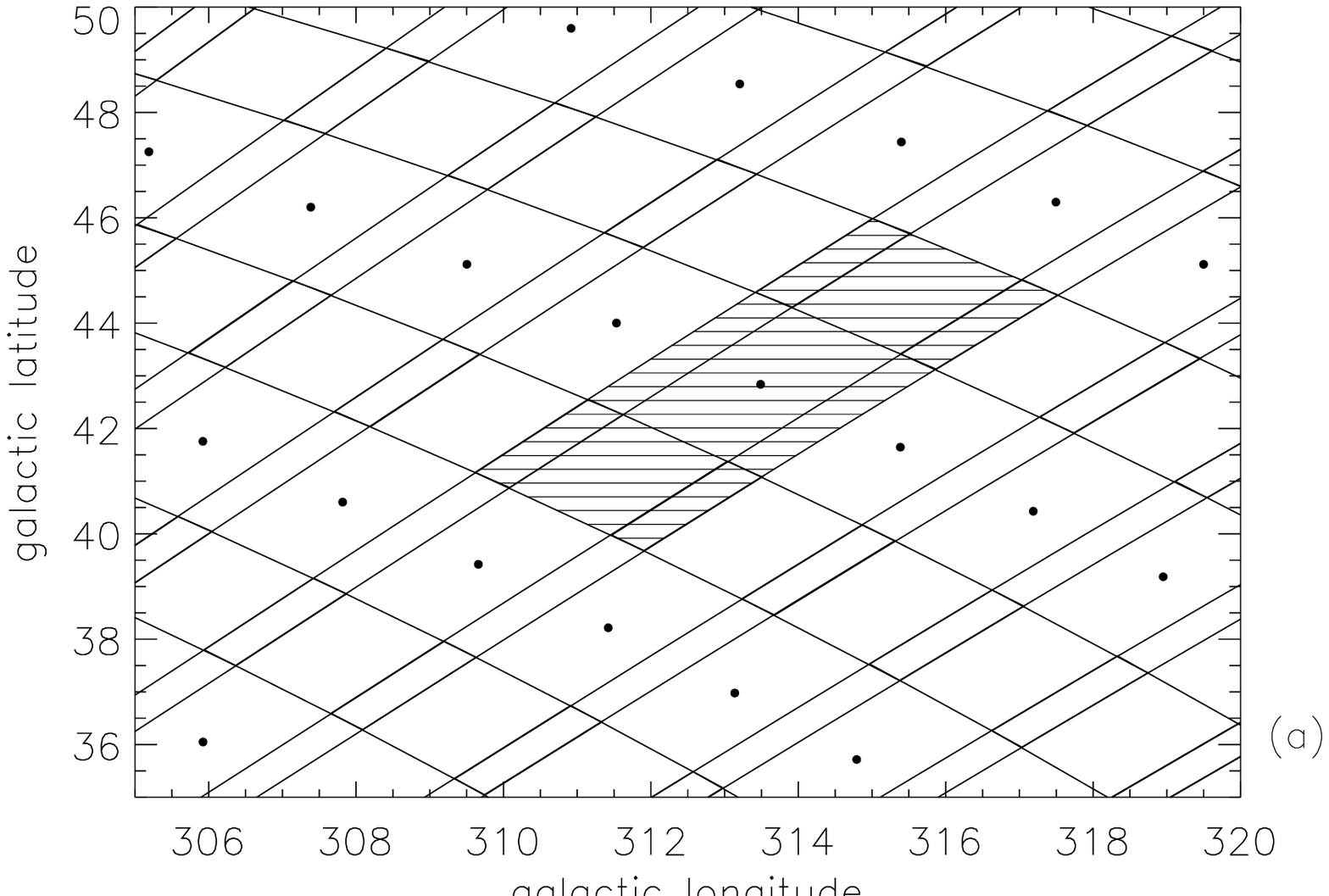}{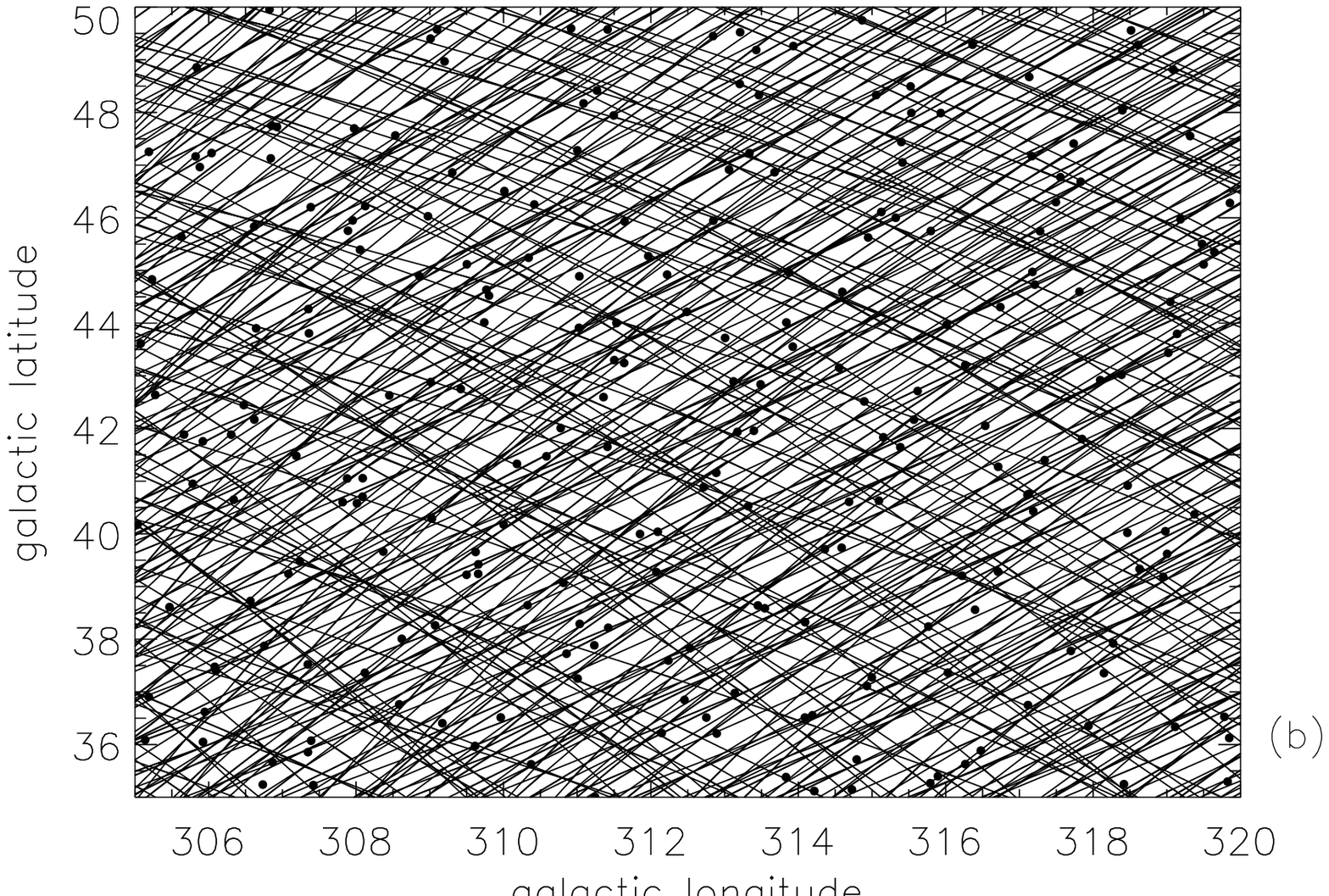}
\caption{The pattern of the FOVs are plotted in these two figures.
The pattern of FOVs from Pioneer 10 on day 68 of 1974 is shown in (a)
with points in the center of the FOVs.  One FOV is shaded to show the
regions which overlap neighboring FOVs.  The pattern of FOVs from the
11 days used in constructing the final maps is shown in (b).  Due to
the different trajectories of Pioneer 10 and 11 and the variable
spacecraft spin axis orientation, the FOVs from different days do not
overlap exactly.
\label{fig_fov_pattern}}
\end{center}
\end{figure}

  On 5 days while Pioneer 10 was between 3.26 and 5.15~AU and on 6
days while Pioneer 11 was between 4.06 and 4.66~AU, the IPP mapped the
background light in the sky.  The resolution of a map made on a single
day is determined by the FOV of the IPP and its overlap with
neighboring FOVs.  Figure~\ref{fig_fov_pattern}a gives an example of
the pattern of FOVs using the Pioneer 10 measurements from day 68 of
1974.  One of the FOVs has been shaded to show the overlap of a FOV
with neighboring FOVs.  The resolution of this map is variable, with
each parallelogram being a resolution element.  In order to actually
achieve this theoretical resolution, an algorithm must be used to
extract the information in the overlapping regions.  In fact, the
resolution of the IPP measurements can be increased significantly by
using measurements made on different days.
Figure~\ref{fig_fov_pattern}b gives the pattern of FOVs for the 11
days used in creating the final high-resolution maps (see below).  The
FOVs from different days do not overlap exactly as the Pioneer 10 and
11 missions were launched on different trajectories and the spin axis
of each spacecraft changed direction slowly as a function of distance
from the Sun.

\subsection{Map Generation Algorithm}

  The algorithm used to create the final maps is similar to that by
Aumann, Fowler, \& Melnyk (1990).  Their algorithm is called the
Maximum Correlation Method (MCM) and it was able to improve the
resolution of {\em IRAS} maps by a factor of $\sim6.5$, from
$\sim4\arcmin$ to $\sim36\arcsec$.  Our algorithm was similar to MCM
and worked in the following manner.  An initial guess at the final
image (zeroth iteration) was taken as a positive flat image with
$0\fdg 25 \times 0\fdg 25$ pixels.  The next iteration was calculated
from
\begin{equation}
p_{ij}^{k+1} = \left( \frac{1}{N} \sum_{m=1}^N C_m \right) p_{ij}^k
\end{equation}
where $p_{ij}^k$ is the surface brightness of the $k$th iteration
image at pixel coordinates ($i$,$j$), $C_m$ is the correction factor
for the $m$th IPP measurement (the surface brightness in particular
FOV) which includes $p_{ij}$, and the sum was done over the $N$ IPP
measurements which include $p_{ij}$.  The value of $C_m$ was
calculated from
\begin{equation}
C_m = I_m \left( \frac{1}{Q} \sum^Q p_{ij}^k \right)^{-1}
\end{equation}
where $I_m$ is the $m$th IPP measurement and the sum was done over the
$Q$ pixels which are included in the $m$th IPP FOV.  The error,
$\sigma_{ij}^k$, in $p_{ij}^k$ was calculated from
\begin{equation}
\sigma_{ij}^k = \frac{1}{N} \sqrt{\sum_{m=1}^N \left( I_m - \left[
  \frac{1}{Q}\sum^Q p_{ij}^k \right]_m \right)^2}.
\end{equation}
With each iteration, the image gives an improved match to the IPP
measurements.  The outcome of this algorithm is to produce an image
which describes all 11 days of the Pioneer measurements.

\begin{deluxetable}{ccrc}
\tablewidth{0pt}
\tablecaption{IPP Usable Days \label{table_11days}}
\tablehead{\colhead{Spacecraft} & \colhead{year} & \colhead{day} &
           \colhead{R\tablenotemark{a}} \\
           \colhead{} & \colhead{[years]} & \colhead{[days]} &
           \colhead{[AU]} }
\startdata
Pioneer 10 & 1972 & 354 & 3.26 \nl
Pioneer 10 & 1973 & 149 & 4.22 \nl
Pioneer 10 & 1973 & 237 & 4.64 \nl
Pioneer 10 & 1973 & 279 & 4.81 \nl
Pioneer 11 & 1974 &  57 & 3.50 \nl
Pioneer 10 & 1974 &  68 & 5.15 \nl
Pioneer 11 & 1974 & 106 & 3.81 \nl
Pioneer 11 & 1974 & 148 & 4.06 \nl
Pioneer 11 & 1974 & 178 & 4.22 \nl
Pioneer 11 & 1974 & 236 & 4.51 \nl
Pioneer 11 & 1974 & 267 & 4.66 \nl
\enddata
\tablenotetext{a}{Sun-spacecraft distance, R, taken from NSSDC WWW pages.}
\end{deluxetable}

  The 11 days of IPP background light measurements that were used in
constructing the final high-resolution maps are tabulated in
Table~\ref{table_11days}.  While the 11 days overall possessed usable
data, a large number of individual measurements were seen to be of
poor quality.  There are a number of sources for the poor quality
data: incorrect subtraction of bright stars, scattered sunlight, and
corrupt data rolls (\cite{tol81}).  The poor quality data were removed
from consideration using four criteria.  First, data contaminated with
scattered sunlight (data taken within $70\arcdeg$ of the sun for
Pioneer 10 and within $45\arcdeg$ for Pioneer 11) were removed.
Second, all data with negative values were removed as these were the
result of interuptions in the datastream of between the spacecraft and
ground station.  Third, the data were divided into $5\arcdeg \times
5\arcdeg$ boxes and data inside each box deviating over 3 standard
deviations from the average in either their blue measurements, red
measurements, or red/blue ratio were removed.  Fourth, a small number
of points were removed by visual inspection.  Approximately 25\% of
the IPP measurements were of poor quality.

\begin{figure}[tbp]
\begin{center}
\plotone{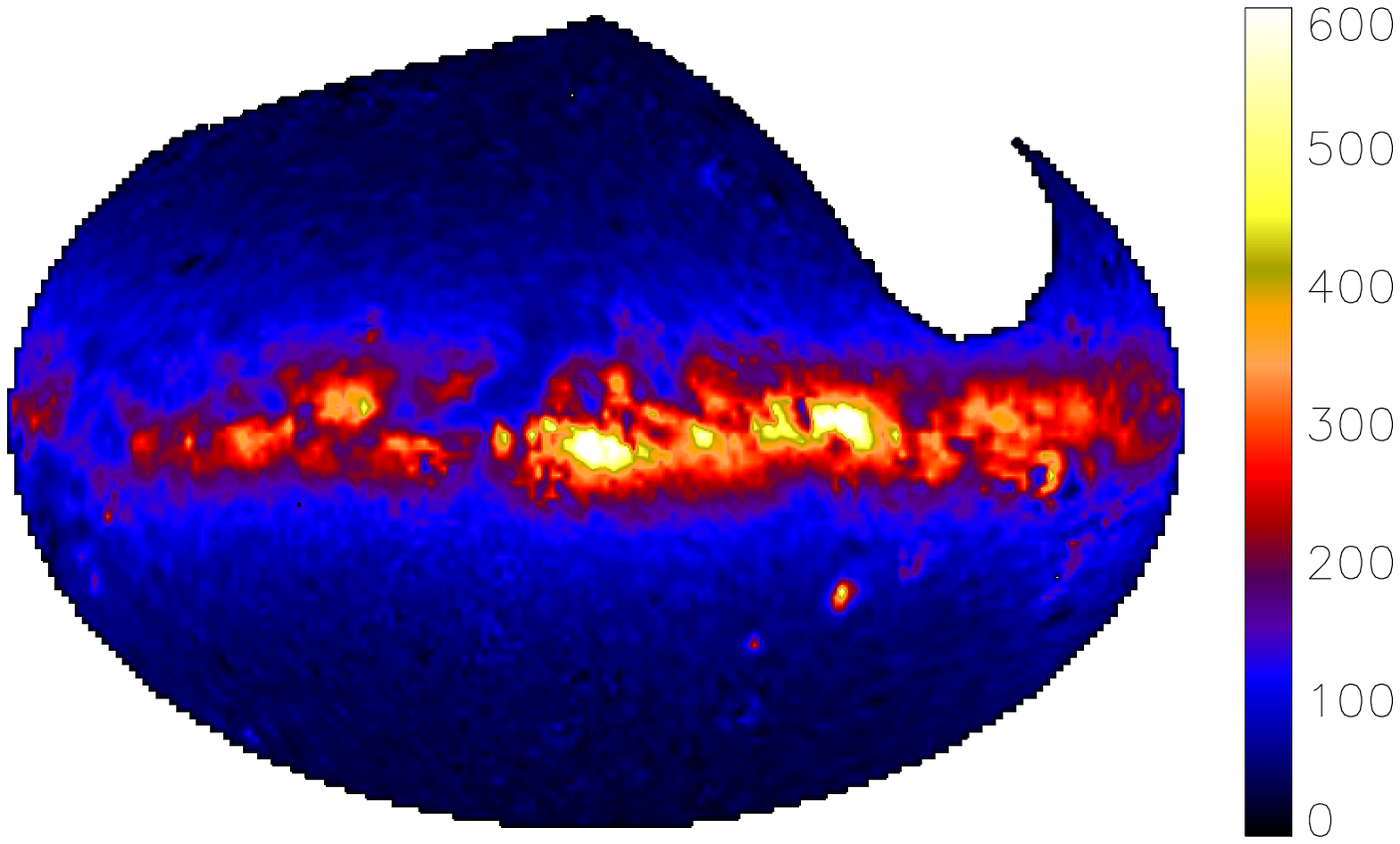}
\caption{The Aitoff projection (galactic longitude of zero in the
center) of the Pioneer blue image of the sky is displayed.  The
resolution of the displayed map is $0\fdg 5$ by $0\fdg 5$.  The large
hole corresponds to the Sun's location as seen from Pioneer 10/11.
The intensity units are \steng.  \label{fig_B_aitoff}}
\end{center}
\end{figure}

  The resulting good data were used as the input for the algorithm
described above to produce the final high-resolution maps.  The
algorithm was iterated 10 times, until little change was detected.
The best iteration map to use depended on the region being
investigated.  For low galactic latitude regions where the amplitude
of real structure in the maps is much larger than the noise amplitude,
the 10th iteration gave the best map.  For high galactic latitude
regions where the real structure amplitude is smaller than the noise
amplitude, the 1st iteration gave the best map.

  The 10th iteration map for the blue is displayed in
Figure~\ref{fig_B_aitoff}.  The 10th iteration red map is similar to
the blue map.  Typical uncertainties were 2\% for the blue and 3\% for
the red.  Figure~\ref{fig_B_aitoff} can be compared directly with the
blue background as seen by the Hipparcos star mapper.  A map of this
background is presented in Figure~6 of \cite{wic95}.  The comparison
is good both in overall strength and morphology.  From this
comparison, the uniqueness of the Pioneer maps was quite apparent as
the Pioneer blue map lacks the substantial zodiacal light seen in the
Hipparcos star mapper blue map.  

  As we were only concerned with high latitude regions, we will use
the 1st iteration map for the rest of this paper and save the higher
iterations for later work.  The 1st iteration blue and red maps are
just smoother versions of the 10th iteration maps.

\section{Photometric Star and Galaxy Counts \label{sec_counts}}

  In order for this investigation to succeed, the contribution to the
Pioneer blue and red measurements from stars and galaxies fainter than
$m_V = 6.5$ was needed.  We have tackled this problem by constructing
a Master Catalog from three separate catalogs, each complete in a
subset of the range between 6.5 and $\sim$20th magnitude.  Ironically,
the stars and galaxies with magnitudes between 12 and $\sim$20
(Palomar O \& E) have the best photometric data available due to the
existence of the Automated Plate Scanner Catalog of the Palomar Sky
Survey I (APS Catalog, \cite{pen93}).  In the magnitude range between
9 and 15 ($\sim$V band in the north and $\sim$B in the south), the
Guide Star Catalog (GSC, \cite{las90}; \cite{rus90}; \cite{jen90})
provides data in only one band.  For the magnitude range between 6.5
and 9.5, there exists no good complete photometric catalog.  We have
used a combination of catalogs (see \S\ref{ss_master_cat}) to
construct a Not So Bright Star Catalog (NSBS Catalog) to give the best
currently available positions and magnitudes for stars with magnitudes
between 6.5 and 9.5.

\subsection{Transformations Between Photometric Systems
\label{sec_transformations}}

\begin{figure}[tbp]
\begin{center}
\plotone{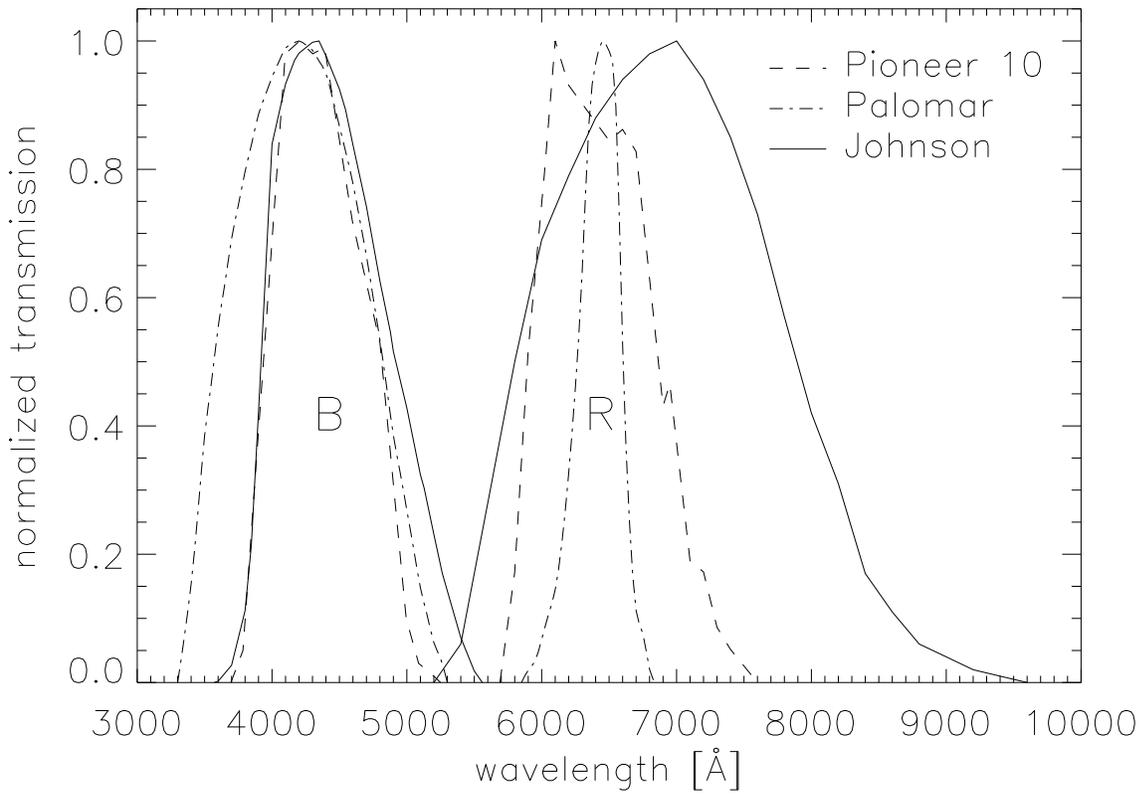}
\caption{The response curves are plotted for the Pioneer blue and red
channels (PB \& PR, Toller 1981), the Johnson B and R (Lamla 1982) and
the Palomar blue and red (O \& E, Minkowsi \& Abell 1963).
\label{fig_resp_curves}}
\end{center}
\end{figure}

  Underlying the construction of the Master Catalog was the
transformation between the Palomar blue \& red (O \& E) magnitudes and
Johnson B \& R magnitudes to Pioneer blue \& red (PB \& PR)
magnitudes.  The normalized response curves, $R(\lambda)$, for the
blue and red bands of all three photometric systems are shown in
Figure~\ref{fig_resp_curves} (\cite{min63}; \cite{lam82};
\cite{tol81}).  The Palomar blue (O) response function was computed
for an airmass of 1.5 (\cite{hay75}) in order to reproduce the
transformation between the Palomar and Johnson systems used in
calibrating the APS Catalog (\cite{hum91}).  For all 6 above bands as
well as the Johnson V band, the band's equivalent wavelength
($\lambda_{\rm eq}$), equivalent bandpass ($\Delta\lambda_{\rm eq}$),
zero magnitude flux (F$_{\lambda}$), and the intensity corresponding
to \steng and \stena units were computed and are tabulated in
Table~\ref{table_mag_details}.  The values of $\lambda_{\rm eq}$ and 
$\Delta\lambda_{\rm eq}$ are obtained by 
\begin{equation}
\lambda_{\rm eq} = \frac{\int \lambda R(\lambda) \, d\lambda} {\int
   R(\lambda) \, d\lambda}
\end{equation}
and
\begin{equation}
\Delta\lambda_{\rm eq} = \frac{\int R(\lambda) \, d\lambda}{R_{\rm
max}},
\end{equation}
respectively.  The flux corresponding to a magnitude of zero,
F$_{\lambda}$, in each band was calculated by summing the product of
the band's response curve and a calibrated spectrum of $\alpha$ Lyrae
(\cite{tug77}).  The calibrated spectrum of $\alpha$ Lyrae was
multiplied by 1.028 before use to account for the fact that $\alpha$
Lyrae's V magnitude is 0.03 (\cite{hof91}).  The intensity
corresponding to one \steng unit and one \stena unit was computed by
summing the product of each band's response curve with the spectrum,
set to 10th magnitude in the V band, of the sun (\cite{loc92}) and
$\alpha$ Lyrae (\cite{tug77}), respectively.  One \sten unit is
defined as intensity equivalent to one 10th V magnitude star of
spectral type X per square degree where X is either G2V or A0V.  The
intensity in mag/$\sq\arcsec$ corresponding to \steng units in the B
bands, V band, and R bands is 28.5, 27.8, and 27.5 mag/$\sq\arcsec$,
respectively. 

\begin{deluxetable}{ccrrcll}
\tablewidth{0pt}
\tablecaption{Photometric Band Details \label{table_mag_details}}
\tablehead{\colhead{System} & \colhead{Band} & \colhead{$\lambda_{\rm eq}$} &
           \colhead{$\Delta\lambda_{\rm eq}$} & 
           \colhead{F$_{\lambda}$\tablenotemark{a}} &
           \colhead{\steng} & \colhead{\stena} \\
           \colhead{} & \colhead{} & \colhead{[\AA]} & \colhead{[\AA]} &
           \colhead{[ergs cm$^{-2}$ s$^{-1}$ \AA$^{-1}$]} &
           \multicolumn{2}{c}{[ergs cm$^{-2}$ s$^{-1}$ \AA$^{-1}$ sr$^{-1}$]}}
\startdata
Johnson &  B & 4467 & 1014 & 6.632\sn{-9} & 1.198\sn{-9}  & 2.174\sn{-9} \nl
Palomar &  O & 4249 & 1168 & 6.343\sn{-9} & 1.087\sn{-9}  & 2.080\sn{-9} \nl
Pioneer & PB & 4370 &  826 & 6.997\sn{-9} & 1.192\sn{-9}  & 2.294\sn{-9} \nl
Johnson &  V & 5553 &  881 & 3.639\sn{-9} & 1.193\sn{-9}  & 1.193\sn{-9} \nl
Johnson &  R & 6926 & 2057 & 1.950\sn{-9} & 8.813\sn{-10} & 6.394\sn{-10} \nl
Palomar &  E & 6412 &  386 & 2.289\sn{-9} & 9.828\sn{-10} & 7.505\sn{-10} \nl
Pioneer & PR & 6441 &  968 & 2.305\sn{-9} & 9.919\sn{-10} & 7.558\sn{-10} \nl
\enddata
\tablenotetext{a}{F$_{\lambda}$ is the flux corresponding to a
magnitude of zero.  See text for details.}
\end{deluxetable}

  The transformations from the Palomar and Johnson systems to the
Pioneer system were accomplished by means of the above band response
functions (Figure~\ref{fig_resp_curves}) and zero magnitude fluxes
(Table~\ref{table_mag_details}) along with an observational grid of
stellar spectra spanning the Hertzsprung-Russell diagram
(\cite{sil92}).  This grid consists of spectra covering 3510-8930~\AA\
with a resolution of 11~\AA\ and includes 72 spectral types spanning
spectral classes O--M and luminosity classes I--V.  Most of the
spectra are for solar metallicity stars, but some are for metal-rich
and metal-poor stars.  The spectra were dereddened and stars of
similar spectral types were averaged to produce the final 72 spectral
type spectra (\cite{sil92}).

\begin{figure}[tbp]
\begin{center}
\plottwo{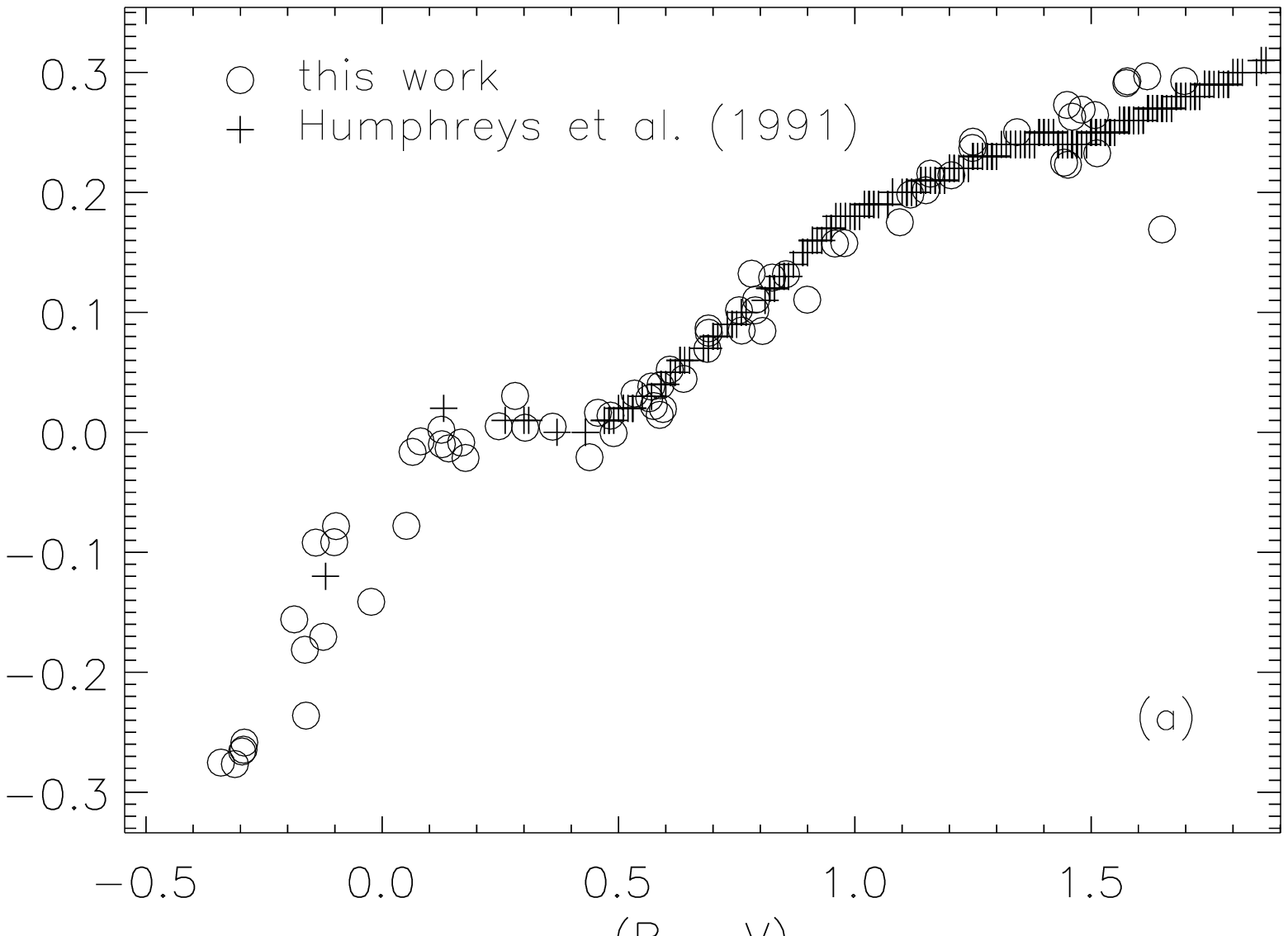}{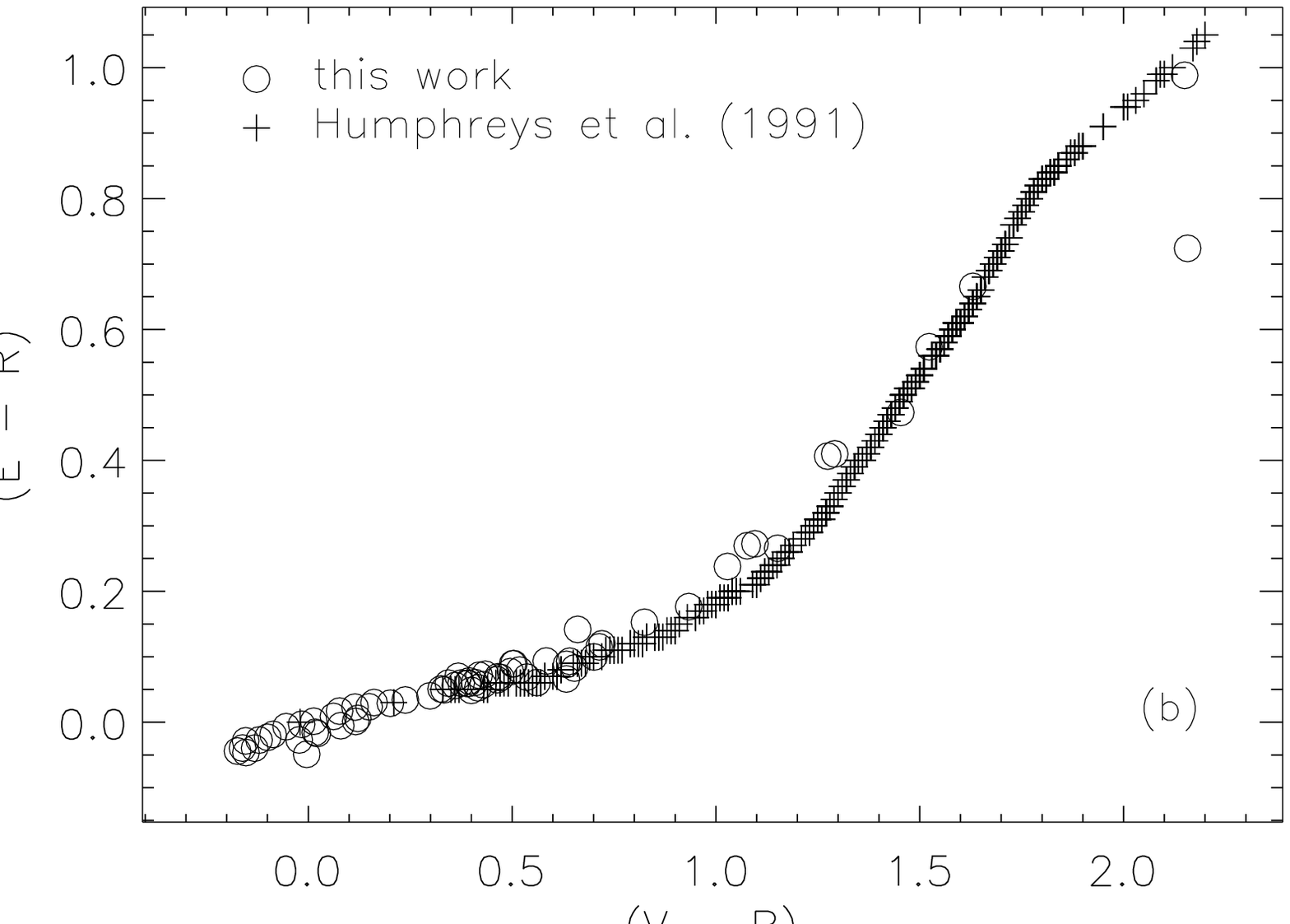}
\caption{The transformation from the Johnson system to the Palomar
system is plotted.  The ($O\!-\!B$) correction is displayed in (a) and
the ($E\!-\!R$) correction is displayed in (b).  Note that the
($O\!-\!B$) and ($E\!-\!R$) corrections derived in this paper agree
quite well with those from Humphreys \etal (1991).
\label{fig_john_to_palo}}
\end{center}
\end{figure}

  In order to check the accuracy of our transformations, the
transformation from the Johnson system to the Palomar system was
computed and compared to the same transformation as determined by
Humphreys \etal (1991).  Figure~\ref{fig_john_to_palo}a displays the
($O\!-\!B$) correction as a function of (\bv) which transforms the
Johnson B magnitude to the corresponding Palomar O magnitude.
Figure~\ref{fig_john_to_palo}b displays the ($E\!-\!R$) correction as
a function of (\vr) which transforms the Johnson R magnitude to the
corresponding Palomar E magnitude.  The agreement between the
($O\!-\!B$) and ($E\!-\!R$) corrections derived in this paper and
those of Humphreys \etal (1991), validates this method for deriving
transformations between photometric systems.

\begin{figure}[tbp]
\begin{center}
\plottwo{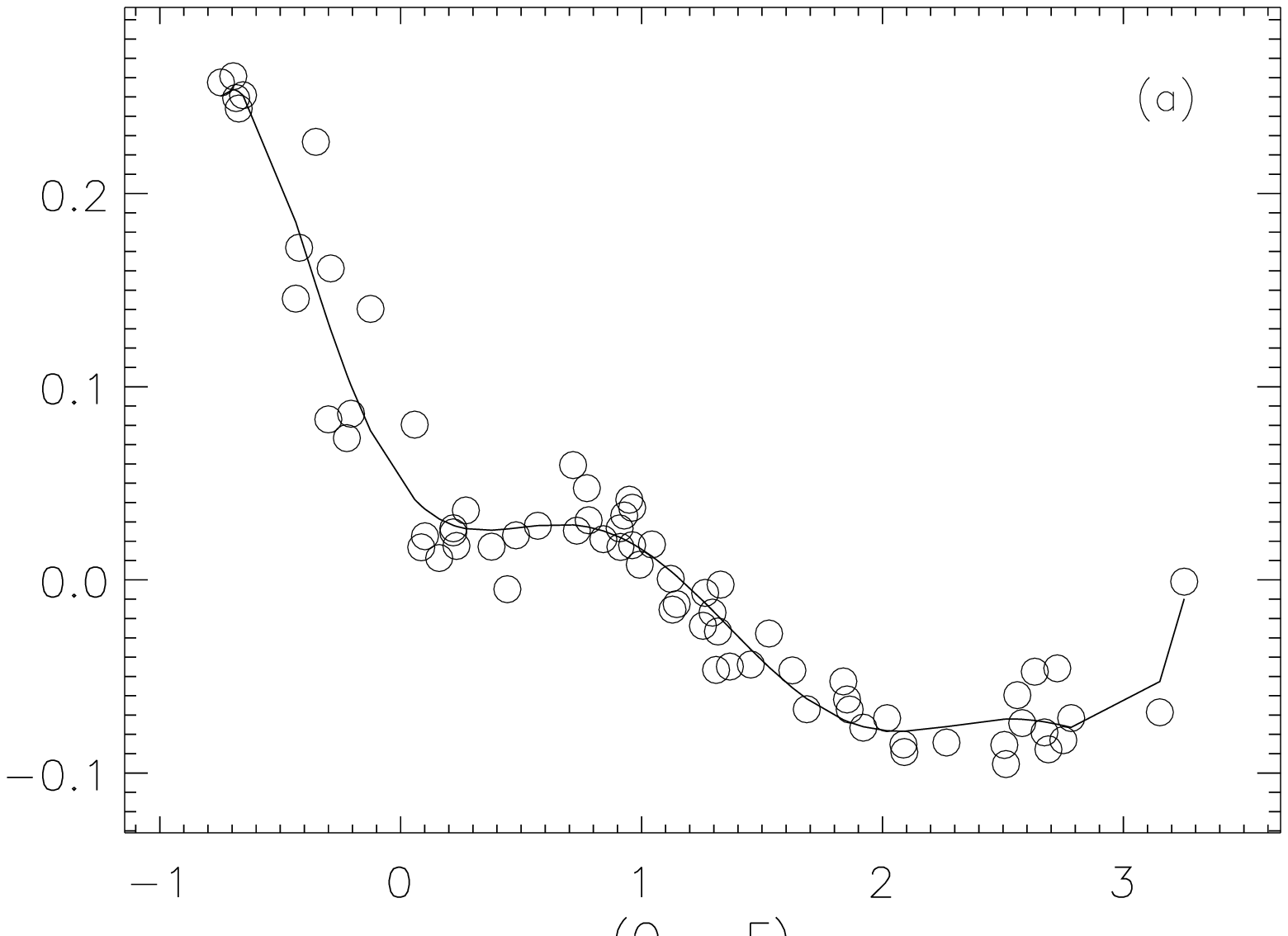}{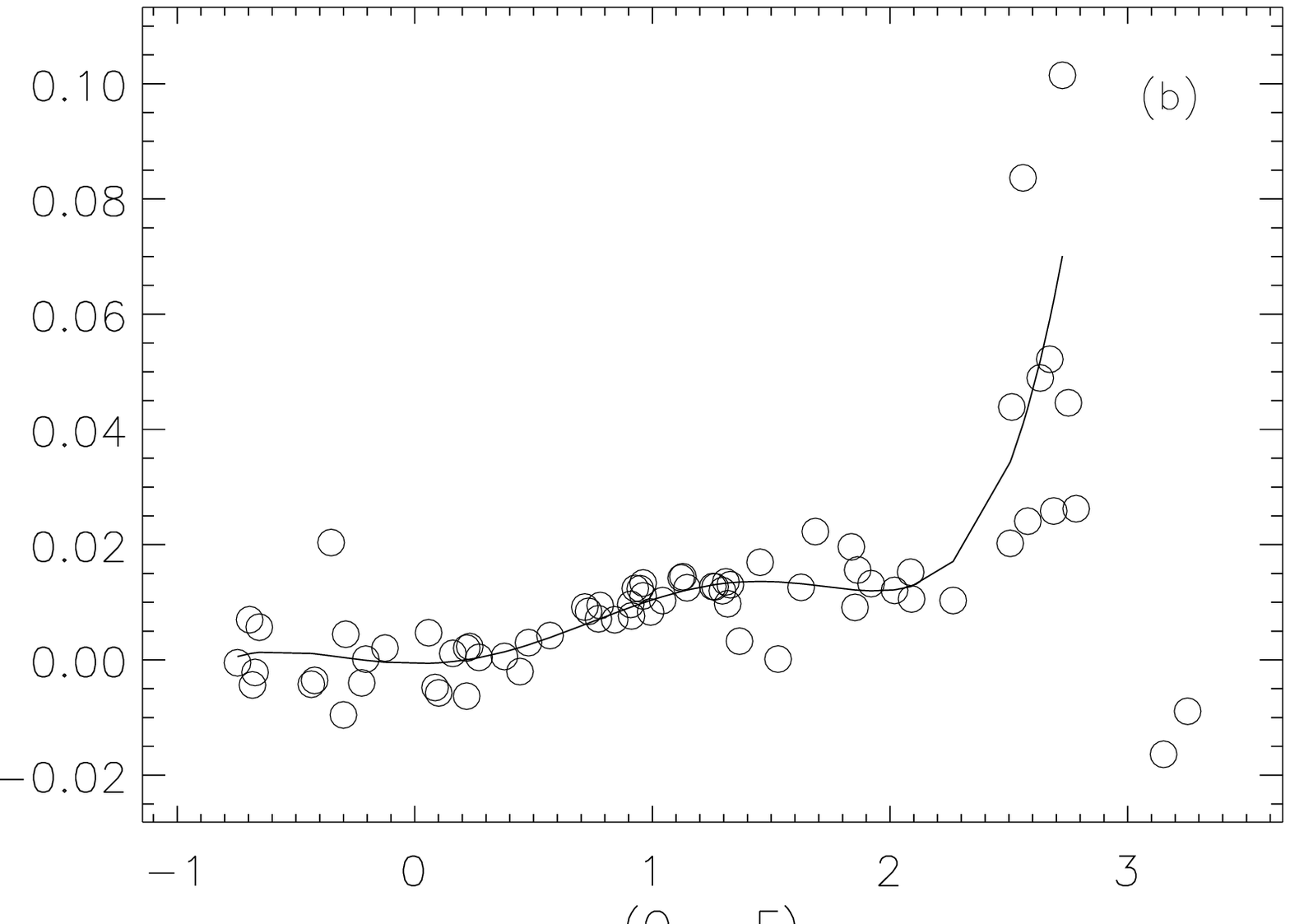}
\caption{The transformation from the Palomar system to the Pioneer
system is plotted.  The ($PB\!-\!O$) correction is displayed in (a)
and the ($PR\!-\!E$) correction is displayed in (b).  The maximum
corrections for both ($PB\!-\!O$) and ($PR\!-\!E$) are small.
\label{fig_palo_to_pion}}
\end{center}
\end{figure}

\begin{figure}[tbp]
\begin{center}
\plottwo{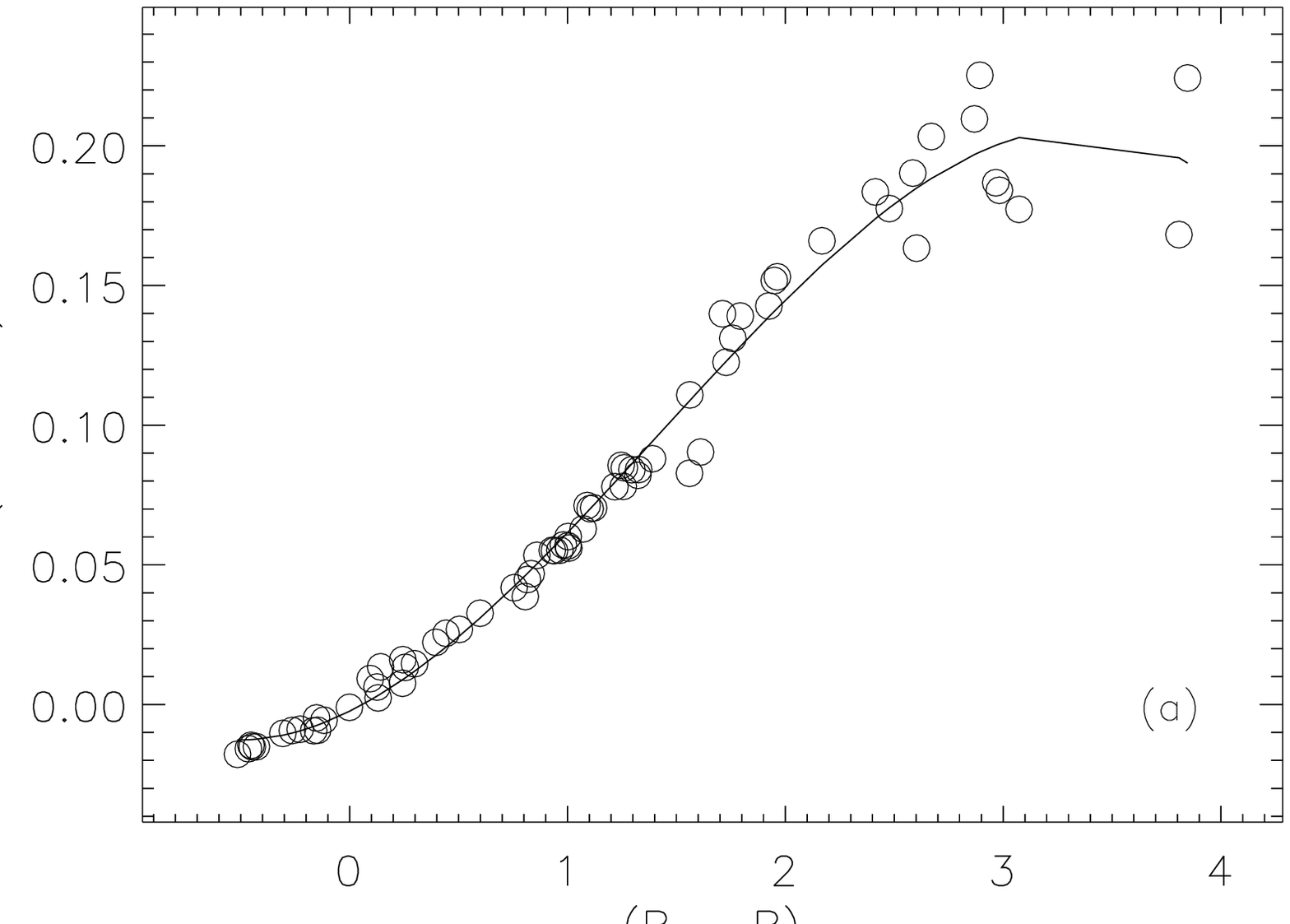}{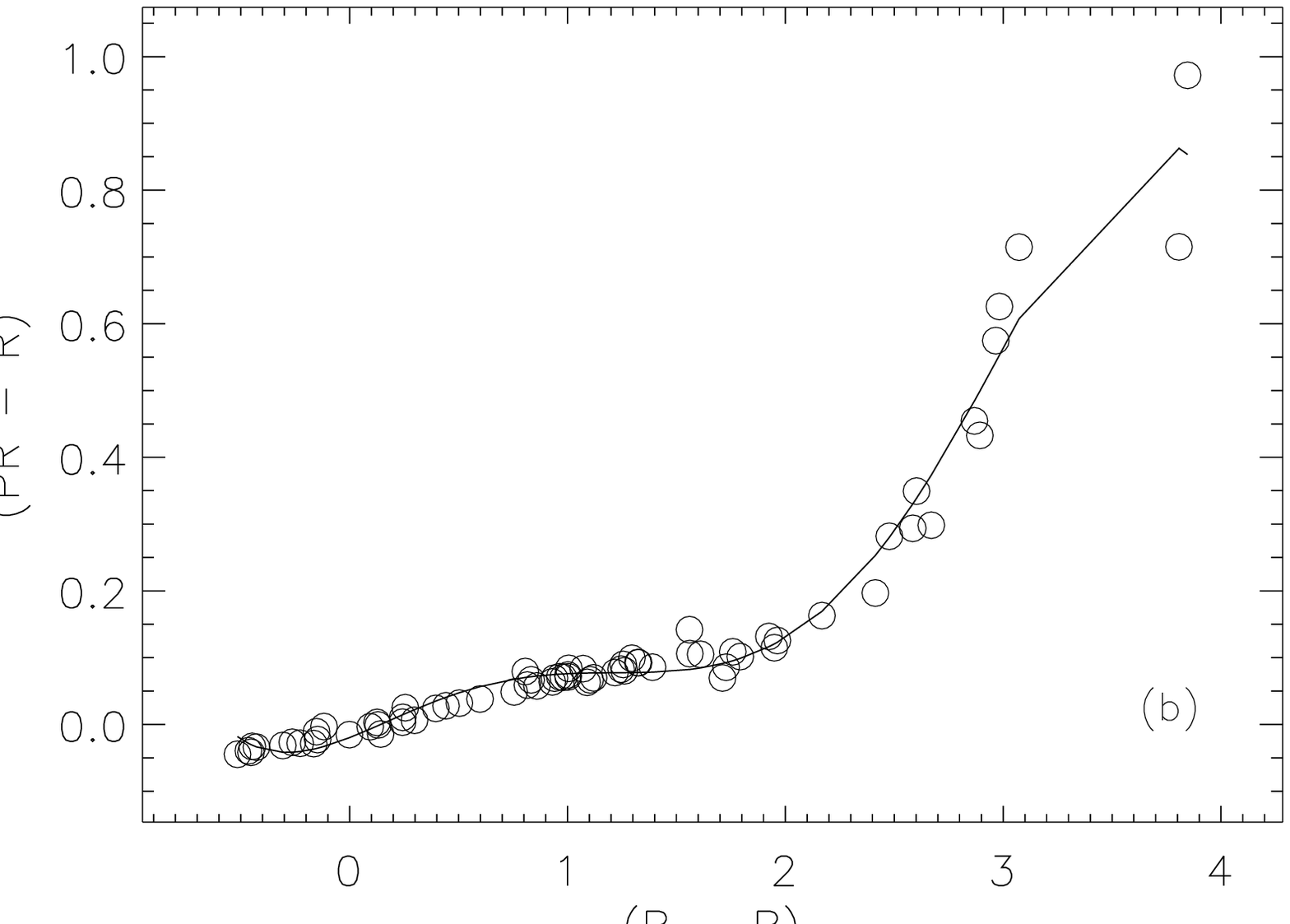}
\caption{The transformation from the Johnson system to the Pioneer
system is plotted.  The ($PB\!-\!B$) correction is displayed in (a)
and the ($PR\!-\!R$) correction is displayed in (b).  While the
maximum correction for ($PB\!-\!B$) is small, the maximum correction
for ($PR\!-\!R$) is large due to the significantly different values of
$\lambda_{\rm eq}$ for the PR and R response curves.
\label{fig_john_to_pion}}
\end{center}
\end{figure}

  The transformation from the Palomar to the Pioneer system is
displayed in Figure~\ref{fig_palo_to_pion} and the transformation from
the Johnson to the Pioneer system is shown in
Figure~\ref{fig_john_to_pion}.  We have fitted the resulting curves
with polynomial functions in order to have an analytic form for the
transformations.  The number of terms in the fitted polynomial was
determined by adding terms until the resulting fitted polynomial
followed the general trend of the points.  The maximum corrections for
($PB\!-\!O$), ($PR\!-\!E$), ($PB\!-\!B$), and ($PR\!-\!R$) are
$\sim$0.25, $\sim$0.10, $\sim$0.20, and $\sim$1.0, respectively.  The
large maximum correction for ($PR\!-\!R$) is due to the large
difference in the $\lambda_{\rm eq}$ value between the Pioneer (6441
\AA) and Johnson (6926 \AA) systems.

\subsection{Master Catalog Construction\label{ss_master_cat}}

  Three star and galaxy catalogs were used in constructing the Master
Catalog.  The three catalogs were the Not So Bright Star Catalog (NSBS
Catalog, see below), the GSC, and the APS Catalog.  As the Master
Catalog includes a large number of objects, we chose to construct it
in small pieces.  The GSC is split into 9537 regions (\cite{jen90})
and we used the same regions for our Master Catalog.

  For stars and galaxies with PB magnitudes between 13 and $\sim$20th
magnitude, the APS Catalog of the Palomar Sky Survey I (POSS I) was
used.  This ambitious survey is using POSS I plates to produce a
catalog of stars and galaxies with positions and Palomar O and E
magnitudes between 12th magnitude and the plate limit ($\sim$20th
magnitude).  Due to difficulties in automated photometry in crowded
fields, the APS Catalog is limited to $|b| > 20\arcdeg$
(\cite{pen93}).  The APS Catalog is also limited by the lack of any
POSS I plates below a declination $-33\arcdeg$ (\cite{min63}).  The
spatial extent of the APS Catalog defines the regions where we can
accurately remove the integrated contributions of stars and galaxies
to the Pioneer data.  The positional accuracy is $0\farcs 5$
(\cite{pen93}).  The O and E magnitudes were calibrated using UBV
photoelectric photometry as detailed in Humphreys (1991).  The
resulting random error in magnitudes was 0.2 for stars and 0.3-0.5 for
galaxies (\cite{pen93}).  The star/galaxy classification was done
using a neural network (\cite{ode92}, 1993).  Regions surrounding
bright stars are not included in the APS Catalog due to scattered
light from the bright stars.  We identified these holes by hand and
stars and galaxies from a nearby region of equal size were copied into
the hole.

  The Guide Star Catalog (GSC) was used to provide data for stars and
galaxies with PB magnitudes between 9.5 and 13.  For each object, the
GSC contains a position and a V ($\left<\lambda\right> \approx 5600$
\AA) or a J ($\left<\lambda\right> = 4500$ \AA) magnitude
(\cite{las90}).  The position error is $0\farcs 2 - 0\farcs 8$ and the
magnitude error is $\sim$0.30 mag (\cite{rus90}).  As the GSC gives
only one magnitude, deriving accurate PB and PR magnitudes for each 
object was not possible.  Since the GSC is complete to 15th magnitude,
a large number of the GSC objects also appear in the APS catalog.
This fact allowed us to statistically transform the GSC object
magnitudes to the Pioneer system.  The average ($PB\!-\!V$) or
($PB\!-\!J$) and ($PB\!-\!PR$) color for stars in common to both the
GSC and APS Catalog was computed and used to transform the magnitudes
of the GSC objects to the Pioneer system.

  The catalog of stars between 6.5 V magnitude and 9.5 PB magnitude
was constructed by combining three different existing catalogs: the
Catalogue of Stellar Identifications (CSI, \cite{och83}), the Michigan
Catalogue of Two-Dimensional Spectral Types for HD Stars (Houk
Catalog, \cite{hou75}; \cite{hou78}; \cite{hou82}; \cite{hou88}), and
the UBV Photoelectric Photometry Catalogue (UBV Catalog, \cite{mer87};
\cite{mer94}).  This produced the Not So Bright Star Catalog (NSBS
Catalog) with each star possessing the most accurate position,
spectral type, PB magnitude, and PR magnitude possible.  The CSI was
used as the base for the NSBS Catalog.  It is complete 
to $m_{\rm V} \approx 9.5$ (\cite{och81}) and is the combination of
many other catalogs, notably the Henry Draper Catalog (HD Catalog,
\cite{can18}; \cite{can25}; \cite{can49}) and the Smithsonian
Astrophysical Observatory Catalogue (SAO Catalog, \cite{sao66}).
Stars already subtracted from the Pioneer data were excluded from the
NSBS Catalog by excluding all stars in the Bright Star Catalogue
(\cite{hof91}) and stars with $m_V < 8$ (\cite{tol87}) in the
Photoelectric Catalog (\cite{bla68}; \cite{och74}).  

\begin{figure}[tbp]
\begin{center}
\plotone{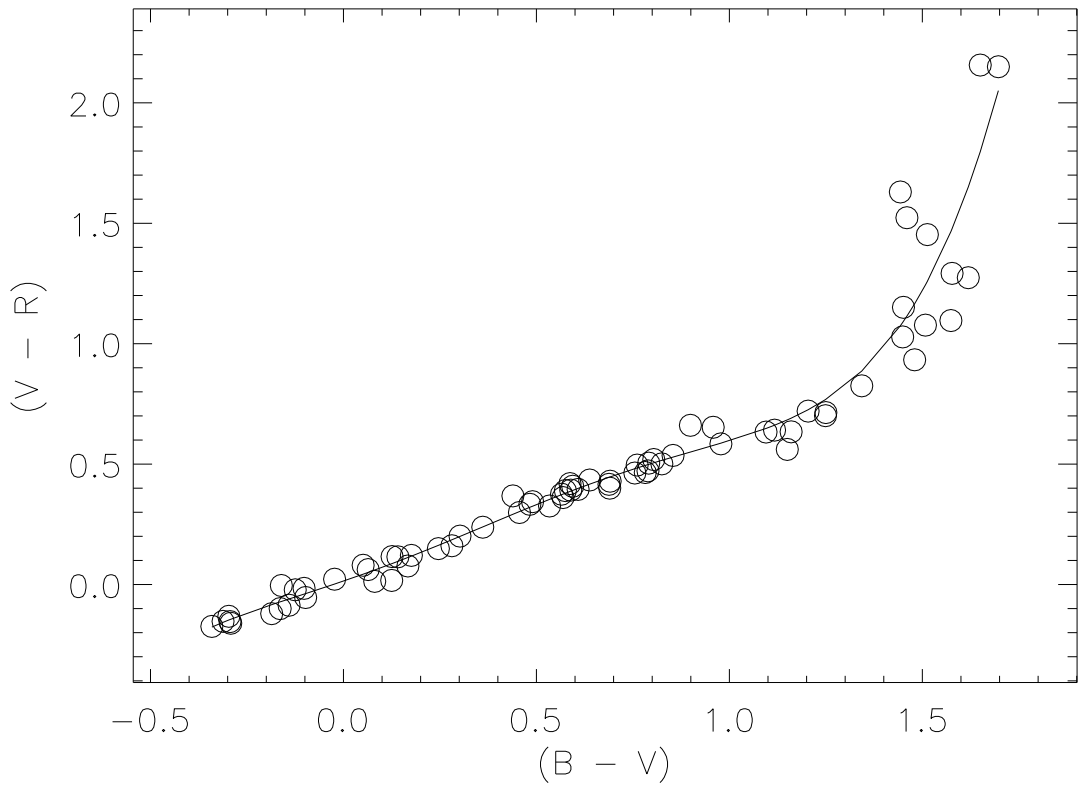}
\caption{The transformation between the Johnson (\bv) to (\vr) color
is plotted above.  The open circles were calculated by summing the
product of the appropriate response curves and the grid of stellar
spectra (Silva \& Cornell 1992).  \label{fig_john_bv2vr}}
\end{center}
\end{figure}

  Information for each star in the NSBS Catalog was gathered in the
following manner.  The star's position was taken from the CSI.  The
star's spectral type was taken (in order of preference) from the Houk
Catalog or the CSI.  The star's PB and PR photometry was determined
from the star's Johnson B and R magnitudes which were calculated using
one of the following algorithms.  If the star's V magnitude and (\bv)
color were available from the UBV Catalog, the R magnitude was
computed by using the fit given in Figure~\ref{fig_john_bv2vr} to
calculate the star's (\vr) color.  If only the star's V magnitude was
available, then the star's (\bv) color was computed from
\begin{equation}
(\bv) = (\bv)_o + E(\bv)
\end{equation}
where (\bv)$_o$ is the star's unreddened color and $E(\bv)$ is the
reddening due to dust.  A similar equation was used to compute the
star's (\vr) color.

\begin{figure}[tbp]
\begin{center}
\plottwo{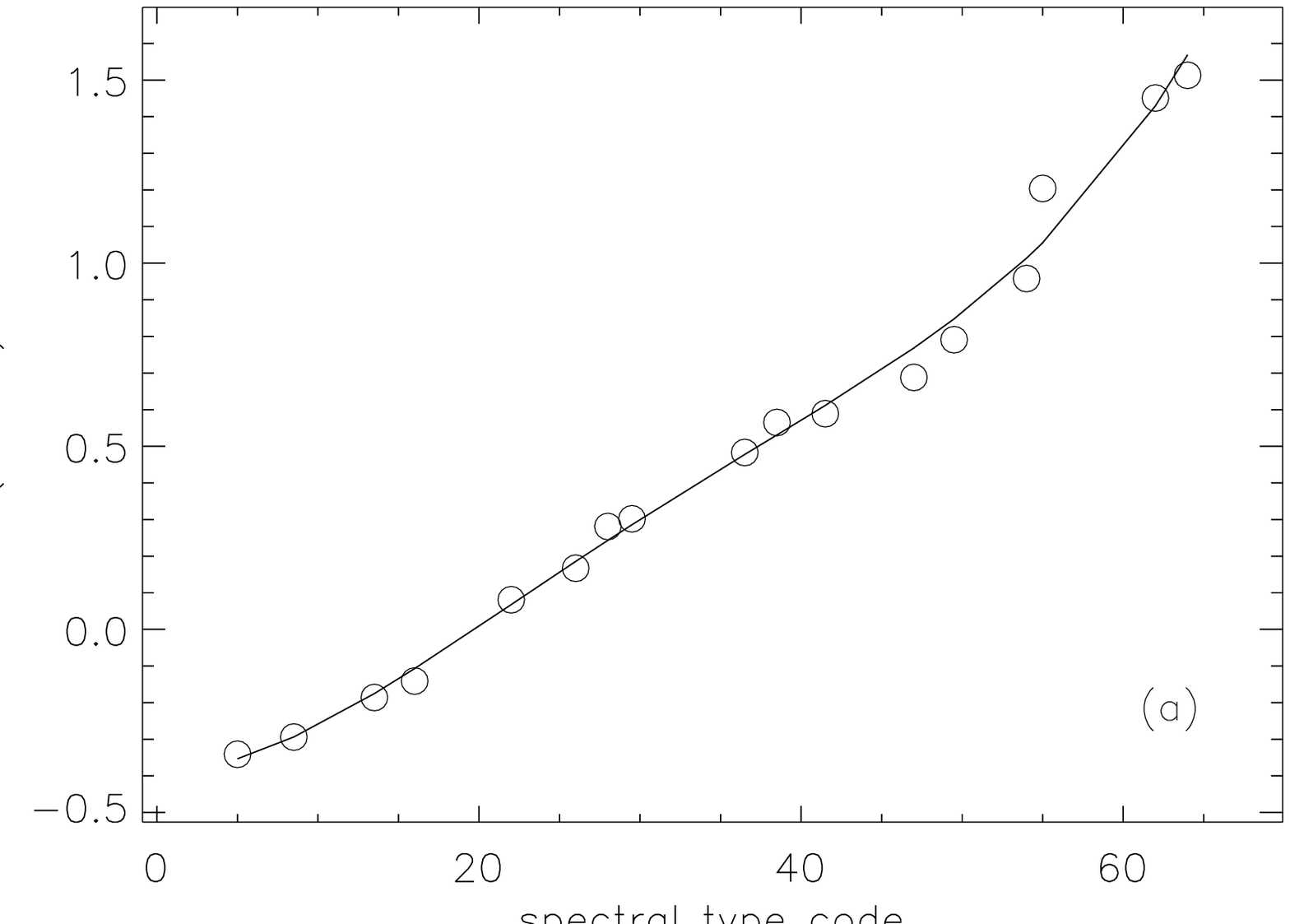}{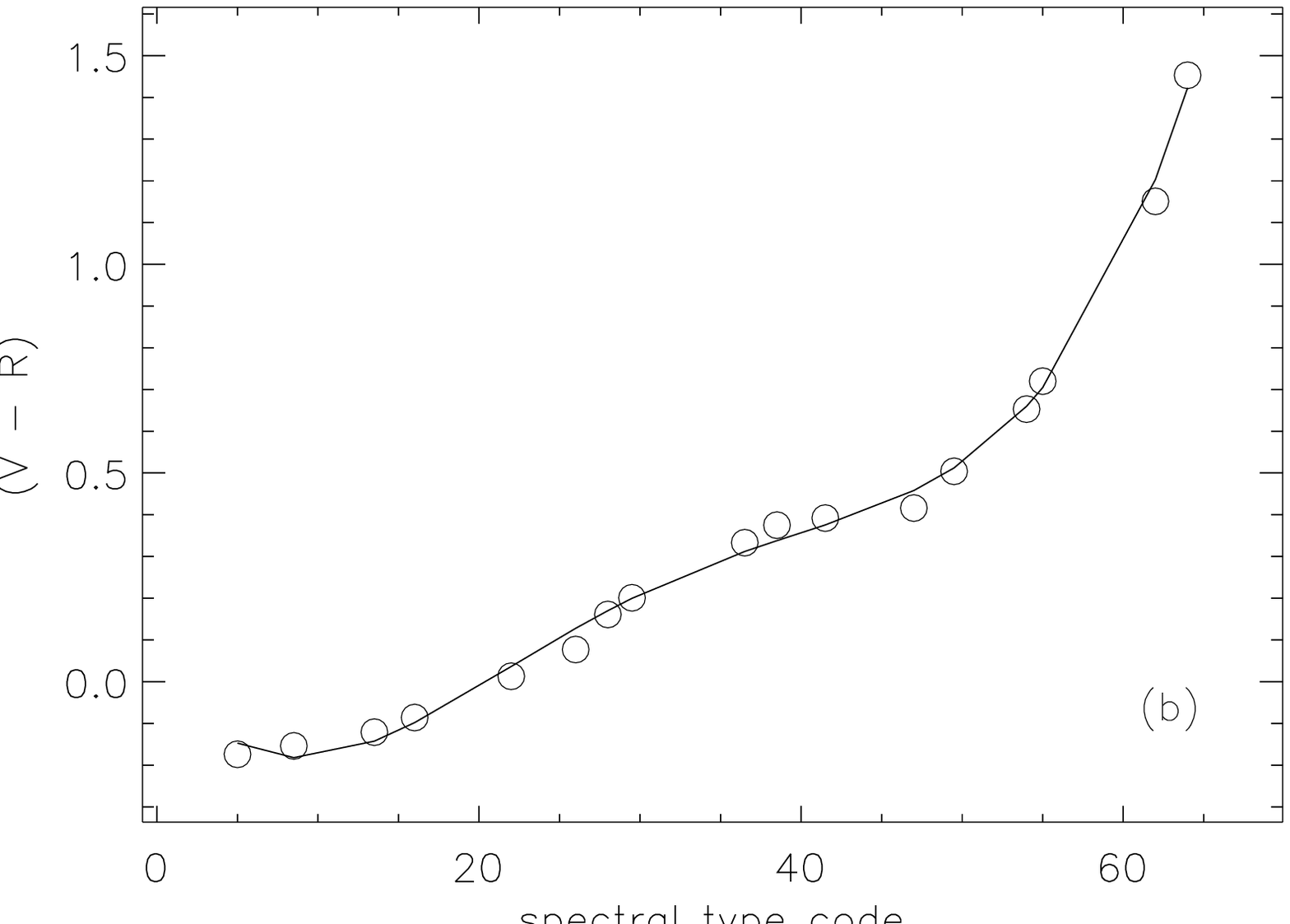}
\caption{The relationship between a luminosity class V star's
temperature class and (a) (\bv) and (b) (\vr).  The spectral type code
(spcode) is a numerical representation of the star's temperature class
with O = 0, B = 10, A = 20, etc. and subclasses worth their values.
For example, a A4 temperature class has a spcode = 24.
\label{fig_mkV2br}} 
\end{center}
\end{figure}

  The unreddened $(\bv)_o$ and $(\vr)_o$ were determined using fits
between spectral type and unreddened color computed using the Johnson
response curves and the grid of stellar spectra (\cite{sil92}).
Figure~\ref{fig_mkV2br} displays the fits to the (\bv)$_o$ and
(\vr)$_o$ colors for luminosity class V.  Fits for luminosity classes
III and I were similar.  Ideally, all the stars in the NSBS Catalog
would have two-dimensional MK spectral types resulting in accurate
unreddened colors. This is true for stars in the HD Catalog (complete
to $m_{\rm pg} \approx 9$ [\cite{rom85}]) with declinations less than
$- 12\arcdeg$ (Houk Catalog) and a few other stars in the CSI with
spectral types from other sources (\cite{och83}).  For the stars with
only one-dimensional temperature classes, the unreddened (\bv)$_o$
and (\vr)$_o$ colors were calculated by taking a weighted average of
the (\bv)$_o$ and (\vr)$_o$ colors corresponding to luminosity classes
V, III, and I.  The weights were the probability of finding a V, III,
or I luminosity class star of the star's temperature class with the
star's V magnitude and absolute value of galactic latitude.  These
probabilities were determined from the Houk Catalog.

  For each star, $E(\bv)$ and $E(\vr)$ values were calculated from
\begin{equation}
E(\bv) = \left[ \frac{E(\bv)}{A_V} \right] A_V = 0.33A_V
\end{equation}
and
\begin{equation}
E(\vr) = \left[ \frac{E(\vr)}{A_V} \right] A_V = 0.17A_V,
\end{equation} 
respectively (\cite{whi92}).  The values for $E(\bv)/A_V$ and
$E(\vr)/A_V$ correspond to an average Milky Way extinction curve, $R_V
= 3.05$ (\cite{whi92}).  $A_V$ is calculated using
\begin{equation}
A_V = \alpha d
\end{equation}
where $\alpha = 0.6~\mbox{mag kpc}^{-1}$ and $d$ is the star's
distance.  In order to determine the value of $\alpha$ given above, we
extracted all 19,746 stars in the NSBS Catalog with two-dimensional
spectral types and an observed (\bv).
Using these stars' two-dimensional spectral types, we calculated their
(\bv) colors from the algorithm described above with a range of
$\alpha$ values.  For an $\alpha = 0.6~\mbox{mag kpc}^{-1}$, the
calculated and observed (\bv) colors were equal with no galactic
latitude dependence of $\alpha$ necessary.  The distance was
computed by solving
\begin{equation}
m_V - M_V = 5 \log \left( \frac{d}{10~\mbox{pc}} \right) + \alpha d
\end{equation}
where $m_V$ is the observed V magnitude and $M_V$ is the absolute
magnitude appropriate for the star's spectral type (\cite{sch82}).
Again, for stars without two-dimensional spectral types, the computed
$E(\bv)$ and $E(\vr)$ values were a weighted average of color excesses
for V, III, and I luminosity classes.  Finally, the star's B and R
magnitudes were transformed to Pioneer PB and PR magnitudes using the
transformations given in \S\ref{sec_transformations}.

  We were only interested in the integrated star/galaxy intensity and
chose to do the integration over $5\arcdeg \times 5\arcdeg$ sized
regions (see section~\ref{sec_detect_red}).  Since the integration was
done over regions larger than an individual Master Catalog region
which contains $\sim$20,000 objects, the error in the integrated
intensity was dominated by systematic errors.  The random error in the
integrated intensity due to random errors in the fluxes of individual
objects was very small due to the large number of objects contributing
to the integrated intensity.

  The only systematic error found was associated with the lack of
two-dimensional spectral types for many of the stars in the NSBS
Catalog.  By using the population statistics from the Houk Catalog
(see above), the magnitude of this error was greatly reduced.  Using
the Master Catalog regions with spectral types from the Houk Catalog,
we determined the remaining systematic error.  The integrated flux for
the stars in these Master Catalog regions with PB magnitudes between
5.5 and 9.5 was determined two ways, using the stars' two-dimensional
spectral types and using only the stars' one-dimensional spectral
types.  On average, the integrated PB star flux using the
one-dimensional spectral types was $2.5 \pm 1.5\%$ too high and the
integrated PR flux was $3.3 \pm 2.5\%$ too low.  Since there was a
systematic zero point error, the PB and PR magnitudes for stars
without two-dimensional spectral types in the NSBS Catalog were
corrected by adding 0.0273 and -0.0349 to their PB and PR magnitudes,
respectively.  This left a random error of $1.5\%$ and $2.5\%$ in the
integrated PB and PR star flux from the NSBS Catalog, respectively.

\begin{figure}[tbp]
\begin{center}
\plottwo{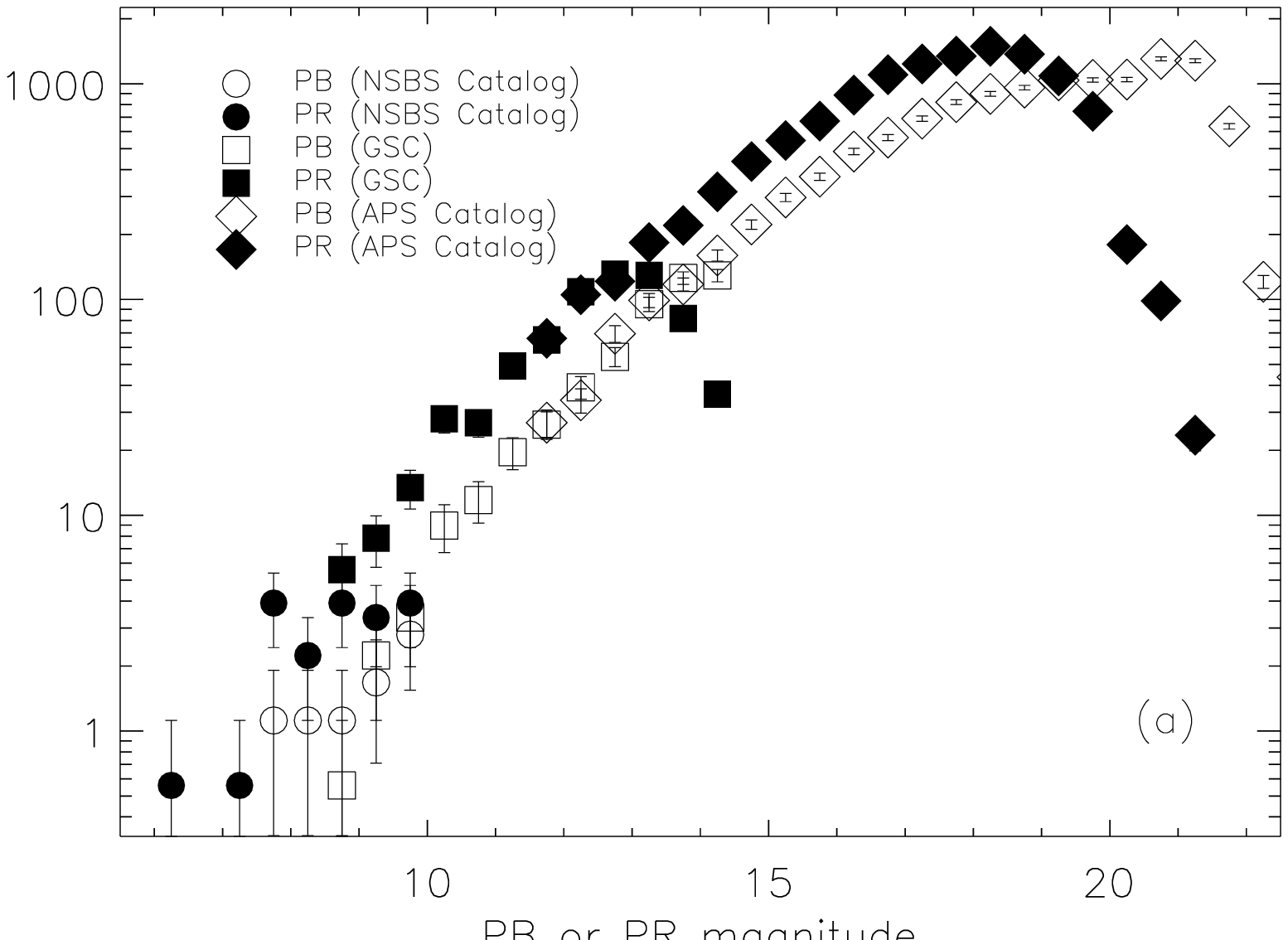}{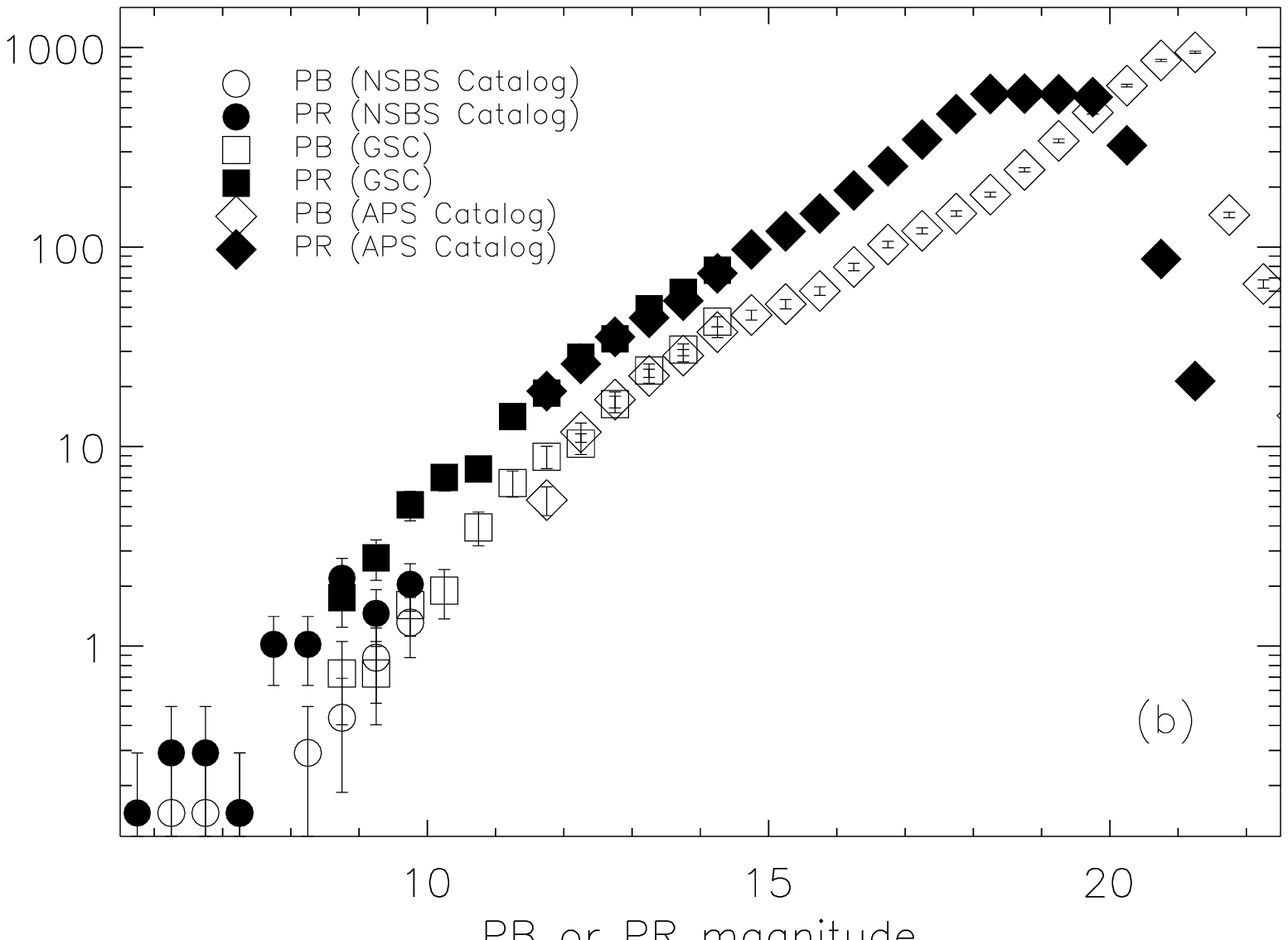}
\caption{The combined star/galaxy counts for the NSBS Catalog, the
GSC, and the APS Catalog are displayed for two Master Catalog regions.
A region at $b = 19\fdg 6$ (GSC \#2947) is displayed in (a) and a
region at $b = 74\fdg 6$ (GSC \#2526) in (b).  The counts were
normalized to reflect the counts for one square degree and one
magnitude wide bins.  The error bars were computed assuming Poisson
statistics.
\label{fig_nga_counts}}
\end{center}
\end{figure}

  In order to show that the construction of the Master Catalog worked,
we display the star/galaxy counts for two Master Catalog regions in
Figure~\ref{fig_nga_counts}.  Figure~\ref{fig_nga_counts}a shows the
counts for a low latitude region and Figure~\ref{fig_nga_counts}b
shows the counts for a high latitude region.  These two figures show
the counts from the NSBS Catalog, the GSC, and the APS Catalog in the
regions where these catalogs are stated to be valid.  The overlap of
the three catalogs is good and within the expected uncertainties.
Additional confirmation of the Master Catalog comes from comparison of
our star counts with that of the SKY model (\cite{coh94}, 1995).  The
star counts predicted from the SKY model agree within Poisson
statistics with our observed Master Catalog counts (\cite{coh97}).
See Toller (1981) for an excellent review of previous star count work.

\begin{figure}[tbp]
\begin{center}
\plottwo{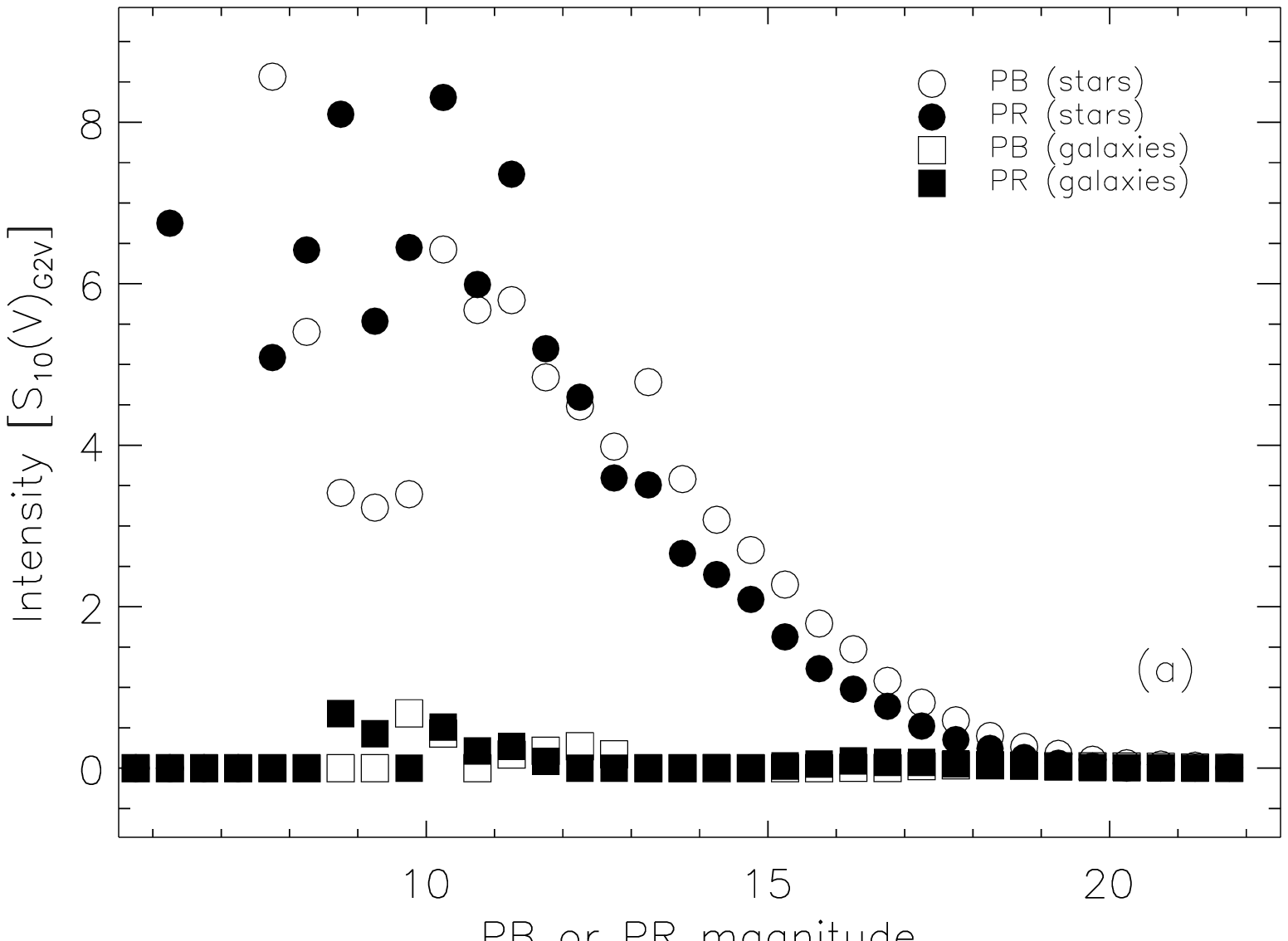}{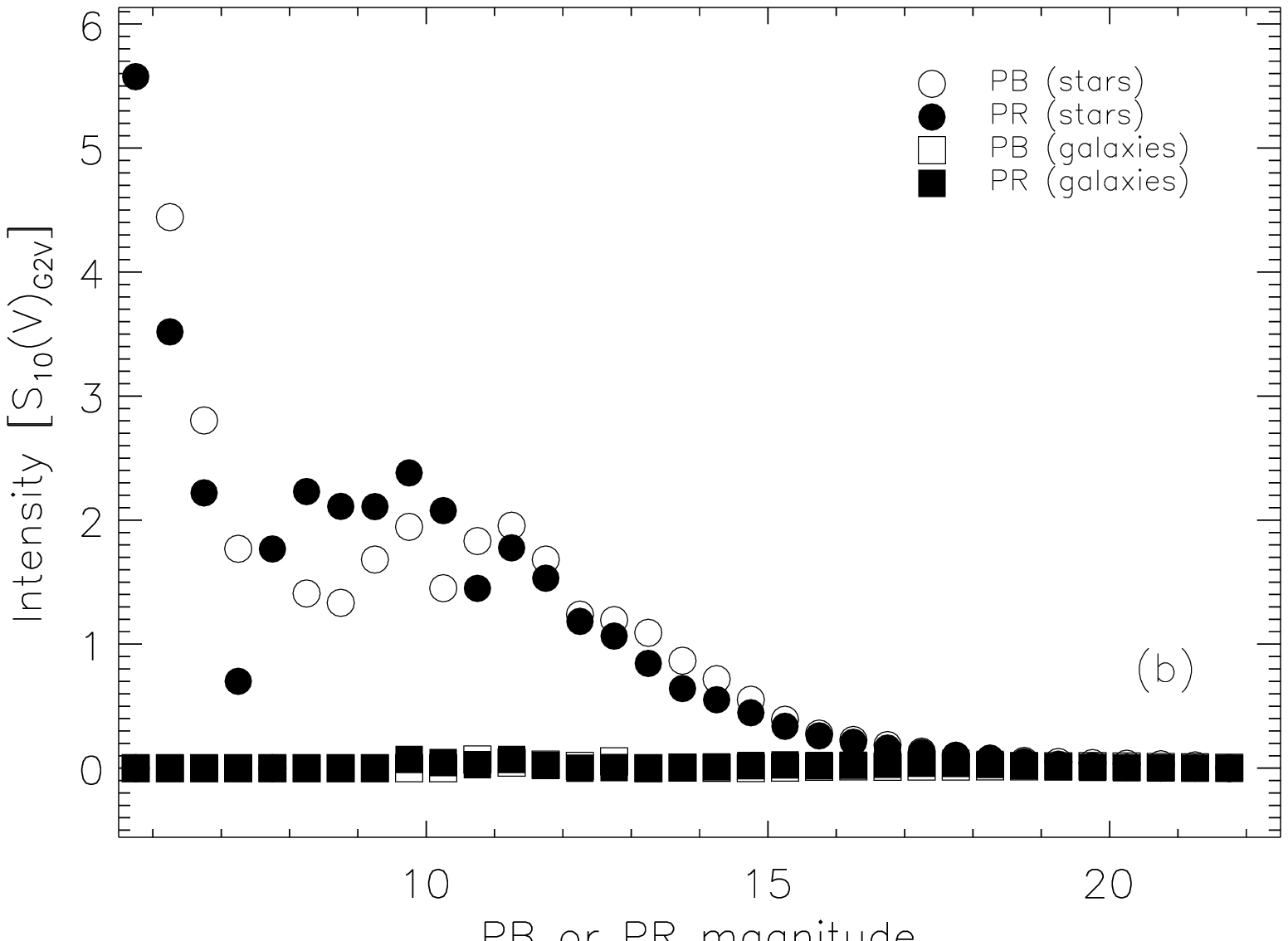}
\caption{The contribution to the total intensity from each half
magnitude bin is plotted for the same two regions as in the previous
figure.  The region at $b = 19\fdg 6$ is displayed in (a) and the
region at $b = 74\fdg 6$ in (b).  The peak contribution comes from
about 10th magnitude and the contribution from objects with magnitudes
$>$ 20 is negligible.  \label{fig_master_S10}}
\end{center}
\end{figure}

  Figure~\ref{fig_master_S10} plots the contribution to the total
intensity from each half magnitude bin for the same regions displayed
in Figure~\ref{fig_nga_counts}.  On average, the maximum contribution
comes from $\sim$10th magnitude objects.  The scatter in the curve for
magnitudes $< 10$ is due to the relatively small number of bright
stars per Master Catalog region.  The contribution from stars and
galaxies with magnitudes $> 20$ is negligible and not the result of
the Master Catalog incompleteness as the curve falls smoothly at
magnitudes well below 20.

\section{Detection of ERE in the Diffuse ISM \label{sec_detect_red}}

  The blue and red intensities of the diffuse ISM were determined by
subtracting the integrated star/galaxy light (ISGL) from the Pioneer
maps.  As the ISGL was computed from the Master Catalog, the spatial
extent of the diffuse ISM intensities was limited to regions of the
sky covered by the APS Catalog.  Currently, the APS Catalog includes
approximately 400 plates from the POSS I in a fairly random pattern
across the sky.  Due to the low resolution of the Pioneer maps and the
size of an individual POSS I plate ($6\fdg 6$, \cite{min63}), regions
with many adjoining plates were needed.  Currently, only two regions
have enough contiguous plates to accurately derive the diffuse ISM
intensities.  Region 1 is located between $(l,b) =
(355\arcdeg,30\arcdeg)$ and $(10\arcdeg,60\arcdeg)$ and region 2 is
located between $(l,b) = (90\arcdeg,-55\arcdeg)$ and
$(120\arcdeg,-20\arcdeg)$.  These two regions have areas of
$315~\sq\arcdeg$ and $820~\sq\arcdeg$ and contain over 110,000 and
746,000 objects, respectively.

  In these two regions, maps of the blue and red intensities for the
ISGL were computed using the Master Catalog.  The fluxes of the stars
and galaxies in the Master Catalog were mapped onto a grid in galactic
longitude and latitude with $0\fdg 25 \times 0\fdg 25$ sized pixels.
The fluxes were converted to \steng units using the conversions given
in Table~\ref{table_mag_details} and solid angles (in steradians)
computed from
\begin{equation}
\Omega = \left( l_2 - l_1 \right) \left( \sin b_2 - \sin b_1
   \right)
\label{eq_area}
\end{equation}
where $l$ is galactic longitude (in radians), $b$ is galactic latitude
(in radians), and the subscripts 1 and 2 refer to the minimum and
maximum values, respectively.  Equation~\ref{eq_area} is valid for
rectangles in galactic longitude and latitude.  The resulting ISGL map
was smoothed to a resolution of $2\arcdeg$ to match the Pioneer maps.

  In computing the ISGL intensities, we did not include a weighting in
accordance with the dwell time of each star within the FOV of
individual IPP measurements for the stars fainter than m = 6.5.  The
corrections for the brighter stars (m $<$ 6.5) which were subtracted
in the original reduction of the IPP measurements
(\S\ref{sec_pioneer}), did include such a correction.  The reason for
this difference in treatment was based on the expected surface density
of stars of different apparent brightness.  The fainter stars are
sufficiently numerous that many are present in the field at any one
time, and on average, a star is leaving the field when another similar
star is entering the field.  The passages of brighter stars through
the field, on the other hand, are sufficiently singular events that
the details of these passages must be taken into account.

\begin{figure}[tbp]
\begin{center}
\plottwo{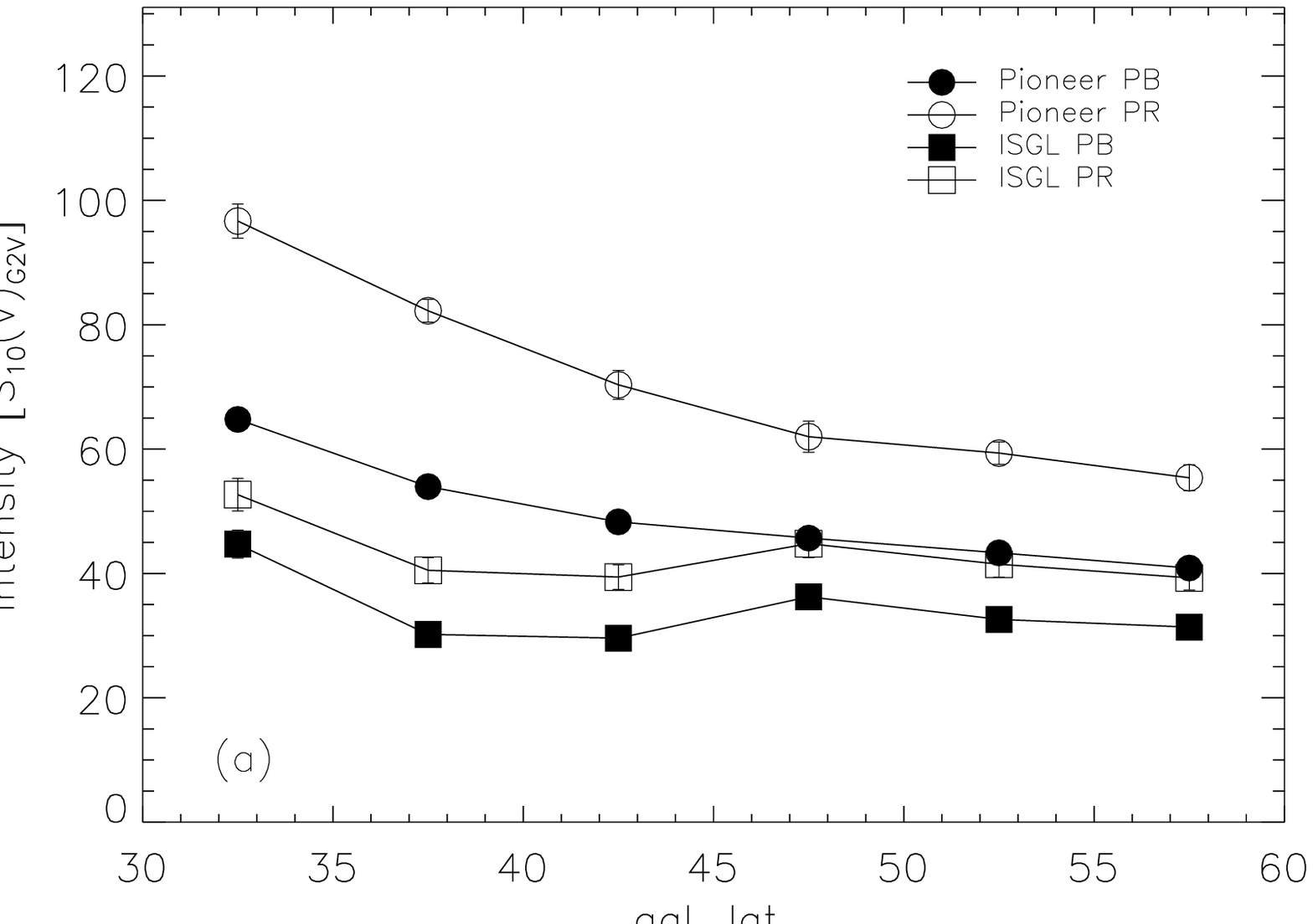}{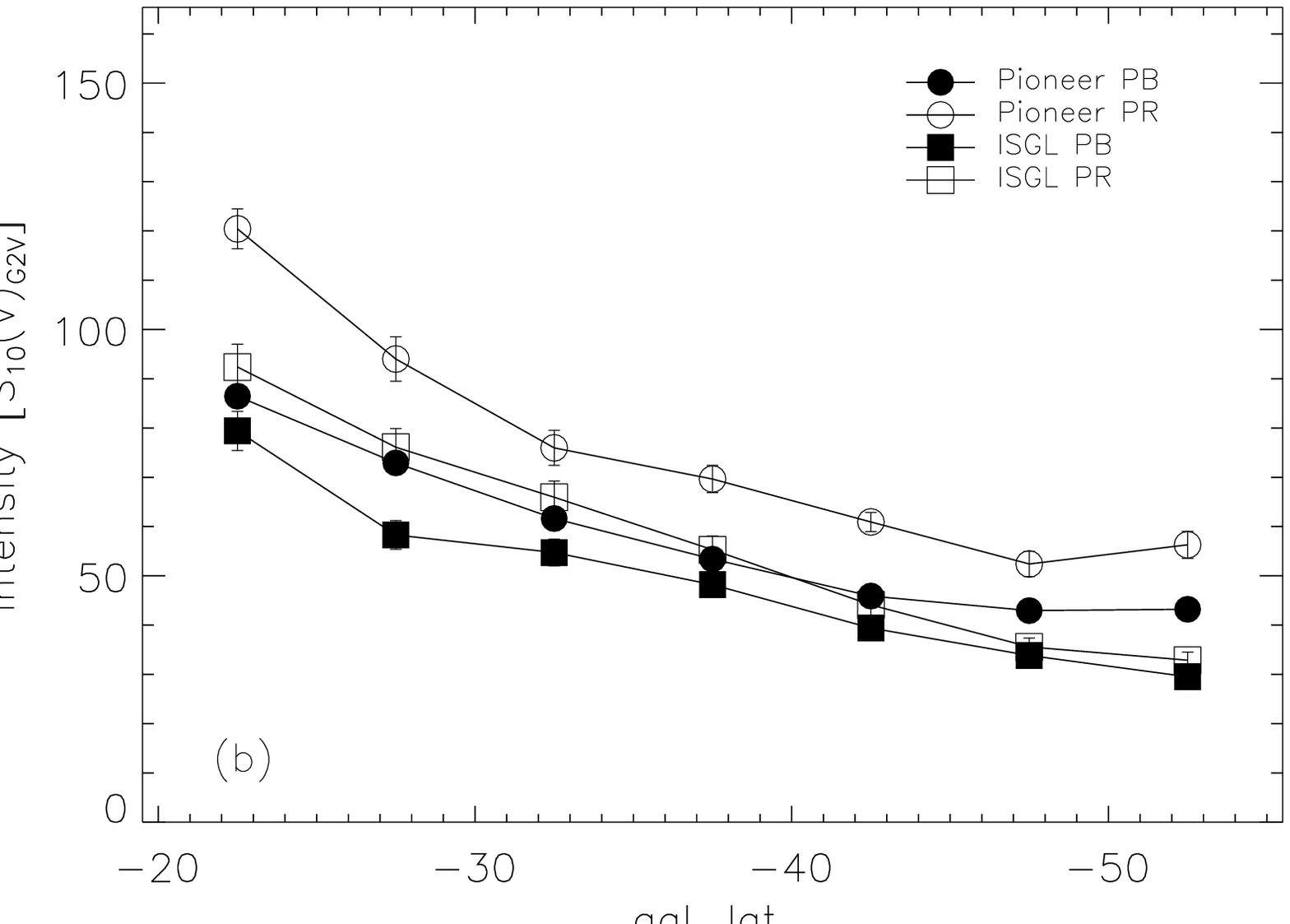}
\caption{The red and blue intensities for the Pioneer measurements and
integrated star/galaxy light (ISGL) are plotted as a function of
galactic latitude for cuts in galactic longitude between $0\arcdeg$
and $5\arcdeg$ (a) and $95\arcdeg$ and $100\arcdeg$ (b).  Each point
corresponds to a $5\arcdeg \times 5\arcdeg$ region.  The Pioneer error
bars were computed using the algorithm described in $\S$2.1.  The ISGL
error bars were assumed to be a conservative 5\% (see
$\S$3.2). \label{fig_cut_pion_isl}}
\end{center}
\end{figure}
  
\begin{figure}[tbp]
\begin{center}
\plottwo{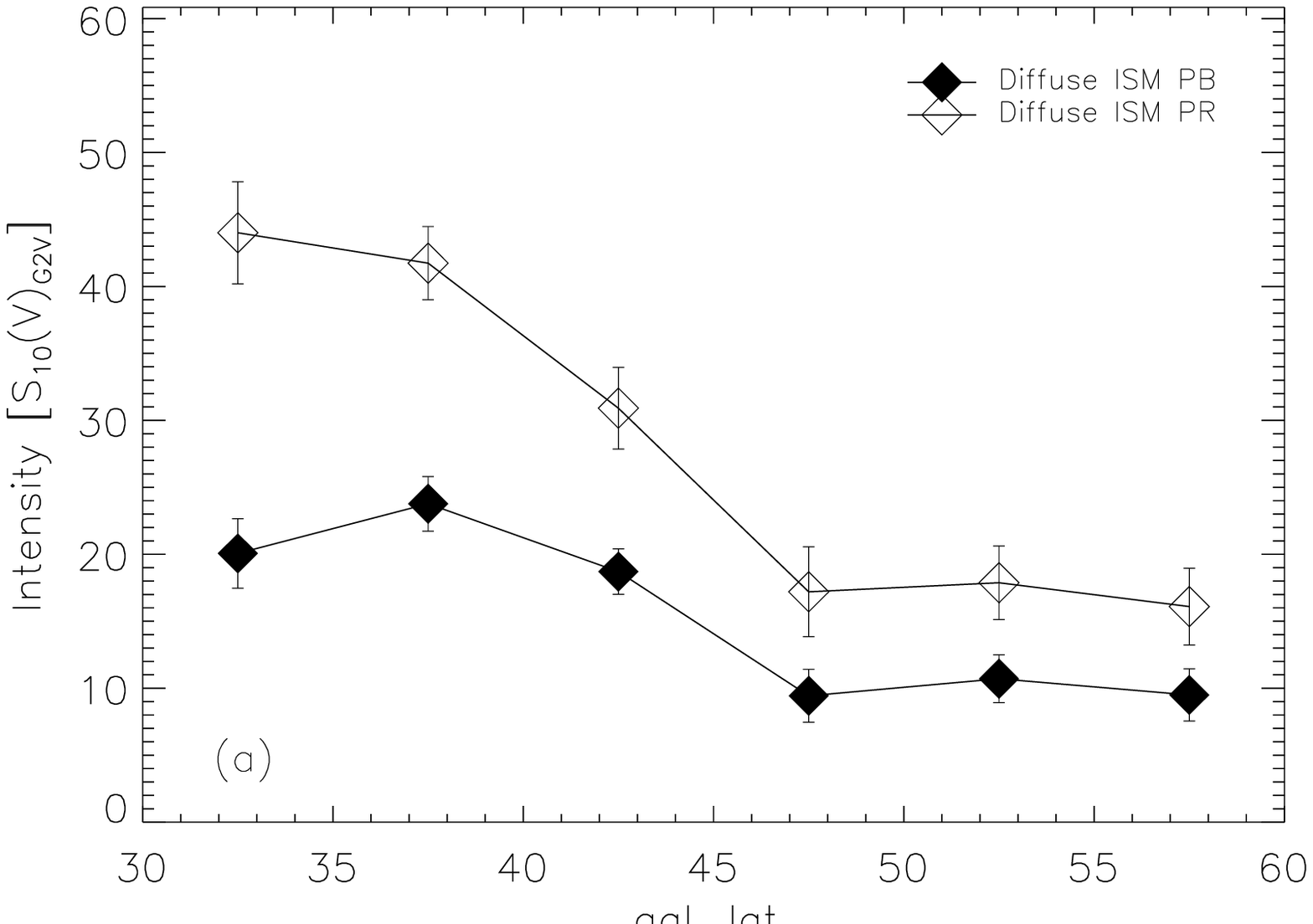}{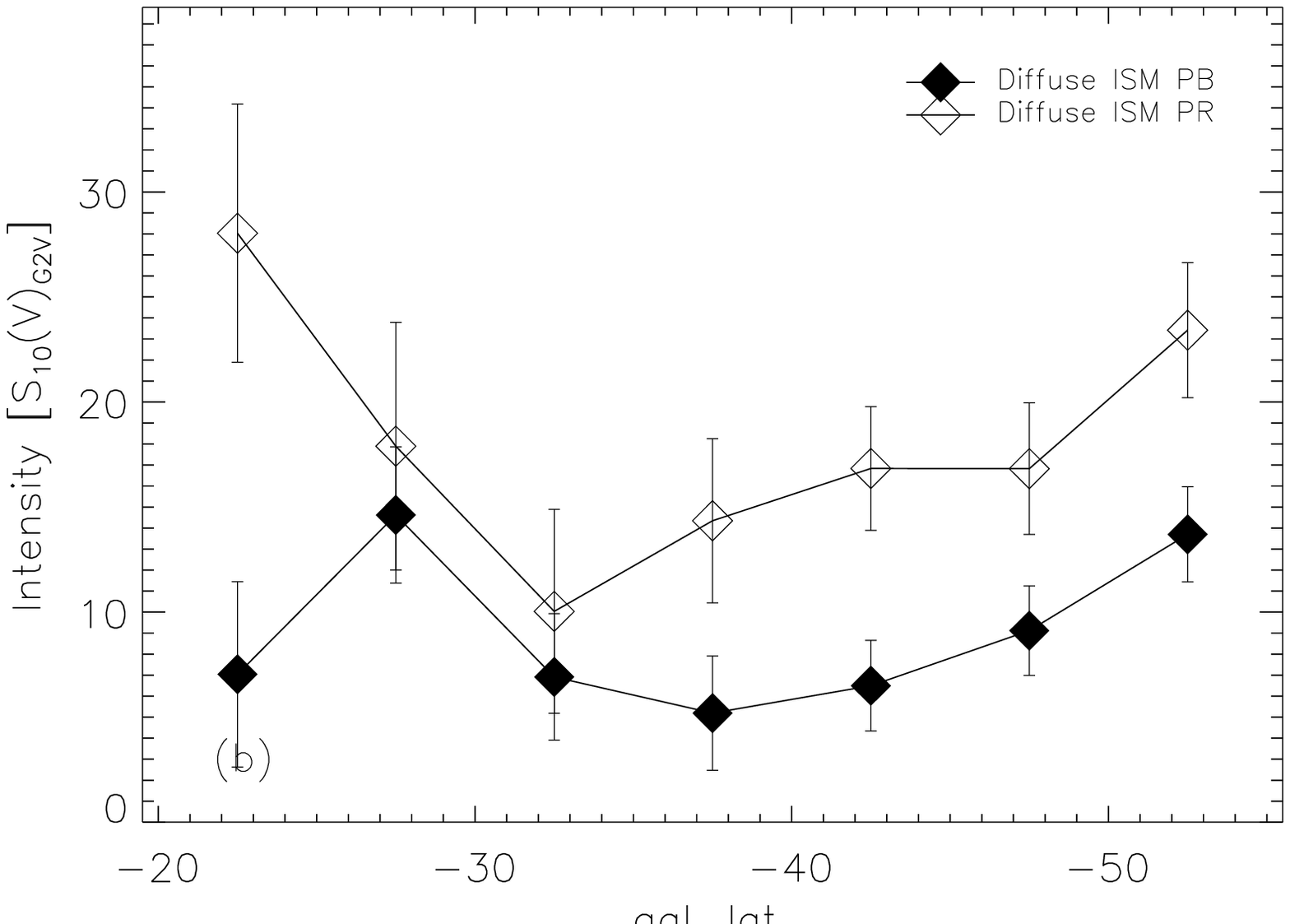}
\caption{The red and blue intensities for the light from the diffuse
ISM are plotted for cuts in galactic longitude between $0\arcdeg$ and
$5\arcdeg$ (a) and $95\arcdeg$ and $100\arcdeg$ (b).  The diffuse ISM
intensities were computed by subtracting the ISGL from the Pioneer
measurements.
\label{fig_cut_diffuse}}
\end{center}
\end{figure}
  
\begin{figure}[tbp]
\begin{center}
\plottwo{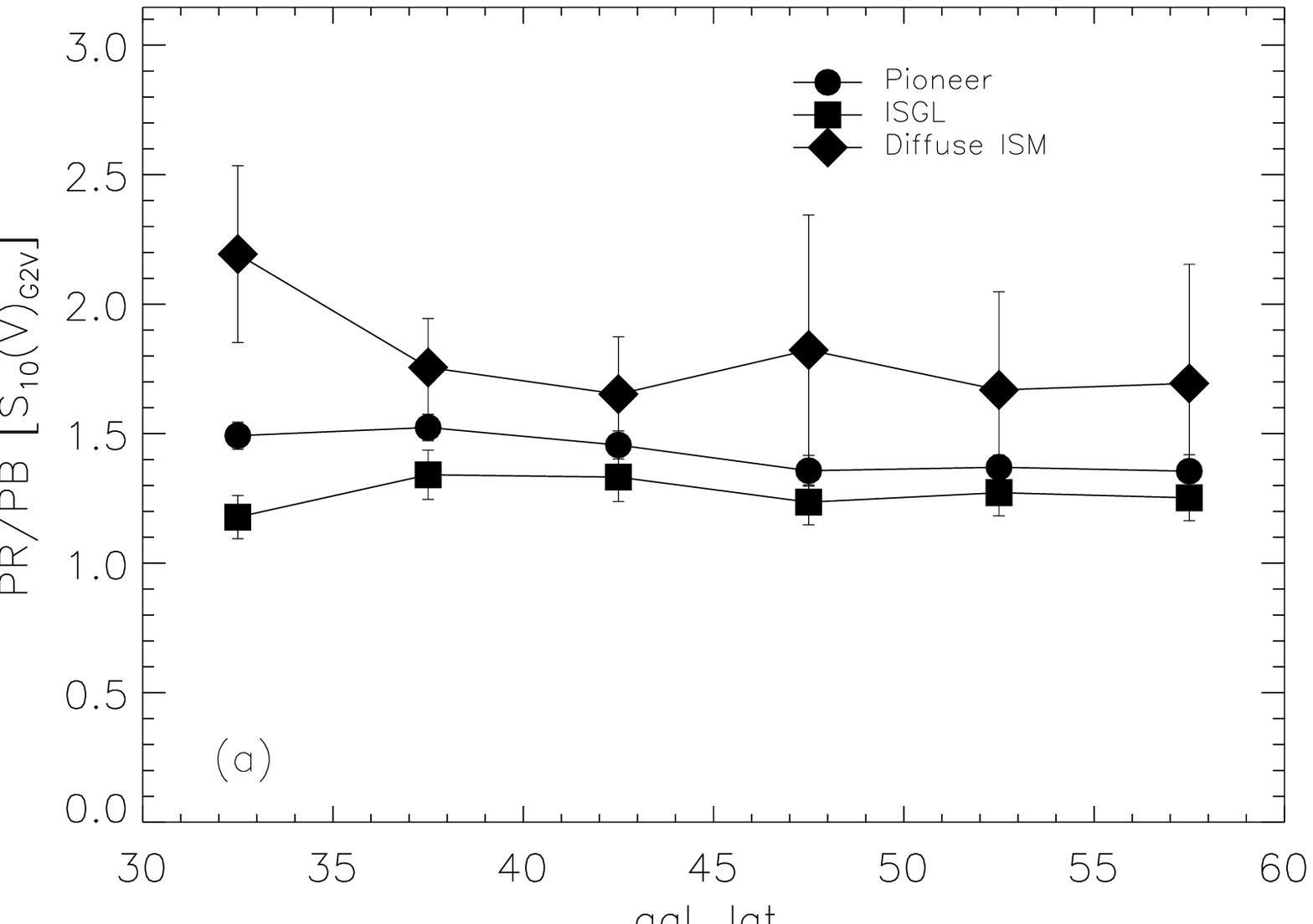}{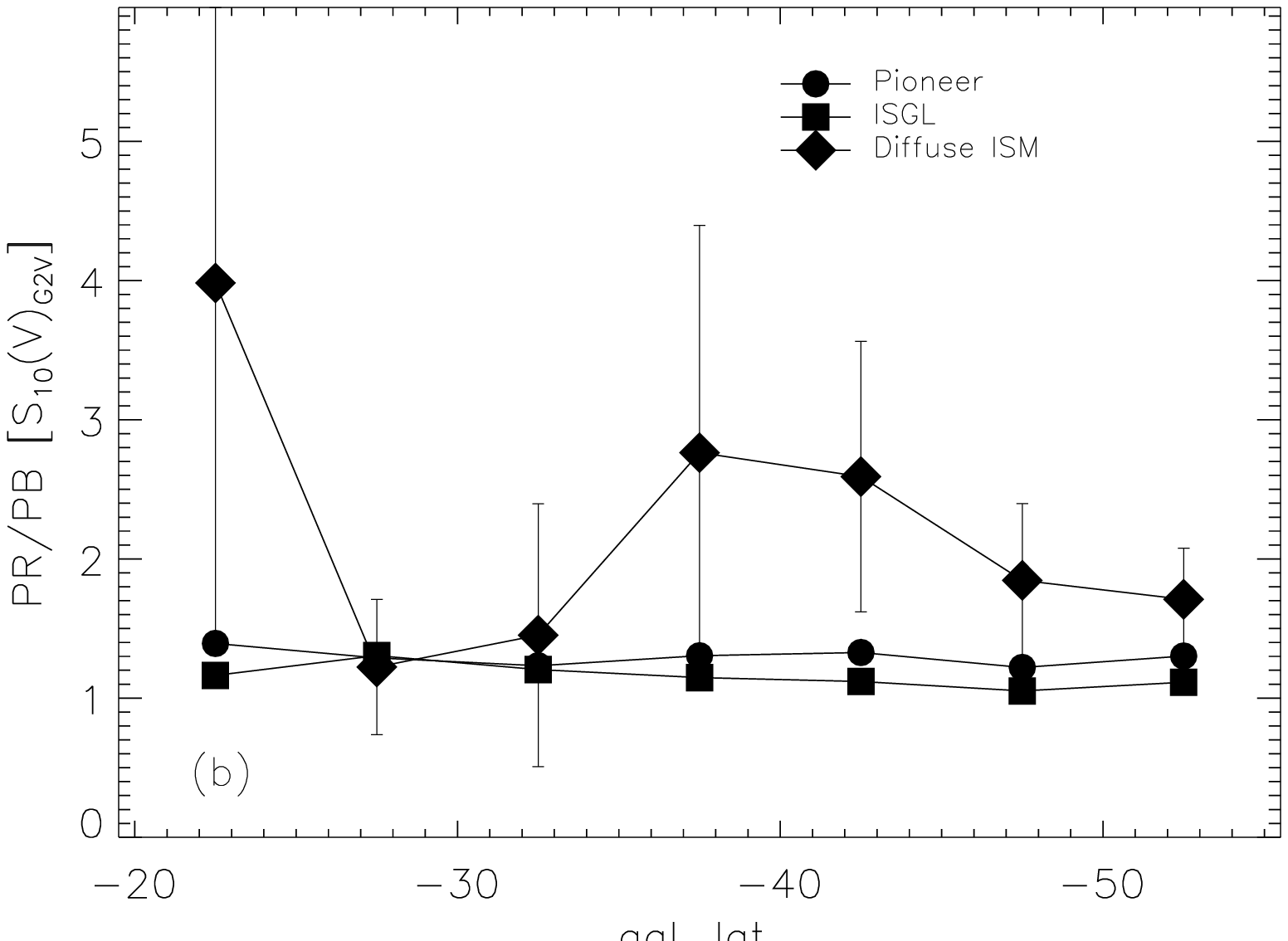}
\caption{Plotted is the red/blue ratio for the Pioneer measurements,
the ISGL, and the diffuse ISM.  The first plot (a) displays the cut in
galactic longitude between $0\arcdeg$ and $5\arcdeg$ and the second
plot (b) the cut between $95\arcdeg$ and $100\arcdeg$.
\label{fig_cut_ratio}}
\end{center}
\end{figure}
  
  Two different cuts in galactic longitude of the Pioneer and ISGL
intensities are displayed in Figure~\ref{fig_cut_pion_isl}.  The first
cut was taken from region~1 between galactic longitudes $0\arcdeg$ and
$5\arcdeg$.  The second cut was taken from region~2 between galactic
longitudes $95\arcdeg$ and $100\arcdeg$.  Other cuts in regions~1 and
2 were examined and found to be similar to those displayed in
Figure~\ref{fig_cut_pion_isl}.  The corresponding diffuse ISM
intensities were determined by subtracting the ISGL intensities from
the Pioneer intensities and are plotted in
Figure~\ref{fig_cut_diffuse}.  Figure~\ref{fig_cut_ratio} displays the
PR/PB ratios (in \steng units) for the Pioneer, ISGL, and diffuse ISM
intensities.  From Figure~\ref{fig_cut_ratio}, it is obvious that the
diffuse ISM is redder (larger PR/PB ratio) than either the Pioneer
measurements or the ISGL.  As the scattered component of the diffuse
ISM (DGL) is bluer (see \S\ref{sec_dgl_model}) than the Pioneer
measurements, this requires that a nonscattered component is present
in the red diffuse ISM intensity.

\subsection{Diffuse Galactic Light Model \label{sec_dgl_model}}

  In order to determine the strength of the nonscattered component of
the diffuse ISM red intensity, the scattered component in the red
(DGL) must be removed.  An accurate calculation of the DGL should
include the effects of multiple scattering, the cloudiness of the
interstellar medium, and the observed anisotropy of the illuminating
radiation field.  Such a model exists and it is the Witt-Petersohn DGL
model (WP model, \cite{wit94b}; \cite{fri96}; \cite{wit97}).  The WP
model treats the Galaxy as a gigantic reflection nebula.  The details
of the WP model and its use in the ultraviolet can be found in
Friedmann (1996) and Witt \etal (1997).  Below, the salient points
of the WP model and its adaptation to the optical are described.

    Application of the WP model to the present problem was as follows.
We derived the radiation field at different heights above and below
the galactic plane from the Pioneer all-sky intensity maps using the
work of Mattila (1980a,b).  The all-sky intensity maps were made by
adding the intensities of the bright stars which were removed during
the Pioneer data reduction (\cite{wei74}) to the Pioneer all-sky maps
after interpolating over the hole caused by the Sun's location (see
Section~\ref{sec_pioneer}).  The total dust optical depth along a
particular line-of-sight was computed from $\tau = C \times N_{HI}$
where the $N_{HI}$ values were from the Bell Laboratories \ion{H}{1}
Survey (\cite{sta92}) and the Parkes \ion{H}{1} survey (\cite{cle79};
\cite{hei79}).  The conversion constant, C, was $7.58 \times
10^{-22}~\mbox{cm}^2$ for PB and $4.71 \times 10^{-22}~\mbox{cm}^2$
for PR.  C was computed using the average Galactic extinction curve
(\cite{whi92}) and $\left< N_{HI}/E(\bv) \right> = 4.93 \times
10^{21}~\mbox{cm}^{-2}~\mbox{mag}^{-1}$ (\cite{dip94}).  The spectrum
of cloud sizes and optical depths, except for scaling to the PB and PR
bands, was the same as that in Witt \etal (1997).

  The WP model uses Monte Carlo techniques to describe the radiative
transfer through dust.  With these techniques, photons are followed
through a dust distribution and their interaction with the dust is
parameterized by the dust optical depth ($\tau$), albedo, and
scattering phase function asymmetry ($g$).  The optical depth
determines where the photon interacts, the albedo gives the
probability that the photon is scattered from a dust grain, and the
scattering phase function gives the angle at which the photon
scatters.

  In the blue (PB), reasonable values of the dust albedo and $g$ are
0.50--0.70 and 0.60--0.80, respectively (\cite{fit76}; \cite{tol81};
\cite{wit82}; \cite{wit84}; \cite{wit90b}).  Putting limits on the red
(PR) values of the dust albedo and $g$ was more difficult as the
objects best suited to studies of the dust albedo and $g$ (i.e.\
reflection nebulae) are just those objects with appreciable ERE
(\cite{wit84}; \cite{wit88}).  Fortunately, there are a few objects
without detectable ERE.  One such object is the Bok globule
investigated by Witt \etal (1990), who found that the dust albedo and
$g$ values in this Bok globule are smaller by $\sim$30\% in the red
than in the blue.  This albedo and $g$ difference between the blue and
the red is similar to that predicted by dust grain models
(\cite{kim96}; \cite{mat96}; \cite{zub96}; \cite{li97}), which predict
either roughly constant or decreasing albedo and $g$ values when
moving from the blue to the red.  Therefore to be conservative, we
have adopted red albedo and $g$ values equal to those in the blue.
This implies that the DGL intensities we computed in the red are the
upper limits.  The adopted dust grain parameters for the model runs
are listed in Table~\ref{table_model_runs}.

\begin{deluxetable}{ccccc}
\tablewidth{0pt}
\tablecaption{WP Model Runs \label{table_model_runs}}
\tablehead{\colhead{} & \multicolumn{2}{c}{PB} &
           \multicolumn{2}{c}{PR} \\
           \colhead{Run} & \colhead{albedo} & \colhead{$g$} &
           \colhead{albedo} & \colhead{$g$} }
\startdata
1 & 0.50 & 0.60 & 0.50 & 0.60 \nl
2 & 0.50 & 0.80 & 0.50 & 0.80 \nl
3 & 0.70 & 0.60 & 0.70 & 0.60 \nl
4 & 0.70 & 0.80 & 0.70 & 0.80 \nl
\enddata
\end{deluxetable}

\begin{figure}[tbp]
\begin{center}
\plottwo{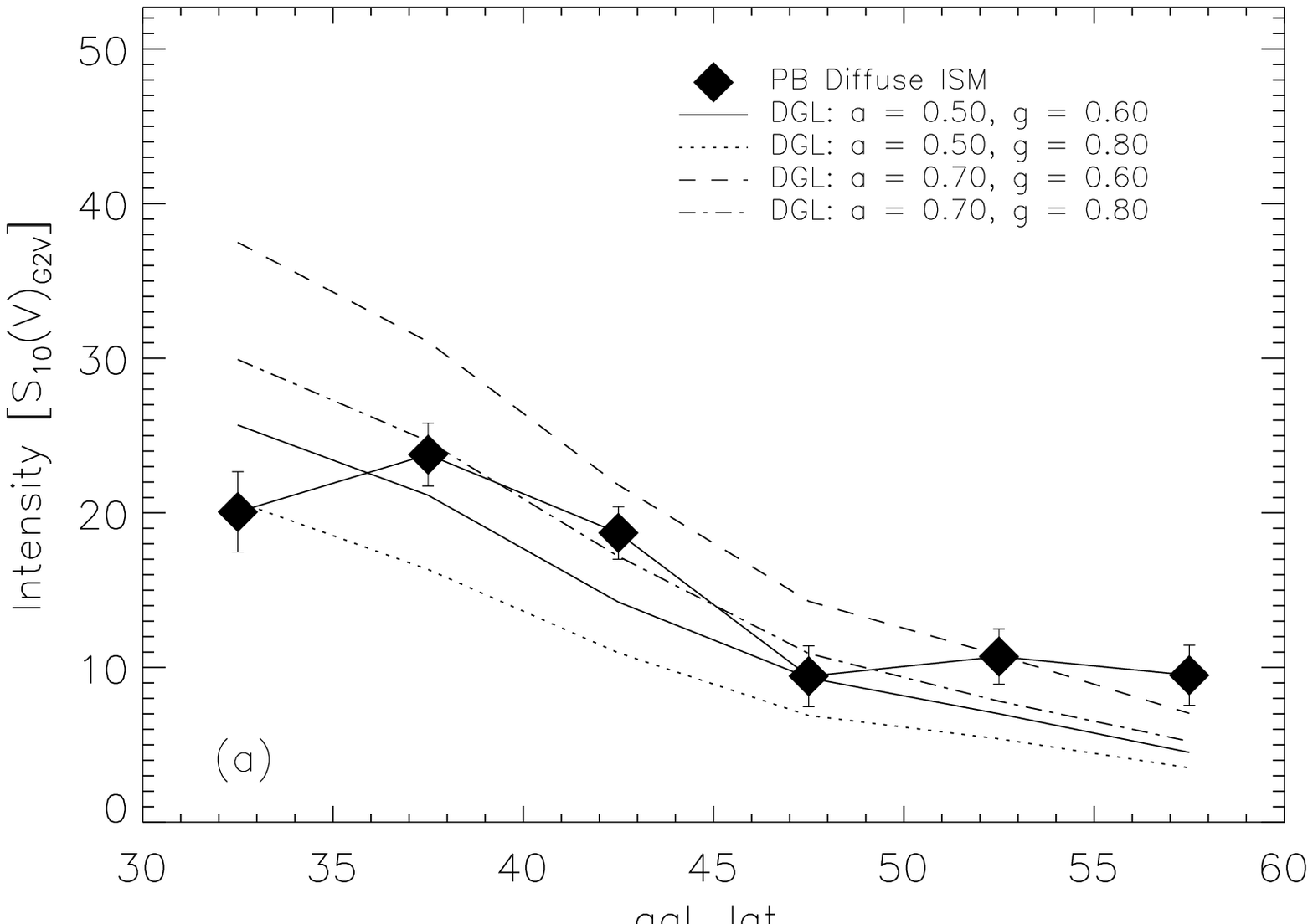}{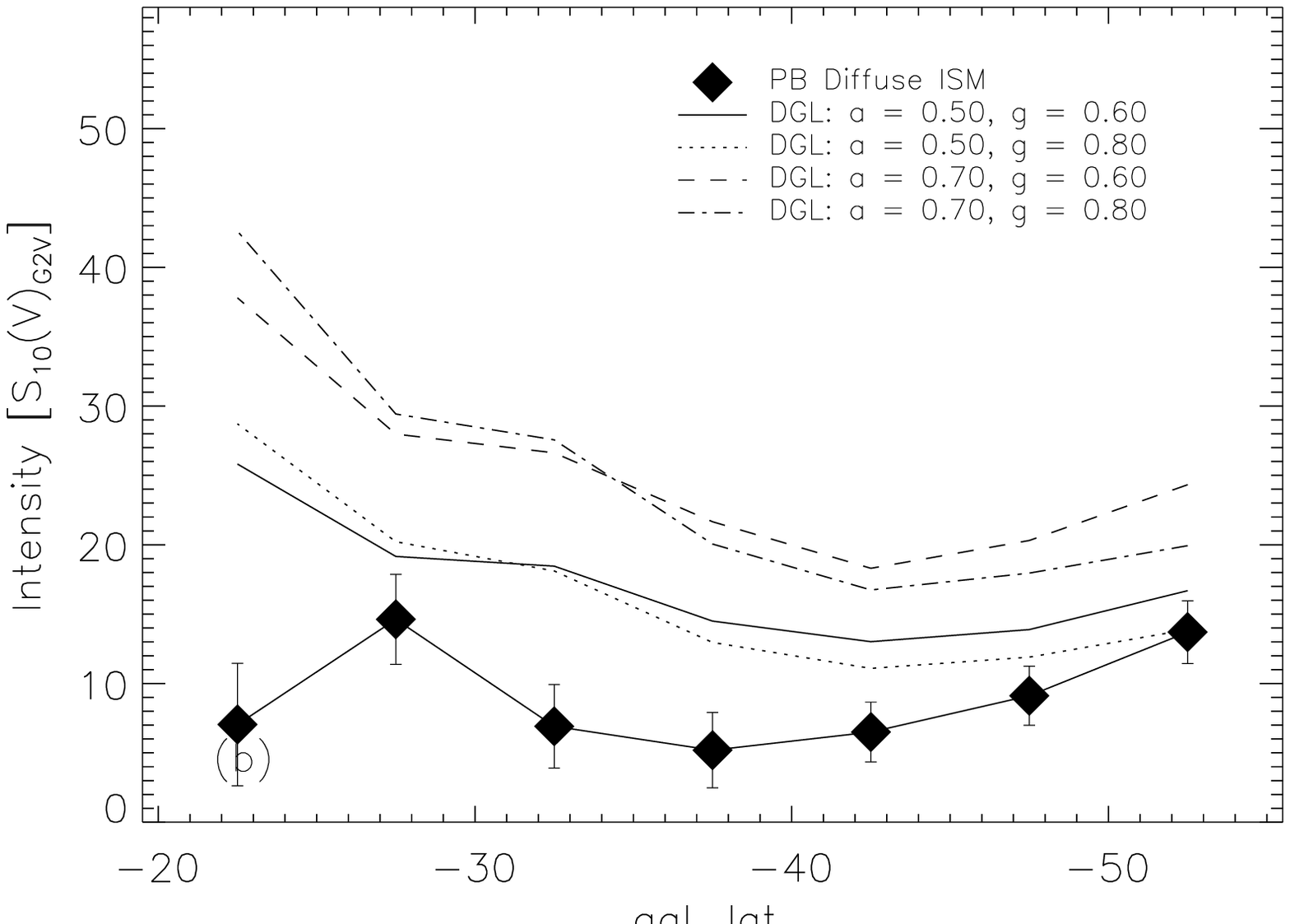}
\caption{The blue intensities for the diffuse ISM as well as the WP
model predictions are plotted for cuts in galactic longitude between
$0\arcdeg$ and $5\arcdeg$ (a) and $95\arcdeg$ and $100\arcdeg$ (b).
\label{fig_cut_PB_dgl}}
\end{center}
\end{figure}
  
\begin{figure}[tbp]
\begin{center}
\plottwo{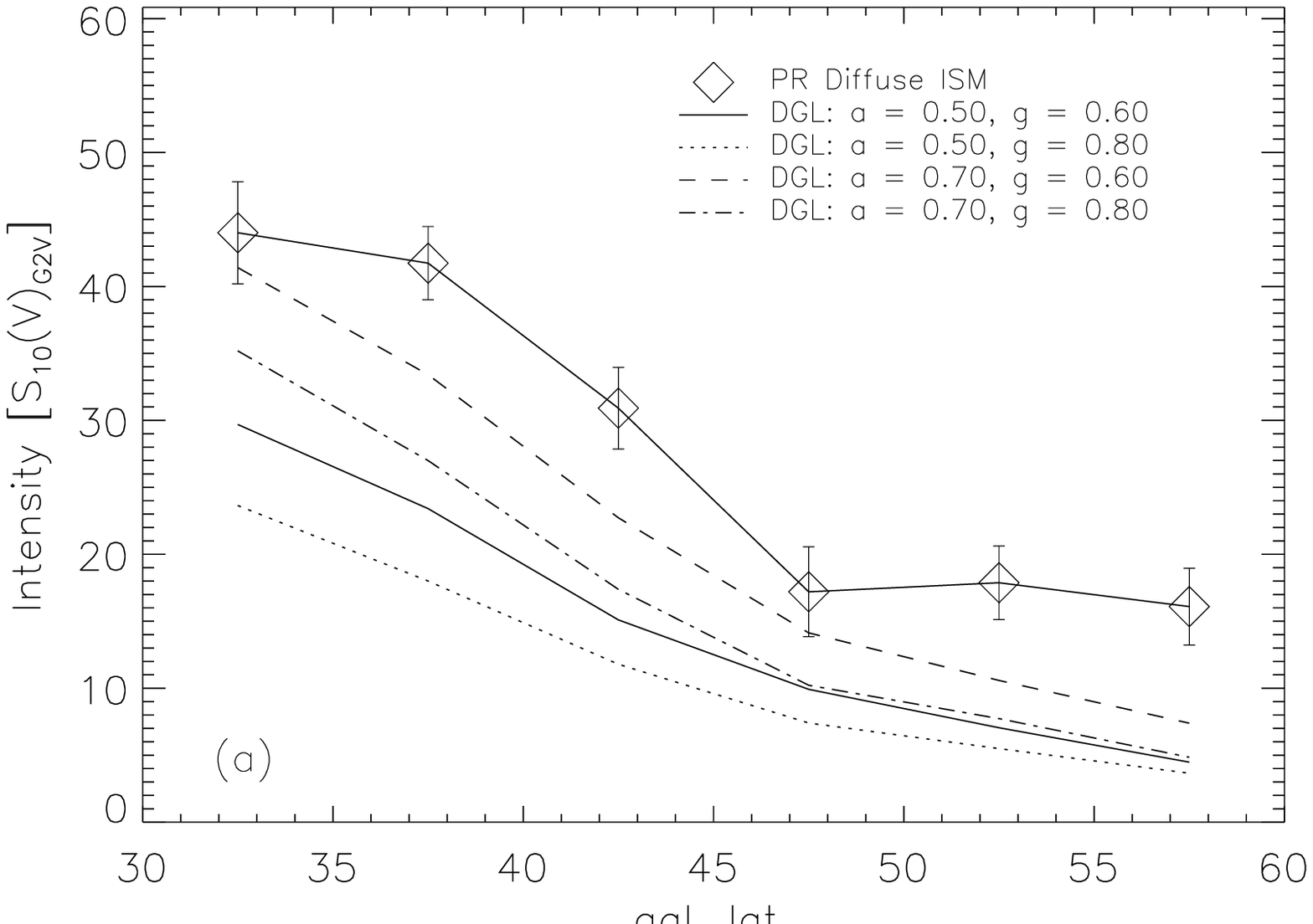}{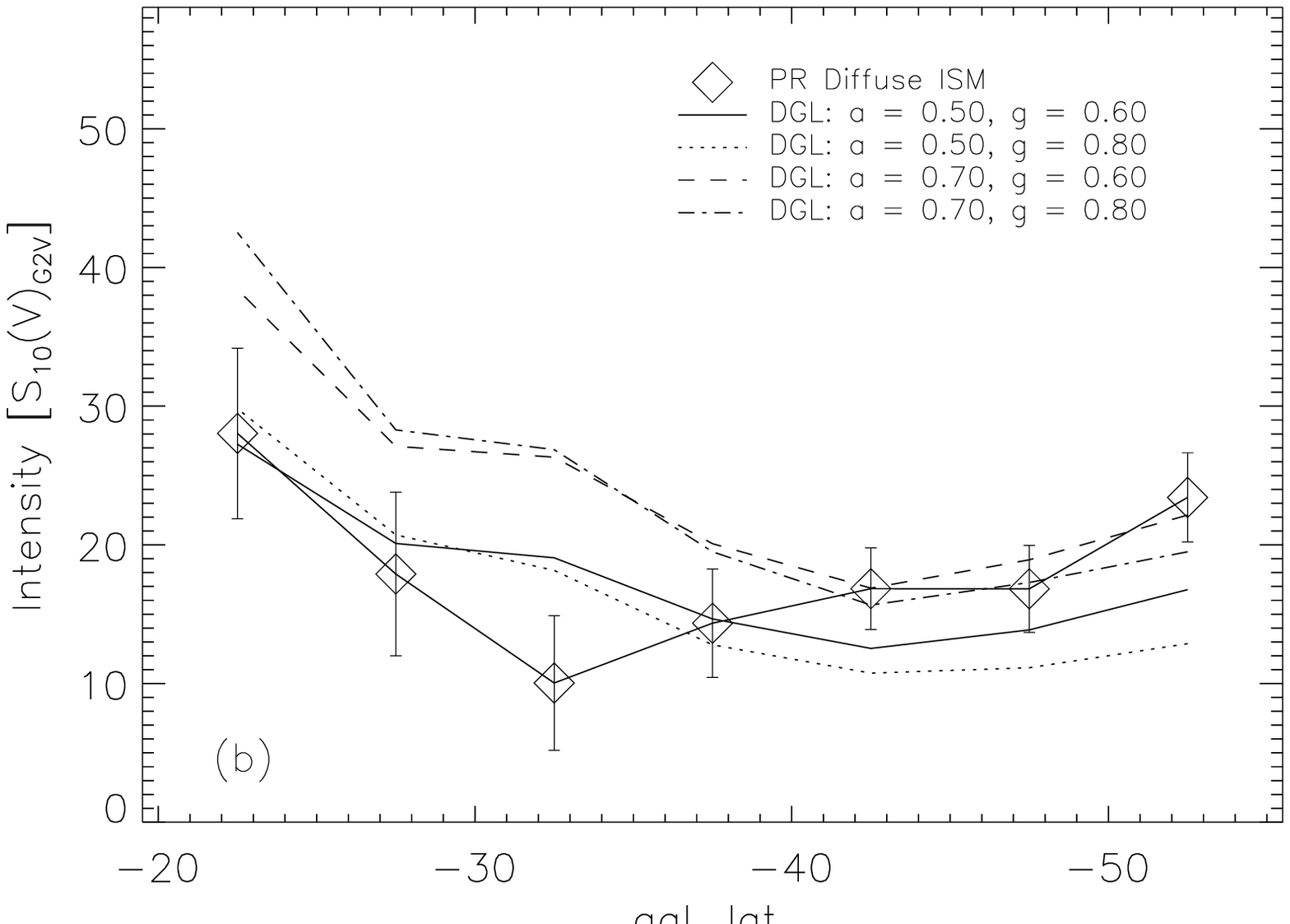}
\caption{The red intensities for the diffuse ISM as well as the WP
model predictions are plotted for cuts in galactic longitude between
$0\arcdeg$ and $5\arcdeg$ (a) and $95\arcdeg$ and $100\arcdeg$ (b).
\label{fig_cut_PR_dgl}}
\end{center}
\end{figure}
  
  A detailed comparison of the diffuse ISM intensities with those
predicted for the DGL by the WP model is presented in
Figures~\ref{fig_cut_PB_dgl} \& \ref{fig_cut_PR_dgl}.  The blue
diffuse ISM intensities in Figure~\ref{fig_cut_PB_dgl}a fall well
within the WP model predictions, confirming that the blue light from
the diffuse ISM is entirely attributable to the DGL.  This is not the
case for Figure~\ref{fig_cut_PB_dgl}b where the blue diffuse ISM
intensities generally have the same galactic latitude dependence as
the WP model predictions, but fall consistently lower.  The
differences between the WP model predictions and the actual blue
diffuse ISM intensities are not surprising as the WP model DGL
predictions apply to an average value of $\left< N_{HI}/E(\bv)
\right>$, average dust grain parameters (albedo and $g$), and a
radiation field derived from the intensities seen at the Earth's
position in the Galaxy.  The radiation field seen by a dust cloud
located high above (or below) the Galactic plane could be different
from what we calculated using the Pioneer maps and the model of
Mattila (1980a,b).  The dust grain properties along particular
lines-of-sight are known to vary (\eg \cite{gor94} and references
therein; \cite{cal95}; \cite{wit97}).  In addition, the Galactic
oxygen abundance gradient found by Smartt \& Rolleston (1997) implies
a dependence in the dust grain properties and $\left< N_{HI}/E(\bv)
\right>$ with galactic longitude since oxygen is a primary grain
component.  In fact, significant variations in $\left< N_{HI}/E(\bv)
\right>$ have been measured by Diplas \& Savage (1994).  The
disagreement between observations and model intensities seen in
Figure~\ref{fig_cut_PR_dgl}b makes it clear that a Galaxy-wide study
of the DGL might reveal important evidence about the dependence of
dust properties on galactic longitude.

\begin{figure}[tbp]
\begin{center}
\plottwo{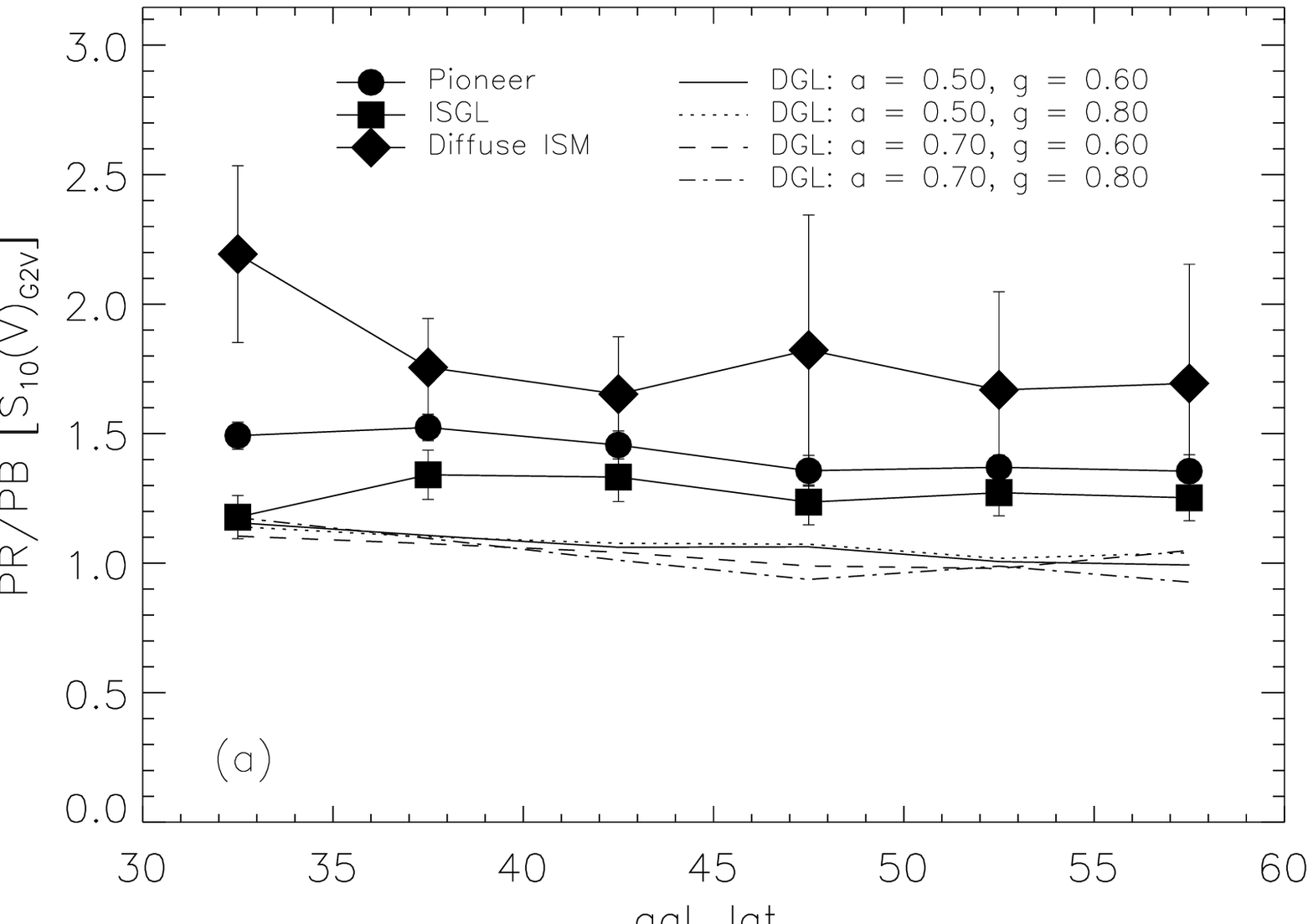}{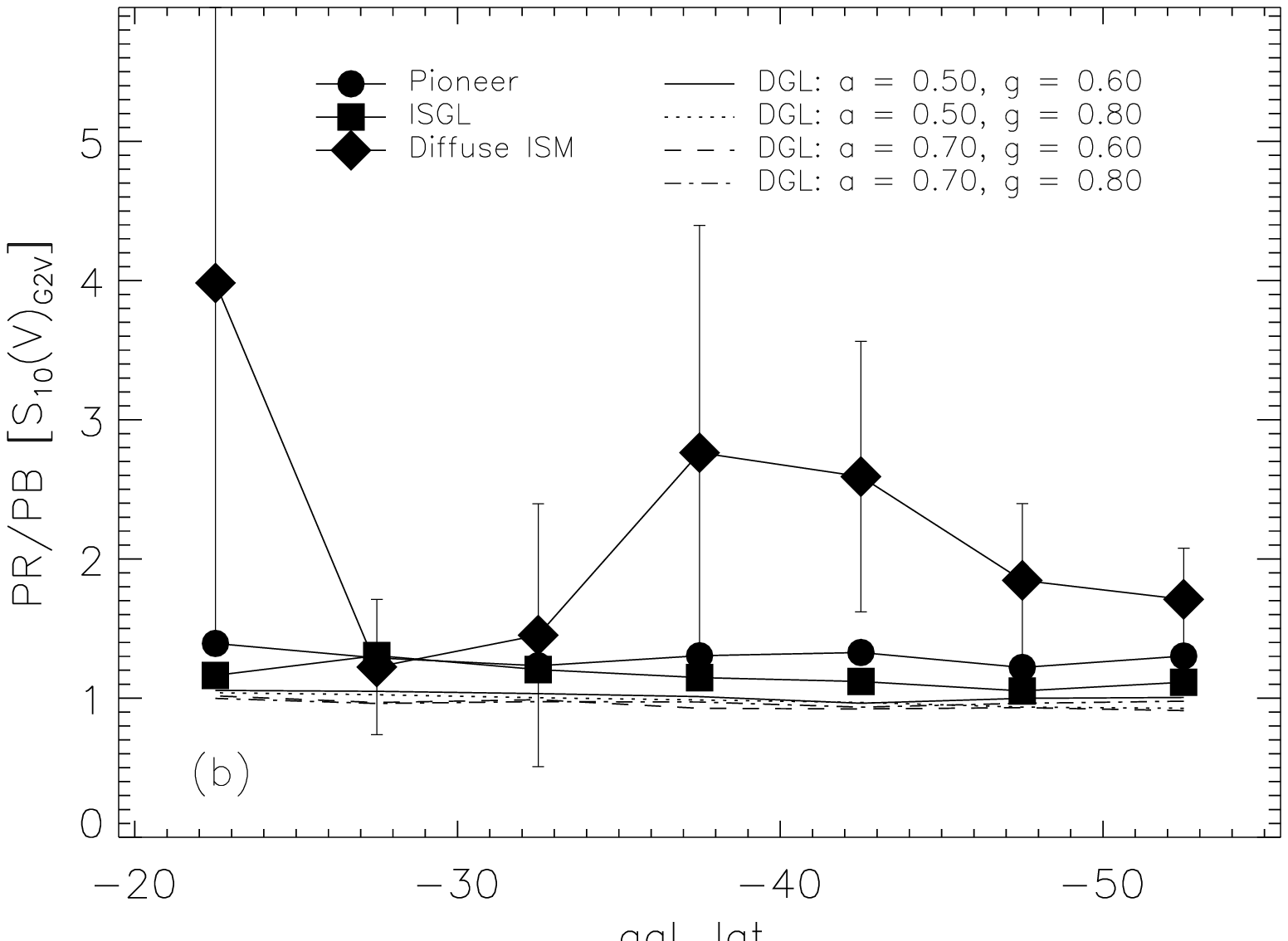}
\caption{Plotted is the red/blue ratio for the Pioneer measurements,
the ISGL, the diffuse ISM, and all four of the WP model runs.  The
first plot (a) displays the cut in galactic longitude between
$0\arcdeg$ and $5\arcdeg$ and the second plot (b) the cut between
$95\arcdeg$ and $100\arcdeg$.
\label{fig_cut_ratio_dgl}}
\end{center}
\end{figure}
  
  While the above three points limit the applicability of the WP model
DGL intensity predictions, the {\em color} of the DGL is fairly
independent of these points because the DGL color is determined almost
entirely by wavelength dependence of the dust grain properties (albedo
and optical depth) between the blue and the red.  The values of the
dust grain properties we used in the DGL model were derived from both
observations and dust grain models.  The dependence of the DGL color
on only the wavelength dependence of the dust grain properties is
illustrated in Figure~\ref{fig_cut_ratio_dgl} which plots the PR/PB
ratio for the Pioneer, ISGL, diffuse ISM, and DGL model intensities.
The PR/PB ratio (color) of all four of the WP model runs is
approximately unity, with a slight gradient in galactic latitude.
This figure clearly shows the presence of excess red intensity over
that expected from scattered red photons as the average observed PR/PB
ratio of the diffuse ISM is approximately two.

  Our results for the presence of an excess red intensity in the
diffuse ISM are very similar to the findings of Guhathakurta \& Tyson
(1989) and Guhathakurta \& Cutri (1994) who found that the (\br)
colors of individual IRAS cirrus clouds were 0.5--2 magnitudes redder
than that expected for scattered light.  The indentification of the
red color with ERE in cirrus clouds has been confirmed by Szomoru \&
Guhathakurta (1998) through optical spectroscopy of cloud edges.  From
Figure~\ref{fig_cut_ratio_dgl}, the color of the diffuse ISM light is
0.5--4.0 times redder than the DGL.  This corresponds to a
($PB\!-\!PR$) color 0.4--1.5 magnitudes redder than that expected for
the DGL.  Our results cover a significantly larger region of the the
sky (1135~$\sq\arcdeg$) than the work of Guhathakurta \& Tyson (1989)
and Guhathakurta \& Cutri (1994) which was for individual regions up
to 1~$\sq\arcdeg$.

\subsection{ERE Intensity}

  In order to determine the excess red intensity strength, the
contribution from the red DGL must be subtracted from the red diffuse
ISM intensity.  As the WP model is not sufficiently complex to
reproduce the blue DGL exactly, we could not use the WP model
predictions of the red DGL.  On the other hand, the {\em color} of the
DGL will not be greatly affected by these deficiencies (described at
the end of \S\ref{sec_dgl_model}).  Therefore, we used the average DGL
color predicted by the WP model runs (see
Figure~\ref{fig_cut_ratio_dgl}) along with the {\em observed blue}
diffuse ISM intensity to calculate the expected intensity of the DGL
in the red.  This results in a predicted red DGL which accurately
reflects the variations of the dust grain properties, $\left<
N_{HI}/E(\bv) \right>$, and radiation field.

\begin{figure}[tbp]
\begin{center}
\plottwo{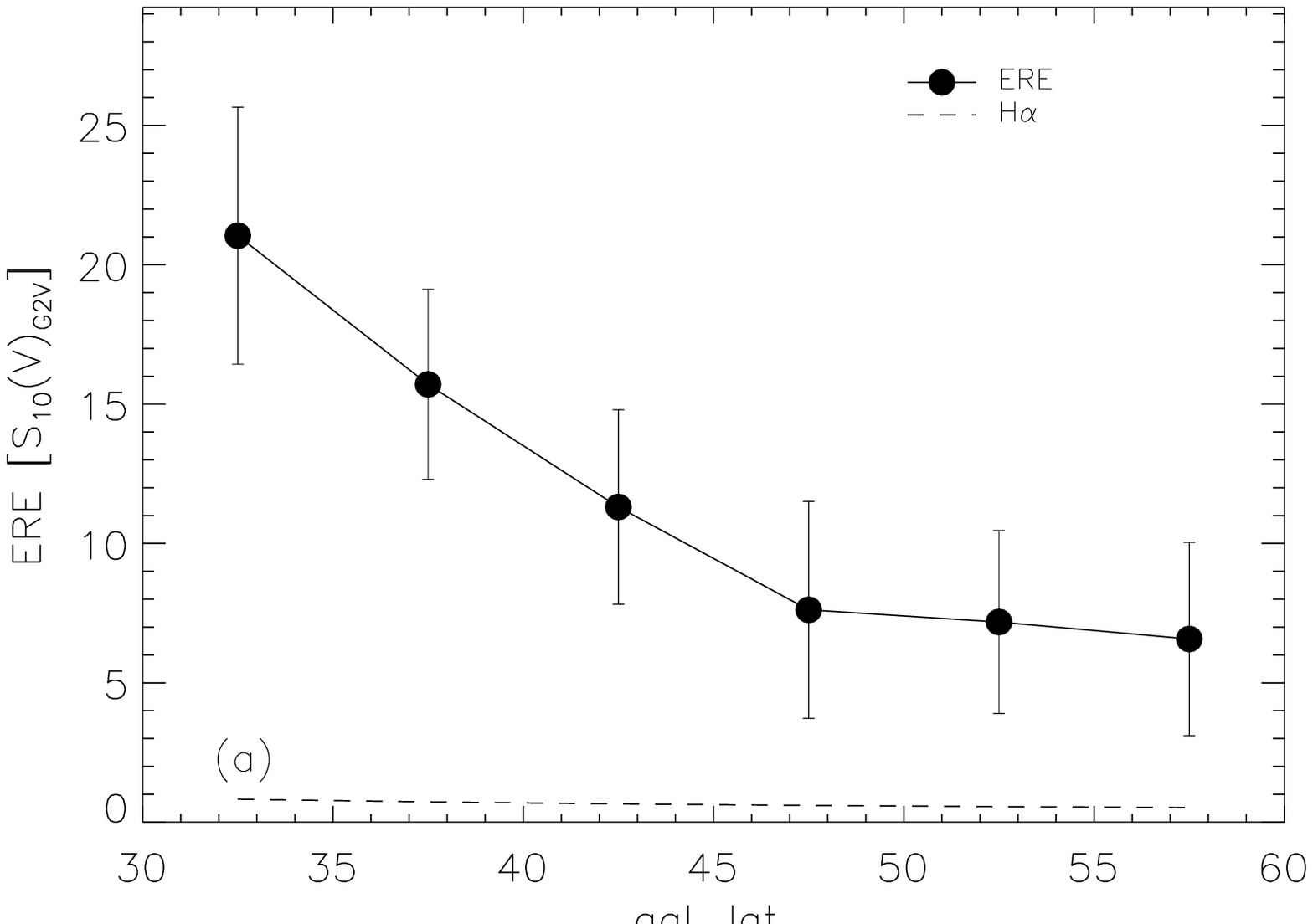}{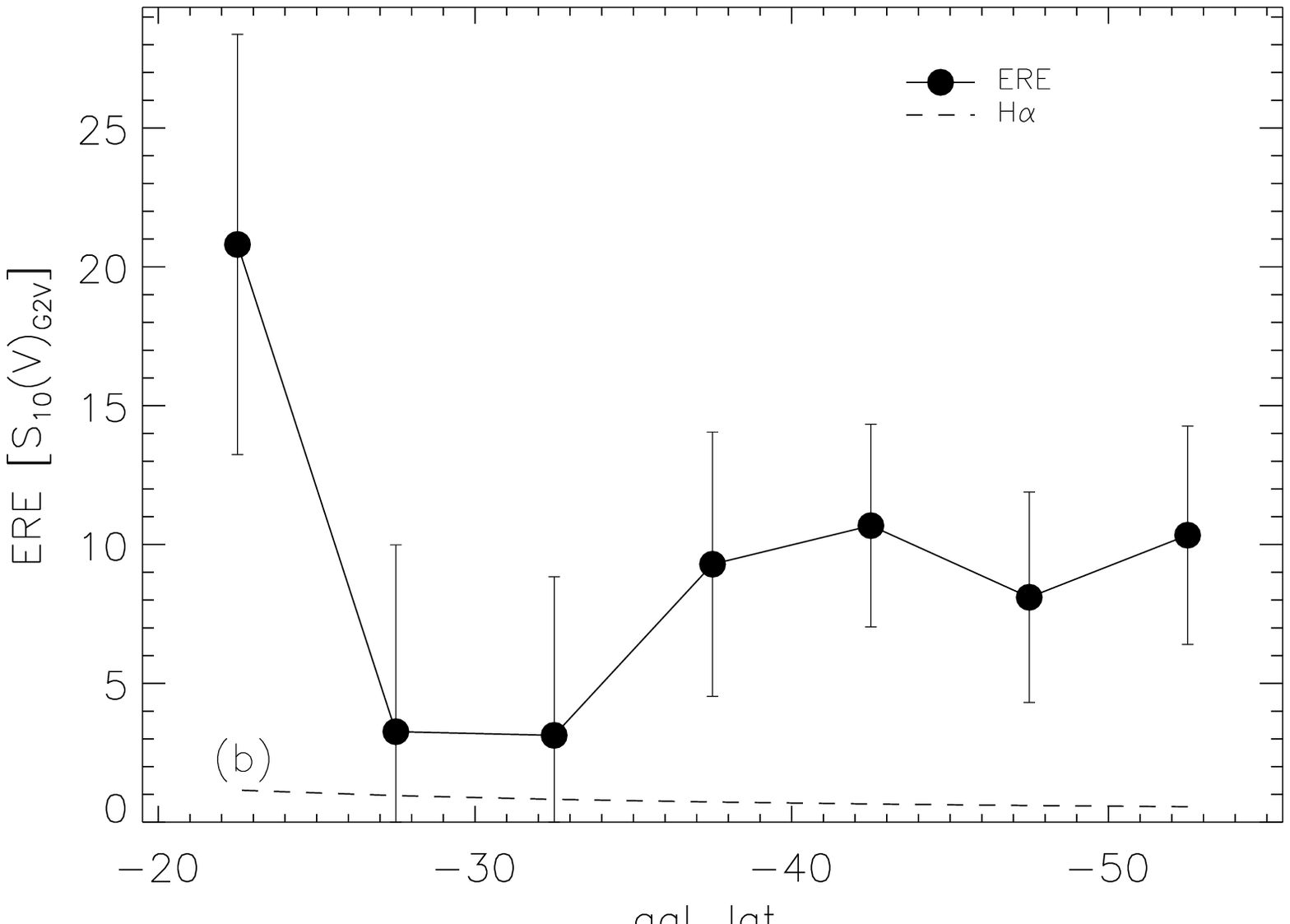}
\caption{The ERE intensities are plotted.  The ERE intensities were
determined by subtracting the predicted red DGL intensities from the
red diffuse ISM intensities.  The expected intensity from diffuse ISM
H$\alpha$ emission (Reynolds 1984) is also plotted using the
conversion of 1 rayleigh = 0.26 \steng.  The first plot (a) displays
the cut in galactic longitude between $0\arcdeg$ and $5\arcdeg$ and
the second plot (b) the cut between $95\arcdeg$ and $100\arcdeg$.
\label{fig_cut_ere}}
\end{center}
\end{figure}
  
  The red DGL intensities calculated using the above prescription were
subtracted from the red diffuse ISM intensities to yield a lower limit
to the red intensity of the nonscattered component of the diffuse ISM.
These red intensities along with the H$\alpha$ intensity expected from
H recombination in the diffuse ISM (\cite{rey84}) for the two cuts
displayed in previous figures are plotted in Figure~\ref{fig_cut_ere}.
Clearly, H$\alpha$ emission cannot explain more than a small fraction
of the nonscattered component of the red diffuse ISM intensity.  This
same conclusion was reached by Guhathakurta \& Tyson (1989) in their
study of Galactic cirrus clouds and has recently been confirmed
through optical spectroscopy of cirrus cloud edges (\cite{szo98}).  We
identify the excess red intensity with ERE, as there are no other
known red emission processes in the diffuse ISM.  The ERE strength we
determined is a lower limit on the true ERE strength as the red DGL
intensity was calculated from the WP model color prediction which was
an upper limit on the true DGL color.

\section{Properties of ERE in the diffuse ISM
\label{sec_ere_correlations}} 

\subsection{ERE Correlations}

  In order to check our identification of the red excess detected in
the previous subsection with ERE, we sought correlations of the ERE
intensity with known tracers of dust: \ion{H}{1} column densities
(\cite{cle79}; \cite{hei79}; \cite{sta92}) and the COBE Diffuse
Infrared Background Experiment (DIRBE) 100, 140, and 240 \micron\
intensities (\cite{hau97}; \cite{tol97}).

\begin{figure}[tbp]
\begin{center}
\plottwo{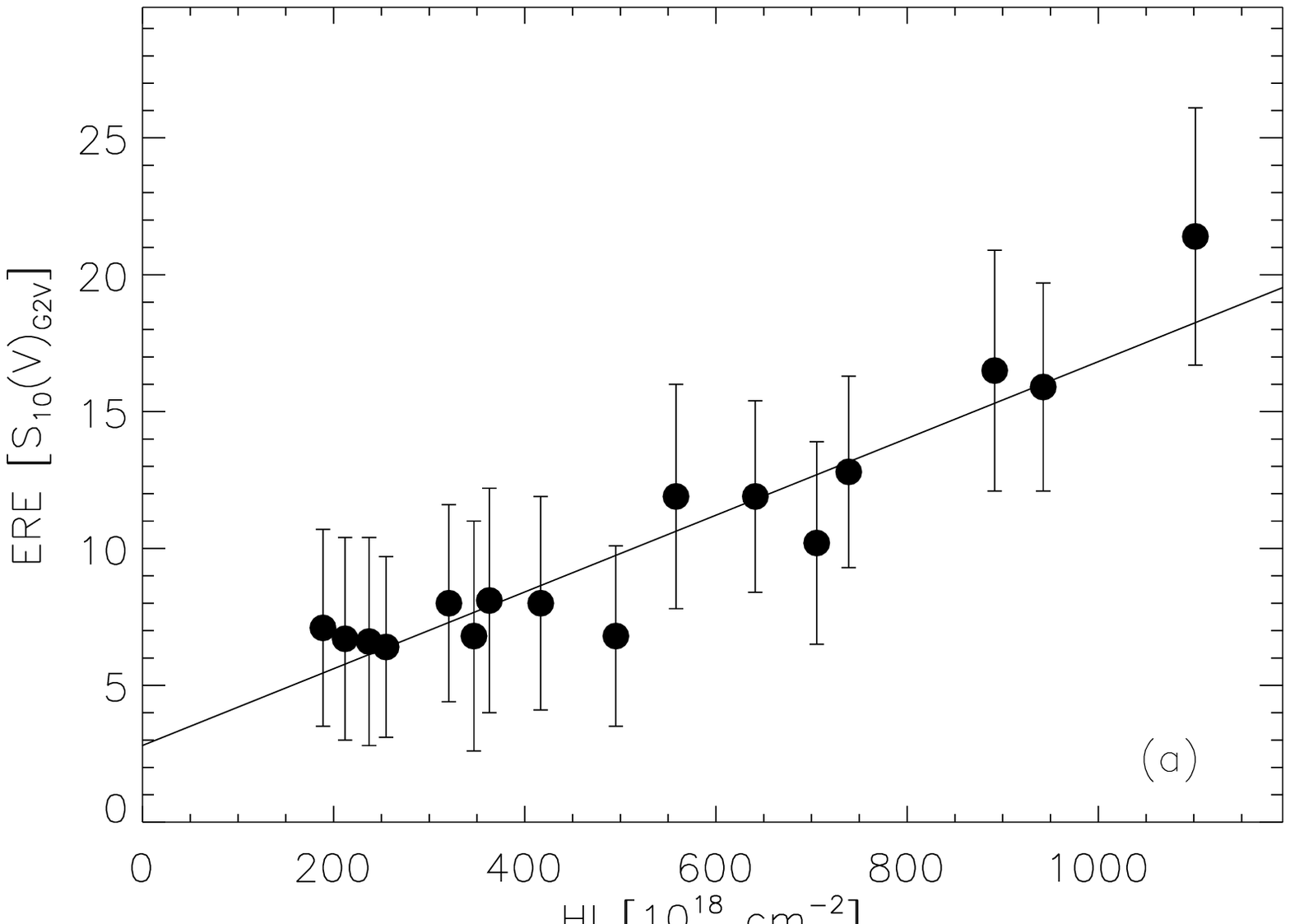}{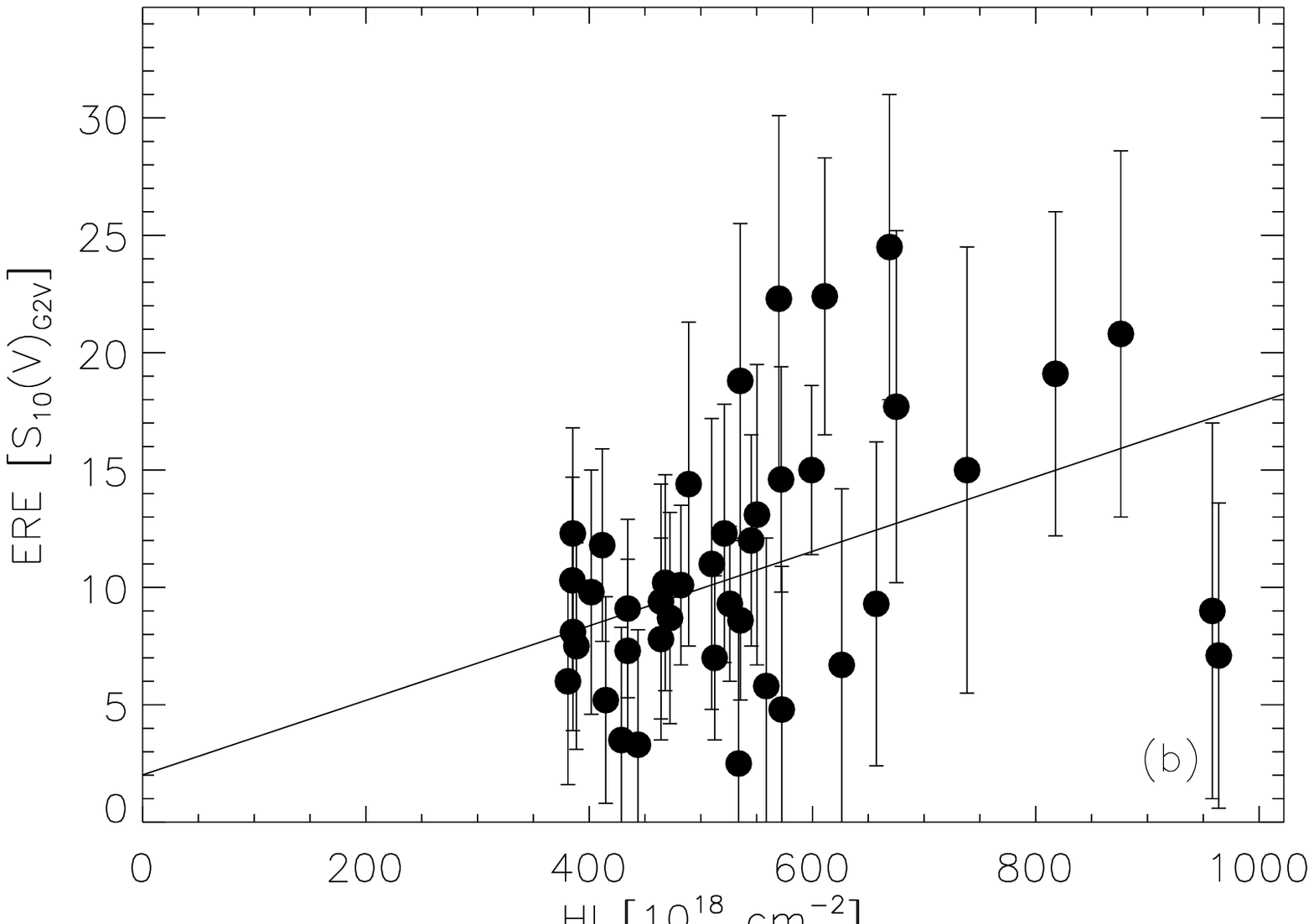}
\caption{This figure displays the correlation between H I and ERE
intensity for regions 1 (a) and 2 (b). \label{fig_cor_hI_ere}}
\end{center}
\end{figure}
  
  The correlations between \ion{H}{1} column densities and ERE
intensities for regions~1 and 2 are displayed in
Figure~\ref{fig_cor_hI_ere}.  The points in
Figures~\ref{fig_cor_hI_ere}a and b were taken from the entire areas
of regions~1 and 2 and include the representative cuts displayed in
Section~\ref{sec_detect_red}.  The best fit line for region 1 was $y =
(2.80 \pm 2.05) + (0.0140 \pm 0.0036)x$.  The best fit line for region
2 was $y = (2.01 \pm 3.41) + (0.0159 \pm 0.0066)x$.  The best fit
lines were determined using the IDL function LINFIT.  The y-intercepts
for both lines are consistent, within the uncertainties, with zero and
the correlations for both regions are equivalent within the
uncertainties.  Combining the two regions yielded a best fit line of
$y = (2.67 \pm 1.71) + (0.0145 \pm 0.0032)x$.  This best fit line
gives the relationship between the ERE energy emitted per H atom and
it is $(1.43 \pm 0.31)$\sn{-29}~ergs~s$^{-1}$~\AA$^{-1}$~sr$^{-1}$ H
atom$^{-1}$.

\begin{figure}[tbp]
\begin{center}
\plottwo{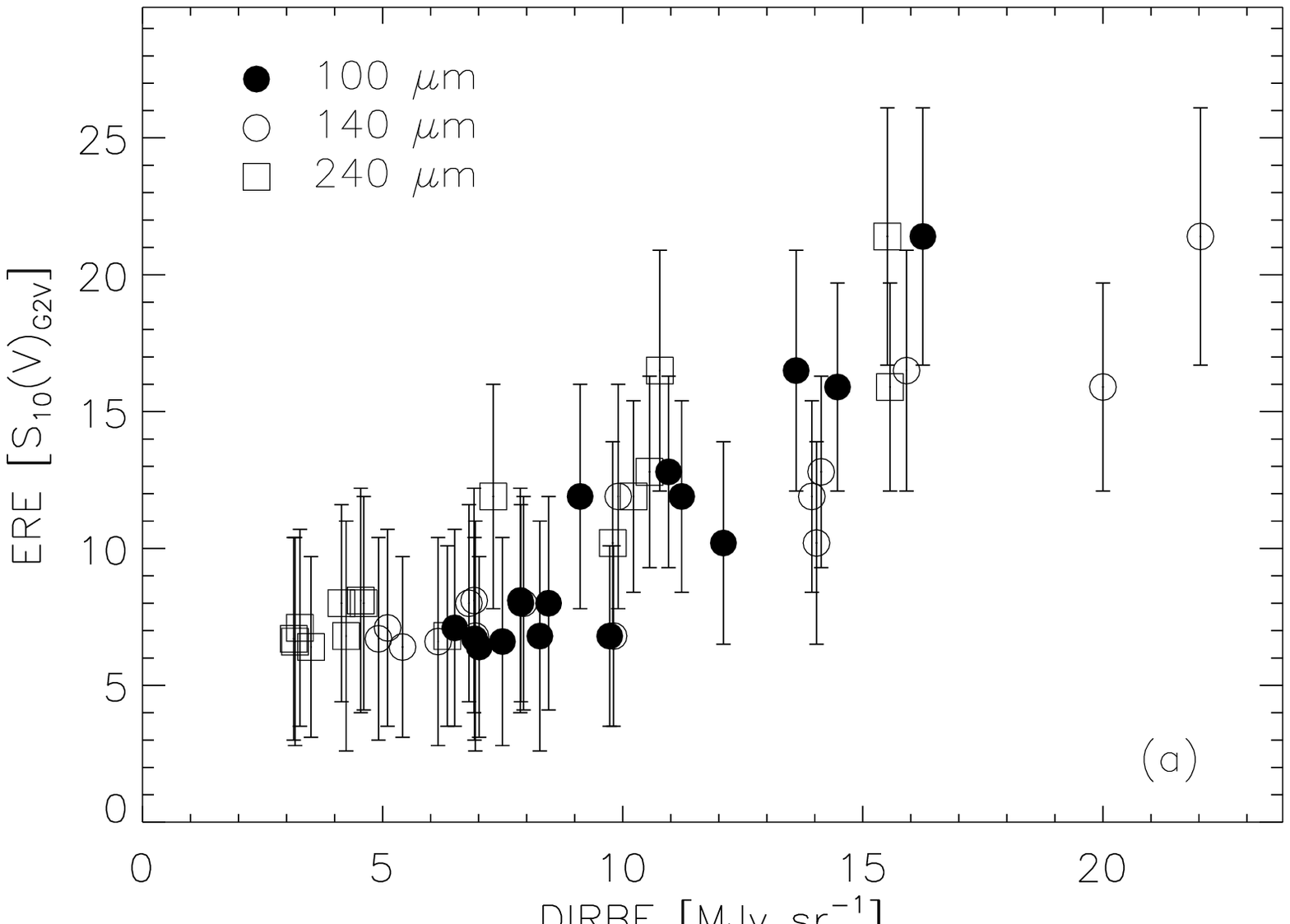}{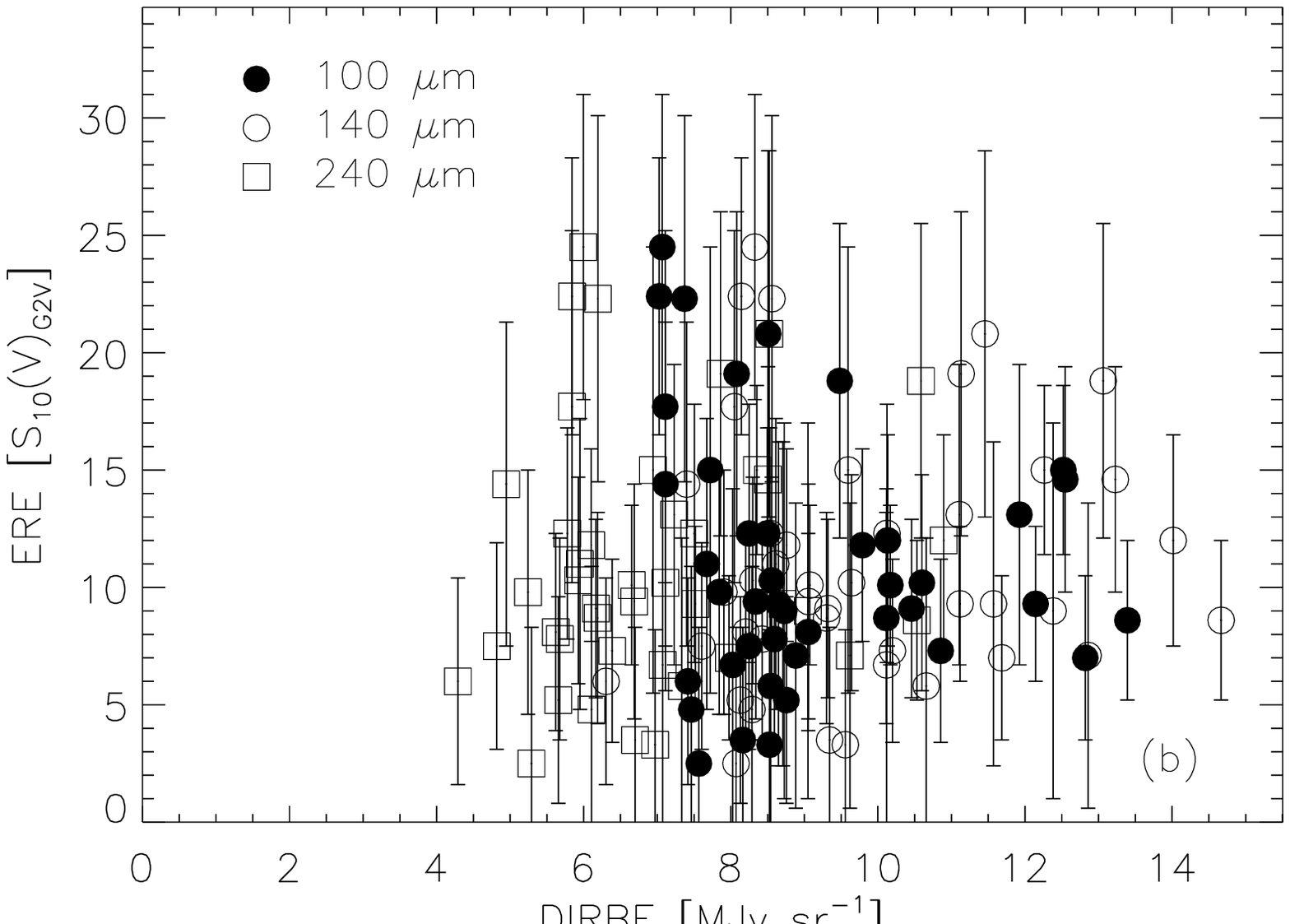}
\caption{This figure displays the correlation between the DIRBE 100,
140, and 240 \micron\ intensities and ERE intensity for regions 1 (a)
and 2 (b).  \label{fig_cor_dirbe_ere}}
\end{center}
\end{figure}
  
\begin{figure}[tbp]
\begin{center}
\plottwo{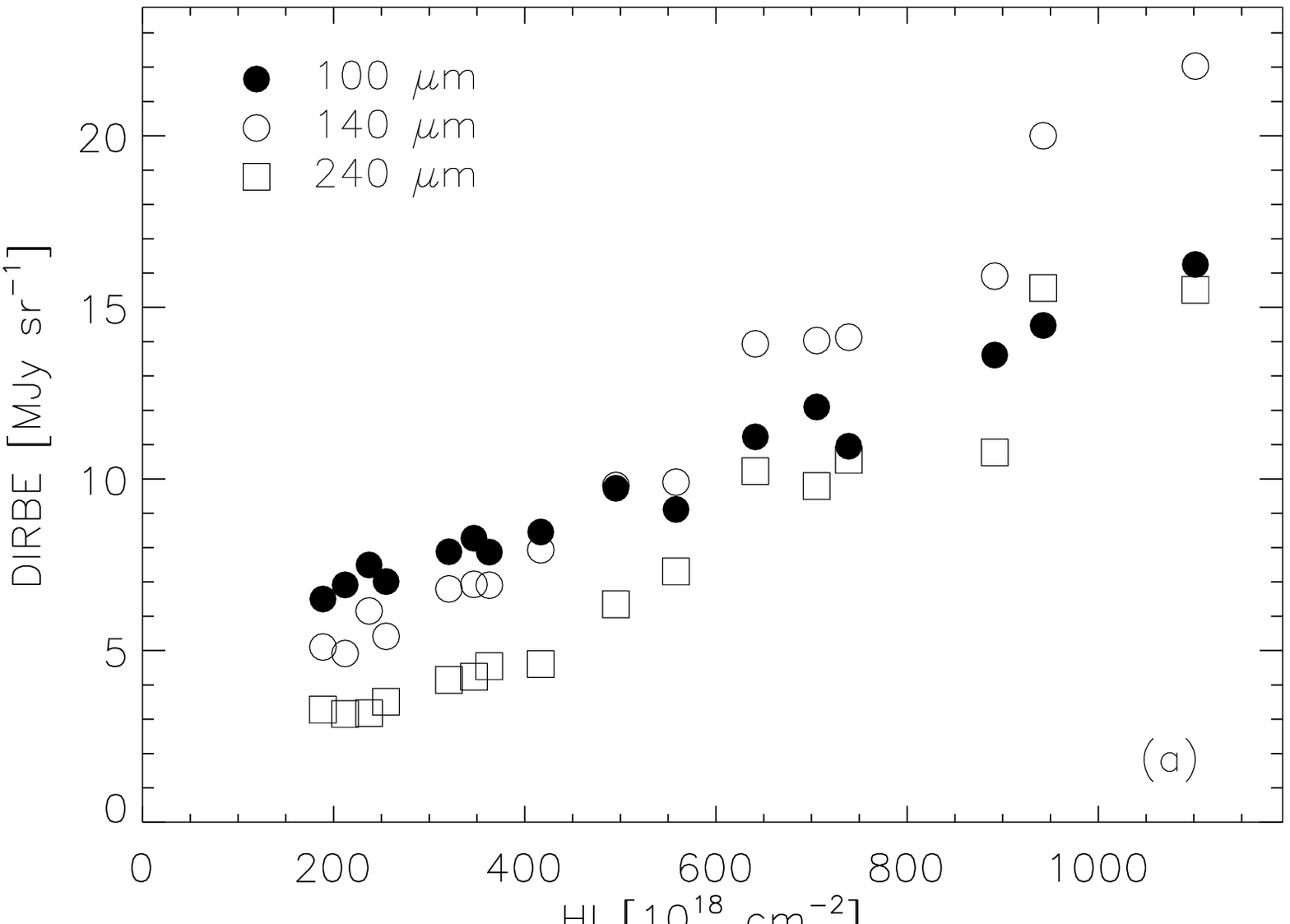}{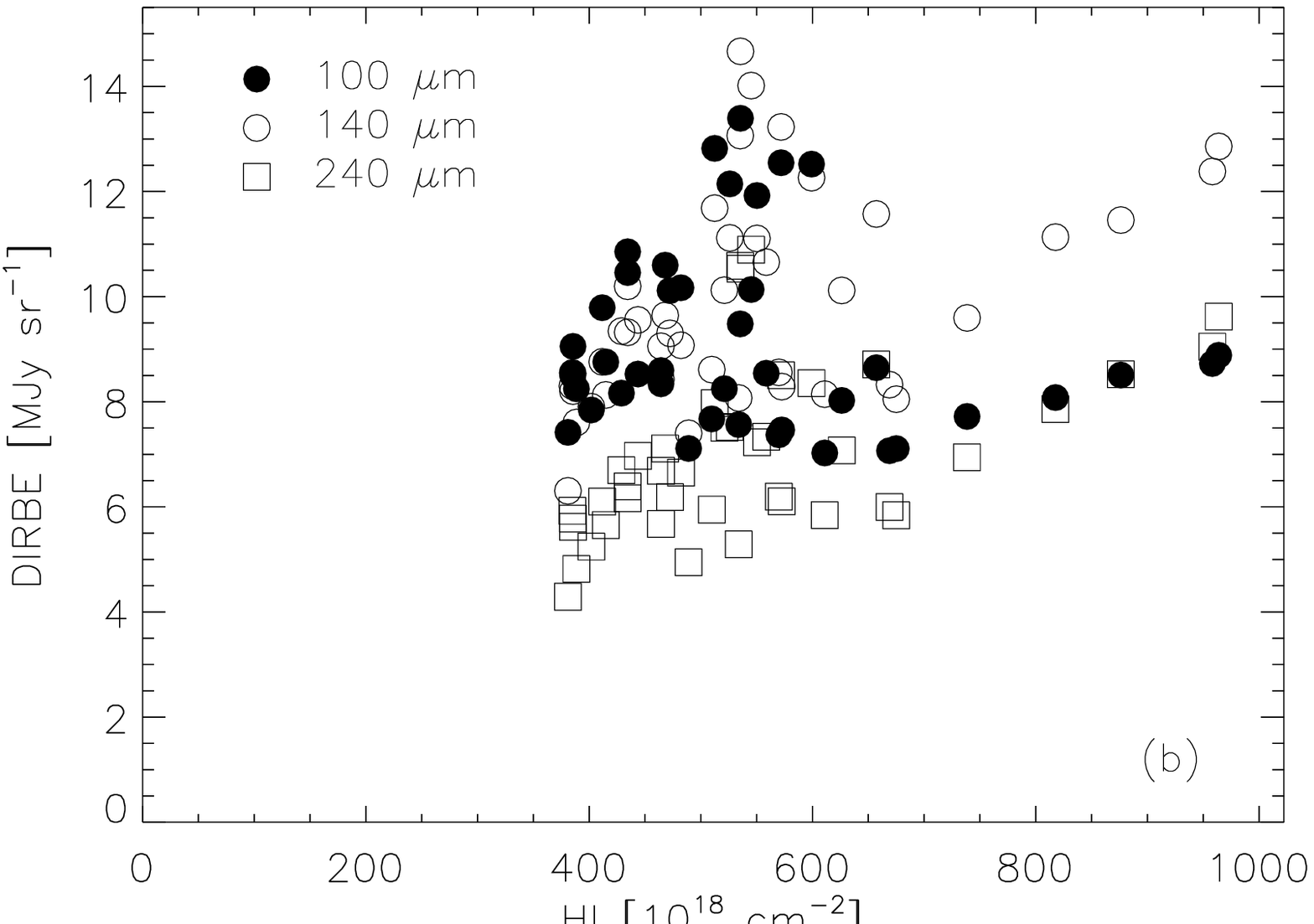}
\caption{This figure displays the correlation between the DIRBE 100,
140, and 240 \micron\ intensities and H I column densities for regions
1 (a) and 2 (b). \label{fig_cor_hI_dirbe}}
\end{center}
\end{figure}
  
  The correlations between the DIRBE 100, 140, and 240 \micron\ and
ERE intensities are plotted in Figure~\ref{fig_cor_dirbe_ere}.  Unlike
region 1, there was no clear relationship between the DIRBE and ERE
intensities for region 2.  This was troubling as the correlations
between the DIRBE and ERE intensities for region 1 was quite good.
The origin of the poor correlations for region 2 is related to the
fact that the DIRBE data were collected from a satellite in near-Earth
space.  While the DIRBE data have had the majority of the zodiacal
dust emission subtracted, there is a noticeable residual left.  This
is illustrated in Figure~\ref{fig_cor_hI_dirbe}b which shows the
correlations between the DIRBE intensities and \ion{H}{1} column
densities for region~2.  The peak in the DIRBE intensities around
$N_{HI} \sim 550 \times 10^{18}~\mbox{cm}^{-2}$ is due to residual
zodiacal dust emission as the ecliptic plane runs through the middle
of region~2.

  The ERE intensity was well correlated with dust tracers for
region~1.  For region~2, the ERE strength was correlated with
\ion{H}{1} column density but not DIRBE intensities.  The poor
correlation between the ERE and DIRBE intensities was explicable when
the presence of residual zodiacal dust emission was taken into
account.  Thus, the identification of the red nonscattered excess
intensity in the diffuse ISM with ERE was greatly strengthened.
Additionally, this is the first time in which it has been possible to
directly correlate the strength of the ERE emission with known dust
tracers.

\subsection{ERE Photon Efficiency}

  The process which produces ERE is thought to be photoluminescence
from a disordered solid like HAC.  Basically, a UV or blue photon
excites an electron from the valence to the conduction band and the
election relaxes back to the conduction band via a number of
(mostly non-radiative) transitions which might include one radiative
transition.  When this radiative transition occurs, an ERE photon is
emitted (\cite{fur94}).  At most, one ERE photon is emitted per
absorbed UV or blue photon.  The photon efficiency is the percent of
photons which are absorbed by the ERE producing material which cause
the emission of ERE photons.

  We calculated the lower limit on the photon efficiency by assuming
all the photons absorbed by dust are absorbed by the ERE producing
material.  The calculation is fairly straightforward, but required a
couple of assumptions.  The calculation begins with the relationship 
between the ERE intensity and \ion{H}{1} column density determined
in the previous section.  The number of ERE photons emitted per
H atom was calculated from
\begin{eqnarray}
\frac{\mbox{ERE photons}}{\mbox{H atoms}} & = & 4 \pi A
  \Delta\lambda_{\rm eq} \left( \frac{hc}{\lambda_{\rm eq}}
  \right)^{-1} \nonumber \\ & = & 5.65 \times
  10^{-14}~\frac{\mbox{photons}}{\mbox{s}~\mbox{H atom}}
\end{eqnarray}
where
\[
A = 1.43\times 10^{-29}~\frac{\mbox{ergs}}{\mbox{s \AA\ sr
  H atom}}, 
\]
$\Delta\lambda_{\rm eq} = 968$~\AA, and $\lambda_{\rm eq} = 6441$~\AA\
(Table~\ref{table_mag_details}).  The number of ERE photons per unit
$\tau_V$ is then
\begin{eqnarray}
\frac{\mbox{ERE photons}}{\tau_V} & = & \frac{\mbox{ERE
  photons}}{\mbox{H atoms}} \frac{N_{HI}}{E(\bv)}
  \frac{E(\bv)}{A_V} 1.086~\frac{A_V}{\tau_V} \nonumber \\ & = & 9.95
  \times10^7~\frac{\mbox{photons}}{\mbox{s}~\mbox{cm}^2~\tau_V}
\end{eqnarray}
where $E(\bv)/A_V = R_V^{-1} = 3.05^{-1}$ (\cite{whi92}) and
$N_{HI}/E(\bv) = 4.93 \times 10^{21}~\mbox{cm}^{-2}~\mbox{mag}^{-1}$
(\cite{dip94}).

  The other half of this calculation was to determine the number of
photons absorbed per unit $\tau_V$.  The photons absorbed per unit
$\tau_V$ was calculated from
\begin{eqnarray}
\label{eqn_abs_photons}
\frac{\mbox{abs. photons}}{\tau_V} & = &
   \int_{912~\mbox{{\scriptsize\AA}}}^{5500~\mbox{{\scriptsize\AA}}}
   \left( 1 - a_{\lambda} \right) \frac{\left( 1 - e^{-\tau_{\lambda}}
   \right)}{\tau_V} I_{ISRF} d\lambda \nonumber \\ & \approx &
   \int_{912~\mbox{{\scriptsize\AA}}}^{5500~\mbox{{\scriptsize\AA}}}
   \left( 1 - a_{\lambda} \right) \frac{\tau_{\lambda}}{\tau_V}
   I_{ISRF} d\lambda \qquad\qquad \left( \tau_{\lambda} \ll 1 \right)
   \nonumber \\ & = & 7.41 \times
   10^8~\frac{\mbox{photons}}{\mbox{s}~\mbox{cm}^2~\tau_V}
\end{eqnarray}
where $a_{\lambda}$ is the wavelength dependent dust albedo
(\cite{gor94} and references therein), $\tau_{\lambda}$ is the
wavelength dependent dust optical depth for $R_V = 3.05$
(\cite{car89}), and $I_{ISRF}$ is the intensity of the interstellar
radiation field at the Earth in units of ergs cm$^{-2}$ s$^{-1}$
\AA$^{-1}$ (\cite{wit73}; \cite{mat83}).  This result is valid in the
optically thin limit ($\tau_{\lambda} \ll 1$) which is valid for
the diffuse ISM far from the galactic plane.  Using the results
calculated above for the number of photons absorbed and emitted, the
ERE photon efficiency in the diffuse ISM was 14\%.  This
corresponds to an {\em energy} efficiency of 5\%.

  This calculation is based on two uncertain assumptions.  The first
was the strength of the interstellar radiation field, $I_{ISRF}$
(\cite{mat83}).  This $I_{ISRF}$ is likely to be too low as the Mathis
\etal (1983) $I_{ISRF}$ was calculated for a smooth diffuse ISM,
whereas the diffuse ISM is known to be clumpy (\cite{elm96}) and the
radiative transfer in a clumpy medium is greatly different from that
in a smooth medium (\cite{wit96}).  Additional discussion of the
appropriate $I_{ISRF}$ can be found in Mathis (1997).  

\begin{figure}[tbp]
\begin{center}
\plotone{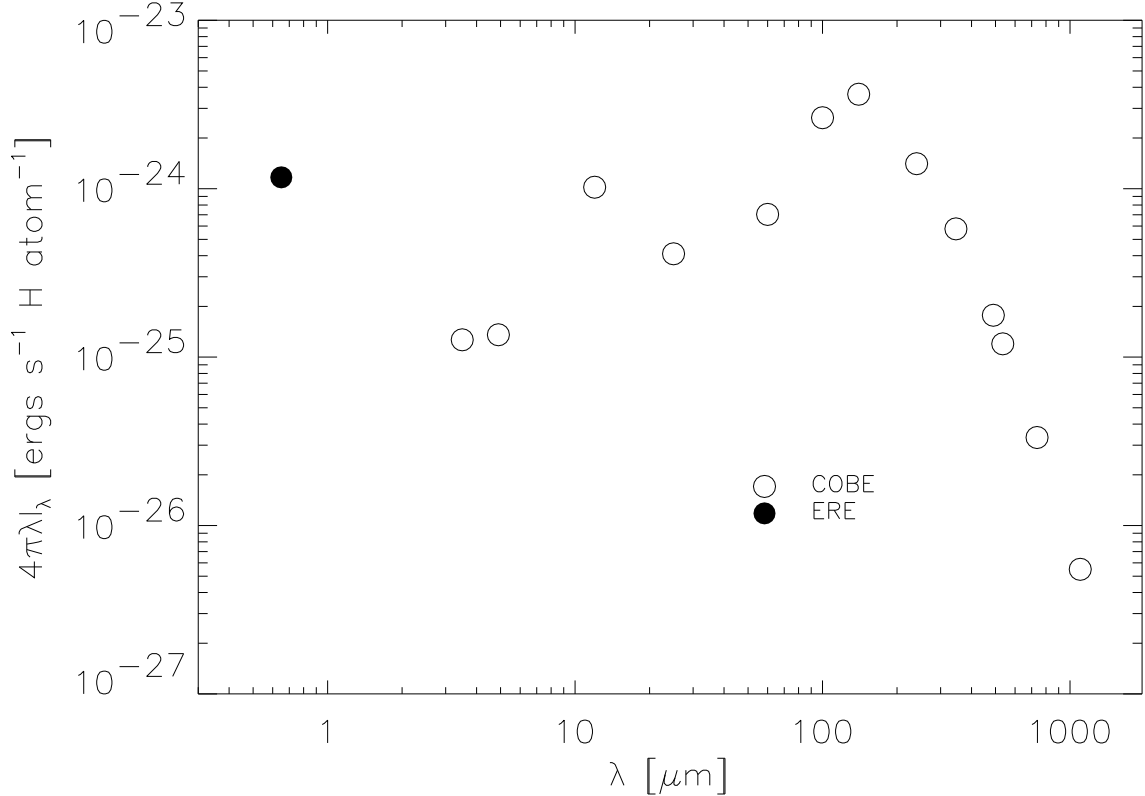}
\caption{The energy emitted per H atom by the diffuse ISM is plotted.
The ERE point is from this work.  The COBE points are from the work of
Arendt \etal (1997) and Boulanger \etal (1996).
\label{fig_diffuse_ISM}}
\end{center}
\end{figure}
  
  We determined the actual level of the $I_{ISRF}$ by equating the
energy absorbed by dust to that emitted by dust.  Except for ERE, the
energy emitted by diffuse ISM dust is emitted in the infrared and the
Cosmic Background Explorer (COBE, \cite{bog92}) contained the
Far-Infrared Absolute Spectrophotometer (FIRAS) and DIRBE which
measured the infrared emission from dust.  Using data from high
latitude regions, the diffuse ISM spectrum per H~atom from 3.3 to
240~$\micron$ was derived by Arendt \etal (1997, as presented in Dwek
\etal [1997]) and from 100~\micron to 1 mm by Boulanger \etal (1996).
In Figure~\ref{fig_diffuse_ISM}, we have plotted the spectrum of the
diffuse ISM including the contribution from ERE as determined in this
work.  Integrating this spectrum gives the emitted energy which is
5.77\sn{-24} ergs s$^{-1}$ H~atom$^{-1}$.  The ERE contributes a quite
significant 3\% of the total energy emitted by dust.  In order to get
the same amount of energy absorbed by dust between 912 \AA\ and 1.2
\micron, the Mathis \etal (1993) $I_{ISRF}$ must be multiplied by
1.33.  Using this elevated $I_{ISRF}$ and given the uncertainties of
all the quantities contributing to this result, we calculated the
lower limit ERE photon efficiency to be $10 \pm 3$\% and the
corresponding energy efficiency to be $4 \pm 1$\%.

  The second assumption was that {\em all} the photons absorbed by
dust were absorbed by the material which produces ERE.  This is
unlikely to be true.  Reducing the number of photons absorbed by the
ERE producing material would raise the ERE efficiency.  Assuming a
100\% conversion efficiency for the ERE producing material would
require that 10\% of the photons absorbed by the dust are converted to
ERE photons.  This makes the ERE producing material an important
component of dust grains.

  Instead of a wavelength independent ERE conversion efficiency,
perhaps it is a question of a high efficiency in a narrow range of
wavelengths.  Such behavior could be caused by a strong dust
absorption feature such as the 2175 \AA\ extinction bump.  The
percentage of the photons absorbed by the 2175 \AA\ bump material as
compared the total number of photons absorbed by dust, was calculated
as the difference between the number of photons absorbed for dust
between 912 \AA\ and 5500 \AA\ with and without a 2175 \AA\ bump .
The number of photons absorbed by bumpless dust was calculated using
equation~\ref{eqn_abs_photons}, a $R_V = 3.05$ $\tau_{\lambda}$ curve
from Cardelli \etal (1989) with the bump term removed, and
$a_{\lambda} = 0.6$.  We calculated that 13\% of the photons absorbed
by dust are absorbed by the 2175 \AA\ absorption feature.  This
percentage is higher than the 10\% ERE efficiency and, thus, there are
enough photons absorbed by the 2175 \AA\ bump material to produce the
observed ERE intensity.  We are not claiming the 2175 \AA\ bump is the
source of the ERE photons, only that this is the kind of absorption
feature needed if the ERE photons are absorbed in a narrow range of
wavelengths.  It is intriguing that a known absorption feature, which
has never been definitively identified with a particular material,
could be the source of the ERE photons.

\section{Discussion and Conclusions \label{sec_discussion}}

  The detection of ERE in the diffuse ISM has direct consequences for
all dust grain models.  With this work, ERE has been detected from
\ion{H}{2} regions to the diffuse ISM effectively spanning the entire
range of environments where dust is present.  Thus, ERE is a
characteristic of dust in general.  None of the current dust grain
models (\eg \cite{kim96}; \cite{mat96}; \cite{zub96}; \cite{dwe97};
\cite{li97}) specifically includes a material which can produce ERE.
As ERE has been detected only in carbon rich planetary nebulae
(\cite{fur92}), this adds ERE to the already long list of dust
features attributed to carbonaceous materials.  This further deepens
the current ``carbon crisis'' (\cite{sno95}, 1996) -- the conflict
between the small amount of carbon available for dust grains and the
large number of observed dust features attributed to carbonaceous
materials.  As the identification of other dust features with
carbonaceous materials is not as strong as that of ERE, the connection
of some dust features with carbonaceous materials needs to be
reevaluated.

\subsection{Anomalous Dust Scattering Properties instead of ERE?}

  A possible (but highly unlikely) explanation of the red excess
observed in the diffuse ISM is anomalous dust scattering properties in
the red.  If the dust albedo increased to unity (perfectly scattering
dust grains) around 6500 \AA, it would almost be possible to explain
the strength of the red excess, because the red excess is
approximately equal to the scattered intensity which was computed for
albedos between 0.5 and 0.7.  In fact, this argument was presented by
Greenstein \& Oke (1977) in their attempt to explain the excess red
intensity of the Red Rectangle.  They concluded that only with most
favorable geometry and a high albedo could they explain the excess.

  A scattering explanation has been effectively ruled out by means of
spectropolarimetric results, which show that the polarization decreases
where ERE increases.  For example, this explanation for the Red
Rectangle was convincingly shown to be false by the spectropolarimetry
of Schmidt \etal (1980).  Schmidt \etal (1980) observed a reduction in
the polarization across the red excess feature which proved that the
excess was due to emission and not scattering.  The emission nature of
the red excess in reflection nebulae has also been confirmed by the
imaging polarimetry of NGC~7023 by Watkin \etal (1991), who found that
the polarization decreased in locations were the ERE was the
strongest.  The two polarimetric observations discussed above are the
only two polarimetric observations which have been done on objects
with ERE.  This suggests that the red albedo is approximately equal
to the blue albedo as derived by Witt \etal (1990) for a Bok globule
without ERE and calculated from dust grain models (\cite{kim96};
\cite{mat96}; \cite{zub96}; \cite{li97}).  Thus, anomalous red dust
scattering properties are extremely unlikely to account for the red
excess in the diffuse ISM.

\subsection{$I_{ERE}/I_{SCA}$ and ERE efficiency}

  The photon efficiency calculated above for the ERE in the diffuse
ISM is 10\%.  The question arises: Is this efficiency
characteristic of ERE in general or only ERE in the diffuse ISM?  It
should be possible to calculate the ERE efficiency knowing the
scattered intensity ($I_{SCA}$), the ERE intensity, and the type of
illuminating radiation field.  From the scattered intensity and the
type of radiation field, the number of photons absorbed by dust in the
UV and blue should be calculable.  In practice, the relationship
between the scattered and absorbed intensity is quite complicated due
to the non-isotropic nature of scattering by dust grains (see
\S\ref{sec_dgl_model}).

\begin{figure}[tbp]
\begin{center}
\plotone{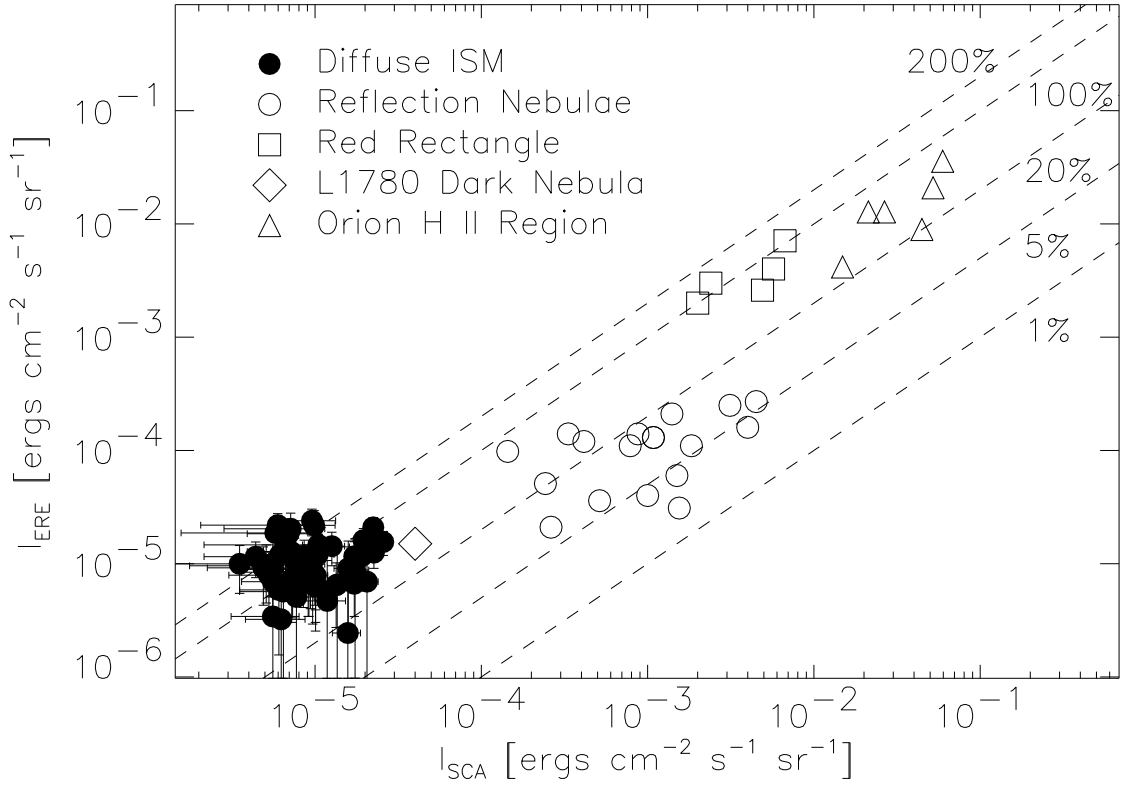}
\caption{This plot shows the ERE versus scattered intensities for
various objects were the ERE intensity has been measured.  The diffuse
ISM points are from this work.  The L1780 dark cloud point is from
Mattila (1979).  The reflection nebulae and the Red Rectangle points
are from Witt \& Boroson (1990).  The Orion H~II region points
were taken from Figures 3 \& 4 of Perrin \& Sivan (1992).  The
dashed lines show $I_{ERE}/I_{SCA}$ ratios of 1\%, 5\%, 20\%, 100\%,
and 200\% and are labeled appropriately.  \label{fig_sca_ere}}
\end{center}
\end{figure}
   
  Figure~\ref{fig_sca_ere} plots the ERE intensity versus the scattered
intensity for all objects with ERE and scattered intensities available
in the literature (\cite{mat79}; \cite{wit90}; \cite{per92}).  This
figure is similar to Figure~1 in Witt \& Boroson (1990) but with a much
larger range in $I_{ERE}$ and $I_{SCA}$ values due to the inclusion of
Orion \ion{H}{2} region points, at the high end, and diffuse ISM
points, at the low end.  The $I_{ERE}/I_{SCA}$ ratio ranges from 0.01 to
over 2.0.  The Red Rectangle, which was long thought to be unique in
its ERE intensity, is seen to have normal $I_{ERE}/I_{SCA}$ ratios.

  Interpreting the scatter in the $I_{ERE}/I_{SCA}$ ratio involves
detailed knowledge of the geometry and type of illuminating radiation
field in each object.  The dust scattering geometry affects
$I_{ERE}/I_{SCA}$ such that the larger the angle of scattering by dust
grains, the lower $I_{SCA}$ for a given radiation field, and the
larger the $I_{ERE}/I_{SCA}$ value.  Essentially, the scattered
intensity is reduced due to the forward scattering nature of dust
grains, while the ERE intensity is unaffected as the dust emits ERE
isotropically.  The illuminating radiation field affects
$I_{ERE}/I_{SCA}$ since the $I_{ERE}$ intensity is proportional to the
radiation field between 912 and 5500 \AA\ and $I_{SCA}$ is
proportional to the number of photons between 5500 and 8000 \AA.
Thus, the bluer the illuminating radiation field, the larger the
$I_{ERE}/I_{SCA}$.
 
\begin{figure}[tbp]
\begin{center}
\plotone{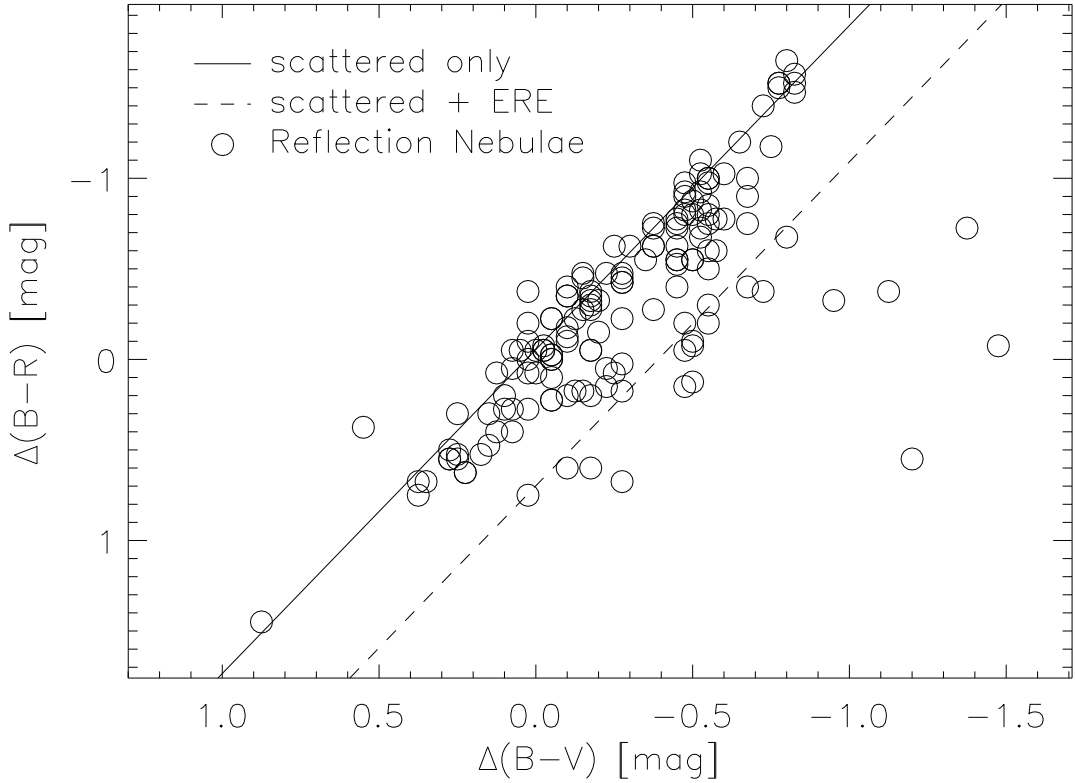}
\caption{The $\Delta$(\bv) and $\Delta$(\br) colors of reflection
nebulae observed by Witt \& Schild (1986) are plotted.  The
$\Delta$(\bv) color is the nebular (\bv) color minus the 
illuminating star's (\bv).  The $\Delta$(\br) color is defined
similarly.  The solid line shows the expected relationship between
$\Delta$(\bv) and $\Delta$(\br) for scattering assuming the dust
albedo is equal at B, V, and R wavelengths.  The dashed line gives the
relationship for $I_{ERE}/I_{SCA} = 1$.
\label{fig_bv_br}}
\end{center}
\end{figure}
  
  The lack of low values of $I_{ERE}/I_{SCA}$ in the diffuse ISM
(0.05--2.0) is likely due to the fact that the diffuse ISM
observations were all at high galactic latitudes where the dust
predominately scatters photons from the Galactic disk at large angles.
Similarly, the high value of $I_{ERE}/I_{SCA}$ (0.3) in the dark
nebula L1780 is due to its location high above the Galactic disk ($b
\approx 36\arcdeg$) and resulting large scattering angle.  The bipolar
geometry of the Red Rectangle, oriented perpendicular to our
line-of-sight, leads to predominately large angle scattering, again
giving large $I_{ERE}/I_{SCA}$ values (0.5--1.5).  The range in
$I_{ERE}/I_{SCA}$ (0.01--0.68) in reflection nebulae is mostly due to
the geometry and this is superbly illustrated as the highest
reflection nebula $I_{ERE}/I_{SCA}$ value (0.68) is for IC~63, an
externally illuminated nebula which scatters photons at $\approx
90\arcdeg$ (\cite{wit89b}; \cite{gor97}).  The real range of
$I_{ERE}/I_{SCA}$ in reflection nebulae is larger than that
represented by the spectroscopic detections of ERE shown in
Figure~\ref{fig_sca_ere}.  Figure~\ref{fig_bv_br} plots the (\bv) and
(\vr) colors observed in 14 reflection nebulae by Witt \& Schild
(1986).  As can be seen from the location of the dashed line, the
maximum $I_{ERE}/I_{SCA}$ is at least 1.0.  On the other hand, the
Orion \ion{H}{2} region has high values of $I_{ERE}/I_{SCA}$
(0.2--0.6) due to the extreme blueness of the illuminating radiation
field which originates from the hot Trapezium stars (late O, early B
spectral types).  Thus, the scatter in $I_{ERE}/I_{SCA}$ could easily
be explained by geometry and radiation field effects with a constant
ERE efficiency of 10\%.  The test of this hypothesis
awaits a detailed calculation of the ERE efficiency in each object
properly accounting for the effects of geometry and illuminating
radiation field.

\subsection{Conclusions}

  Using blue and red all-sky measurements taken by Pioneer 10 and 11
outside the zodiacal dust cloud along with star and galaxy counts to
20th magnitude, we determined the blue and red intensity of the
diffuse ISM in two large regions with areas of $315~\sq\arcdeg$ and
$820~\sq\arcdeg$.  By comparison with a model for the DGL, the blue
diffuse ISM intensity was found to be entirely attributable to DGL.
The diffuse ISM red intensity was found to consist of DGL and ERE in
roughly equal parts.  Thus, the ERE is detected in the diffuse ISM and
shown to be a general characteristic of dust in all dusty
environments.  The ERE in the diffuse ISM is consistent with
photoluminescence with an photon efficiency of $10 \pm 3$\% (lower
limit) which corresponds to a $4 \pm 1$\% energy efficiency.

  We plan to expand on this work by expanding the fraction of the sky
studied.  The APS Catalog is scheduled for completion by the end of
1997 (\cite{hum97}) and will cover $\sim$50\% of the sky.  We
anticipate using the SKY model (\cite{coh94}, 1995) to derive the
faint star counts in regions not covered by the APS catalog after
calibrating the SKY model in those regions where the APS catalog
exists.  This will allow us to investigate the distribution of ERE and
DGL over almost the entire sky.  The all-sky distribution of ERE will
allow us to determine the global ERE photon efficiency as well as any
variations as a function of position in the sky.  The all-sky
distribution of the DGL will allow us to quantify the variations in
the WP model inputs (dust grain albedo and $g$, $\left< N_{HI}/E(\bv)
\right>$, and radiation field) on a Galaxy-wide scale.  In addition,
the comparison between the SKY model and the APS Catalog will yield
valuable information about Galactic structure.

\acknowledgements

  This paper was part of K.\ D.\ Gordon's PhD thesis and, so, thanks
to the thesis committee members Song Cheng, Al Compaan, Steve Federman,
Nancy Morrison, Gary Toller, and Adolf Witt, especially the last
three.  Thanks to Gary Toller and Jerry Weinberg for their help
understanding how to use the Pioneer measurements.  Additional thanks
to Gary Toller for providing the DIRBE maps.  Thanks to Chris
Cornuelle, Jeff Larson, and Roberta Humphreys for providing lots of
help and information about the APS Catalog of the POSS I.  The Pioneer
10 and 11 IPP data were provided by J.\ L.\ Weinberg (Principal
Investigator) and the National Space Science Data Center (NSSDC).
This research has made use of the APS Catalog of the POSS I, which is
supported by the National Science Foundation, the National Aeronautics
and Space Administration, and the University of Minnesota.  The APS
database can be accessed at {\em http://isis.spa.umn.edu/}.  This
research was financially supported under NASA LTSAP grants NAGW-3168 \&
NAG5-3367 to The University of Toledo.  The COBE datasets were
developed by the NASA Goddard Space Flight Center under the guidance
of the COBE Science Working Group and were provided by NSSDC.

\end{document}